# Astrometry and optics during the past 2000 years

Erik Høg     Niels Bohr Institute, Copenhagen, Denmark

2011.05.03:  Collection of reports from November 2008

ABSTRACT:  The satellite missions Hipparcos and Gaia by the European Space Agency will together bring a decrease of astrometric errors by a factor 10000, four orders of magnitude, more than was achieved during the preceding 500 years. This modern development of astrometry was at first obtained by photoelectric astrometry. An experiment with this technique in 1925 led to the Hipparcos satellite mission in the years 1989-93 as described in the following reports Nos. 1 and 10. The report No. 11 is about the subsequent period of space astrometry with CCDs in a scanning satellite. This period began in 1992 with my proposal of a mission called Roemer, which led to the Gaia mission due for launch in 2013. My contributions to the history of astrometry and optics are based on 50 years of work in the field of astrometry but the reports cover spans of time within the past 2000 years, e.g., 400 years of astrometry, 650 years of optics, and the "miraculous" approval of the Hipparcos satellite mission during a few months of 1980.

2011.05.03:  Collection of reports from November 2008. The following contains overview with summary and link to the reports Nos. 1-9 from 2008 and Nos. 10-13 from 2011. The reports are collected in two big file, see details on p.8.

## CONTENTS of Nos. 1-9 from 2008

# CONTENTS of Nos. 10-13 from 2011

# Overview with links to Nos. 1-9

No. 1 -  2008.05.27:

## Bengt Strömgren and modern astrometry:
## Development of photoelectric astrometry
## including the Hipparcos mission

ABSTRACT:  Bengt Strömgren is known as the famous astrophysicist and as a leading figure in many astronomical enterprises. Less well-known, perhaps, is his role in modern astrometry although this is equally significant. There is an unbroken chain of actions from his ideas and experiments with photoelectric astrometry since 1925 over the new meridian circle in Denmark in the 1950s up to the Hipparcos and Tycho Catalogues published in 1997.

www.astro.ku.dk/~erik/Stroemgren.pdf

Contribution to IAU Symposium No. 254 in Copenhagen, June 2008: The Galaxy Disk in Cosmological Context – Dedicated to Professor Bengt Strömgren (1908-1987).

No. 1A - 2008.06.10:

## Bengt Strömgren and modern astrometry ... (Short version)

www.astro.ku.dk/~erik/StroemgrenShort.pdf

The same title as No. 1, but containing the short version posted at the symposium.

No. 2 - 2008.03.31:

## Lennart Lindegren's first years with Hipparcos

ABSTRACT: Lennart Lindegren has played a crucial role in the Hipparcos project ever since he entered the scene of space astrometry in September 1976. This is an account of what I saw during Lennart's first years in astrometry after I met him in 1976  when he was a young student in Lund.

www.astro.ku.dk/~erik/Lindegren.pdf

No. 3 – 2008.05.28:

## Miraculous approval of Hipparcos in 1980

ABSTRACT: The approval of the Hipparcos mission in 1980 was far from being smooth since very serious hurdles were encountered in the ESA committees. This process is illuminated here by means of documents from the time and by recent correspondence. The evidence leads to conclude that in case the approval would have failed, Hipparcos or a similar scanning astrometry mission would never have been realized, neither in Europe nor anywhere else.
www.astro.ku.dk/~erik/HipApproval.pdf

No. 4 - 2007.12.10:

## From the Roemer mission to Gaia

ABSTRACT: At the astrometry symposium in Shanghai 1992 the present author made the first proposal for a specific mission concept post-Hipparcos, the first scanning astrometry mission with CCDs in time-delayed integration mode (TDI). Direct imaging on CCDs in long-focus telescopes was described as later adopted for the Gaia mission. The mission called Roemer was designed to provide accurate astrometry and multi-colour photometry of 400 million stars brighter than 18 mag in a five-year mission. The early years of this mission concept are reviewed.
www.astro.ku.dk/~erik/ShanghaiPoster.pdf
Presented as poster at IAU Symposium No. 248 in Shanghai, October 2007. Only the first three pages appear in the Proceedings.

No. 5 - 2008.05.23, updated 2008.11.25.
Note in 2011: See further update in **www.astro.ku.dk/~erik/History2**

# Four lectures on the general history of astrometry

Overview, handout, abstracts at:  www.astro.ku.dk/~erik/Lectures.pdf
**Brief overview :**
  Lecture No. 1:

### Astrometry and photometry from space: Hipparcos, Tycho, Gaia
 The introduction covers 2000 years of astronomy from Ptolemy to modern times. The Hipparcos mission of the European Space Agency was launched in 1989, including the Tycho experiment. The Hipparcos mission and the even more powerful Gaia mission to be launched in 2011 are described.

 Lecture No. 2:

### From punched cards to satellites: Hipparcos, Tycho, Gaia
 A personal review of 54 years development of astrometry in which I participated.

 Lecture No. 3:

### The Depth of Heavens - Belief and Knowledge during 2500 Years
 The lecture outlines the understanding of the structure of the universe and the development of science during 5000 years, focusing on the concept of distances in the universe and its dramatic change in the developing cultural environment from Babylon and ancient Greece to modern Europe.

 Lecture No. 4, included on 2008.11.25:

### 400 Years of Astrometry: From Tycho Brahe to Hipparcos
 Four centuries of techniques and results are reviewed, from the pre-telescopic era up to the use of photoelectric astrometry and space technology in the first astrometric satellite, Hipparcos, launched by ESA in 1989. The lecture was presented as invited contribution to the symposium at ESTEC in September 2008: **400 Years of Astronomical Telescopes: A Review of History, Science and Technology.** The report submitted to the proceedings is included as No. 8 among "Contributions to the history of astrometry".

No. 6 – 2008.11.25:

# Selected astrometric catalogues

ABSTRACT: A selection of astrometric catalogues are presented in three tables for respectively positions, proper motions and trigonometric parallaxes. The tables contain characteristics of each catalogue to show especially the evolution over the last 400 years in optical astrometry. The number of stars and the accuracy are summarized by the weight of a catalogue, proportional with the number of stars and the statistical weight.
www.astro.ku.dk/~erik/AstrometricCats.pdf

No. 7 – 2008.11.25:

# Astrometric accuracy during the past 2000 years

ABSTRACT: The development of astrometric accuracy since the observations by Hipparchus, about 150 B.C., has been tremendous and the evolution has often been displayed in a diagram of accuracy versus time. Some of these diagrams are shown and the quite significant differences are discussed. A new diagram is recommended and documented.
www.astro.ku.dk/~erik/Accuracy.pdf
The two diagrams, Fig. 1a and 1b, in black/white and colour :
www.astro.ku.dk/~erik/AccurBasic.pdf          www.astro.ku.dk/~erik/AccuracyColour.jpg
www.astro.ku.dk/~erik/AccuracyBW.wmf          www.astro.ku.dk/~erik/AccuracyColour.wmf

No. 8 -  2008.11.25:

# 400 Years of Astrometry: From Tycho Brahe to Hipparcos

ABSTRACT: Galileo Galilei's use of the newly invented telescope for astronomical observation resulted immediately in epochal discoveries about the physical nature of celestial bodies, but the advantage for astrometry came much later. The quadrant and sextant were pre-telescopic instruments for measurement of large angles between stars, improved by Tycho Brahe in the years 1570-1590. Fitted with telescopic sights after 1660, such instruments were quite successful, especially in the hands of John Flamsteed. The meridian circle was a new type of astrometric instrument, already invented and used by Ole Rømer in about 1705, but it took a hundred years before it could fully take over. The centuries-long evolution of techniques is reviewed, including the use of photoelectric astrometry and space technology in the first astrometry satellite, Hipparcos, launched by ESA in 1989. Hipparcos made accurate measurement of large angles a million times more efficiently than could be done in about 1950 from the ground, and it will soon be followed by Gaia which is expected to be another one million times more efficient for optical astrometry.
www.astro.ku.dk/~erik/Astrometry400.pdf
Invited contribution to the symposium in Leiden in October 2008:
**400 Years of Astronomical Telescopes: A Review of History, Science and Technology**

No. 9 -  2008.11.25:

# 650 Years of Optics: From Alhazen to Fermat and Rømer

ABSTRACT: Under house arrest in Cairo from 1010 to 1021, Alhazen wrote his Book of Optics in seven volumes. (The caliph al-Hakim had condemned him for madness.) Some parts of the book came to Europe about 1200, were translated into Latin, and had great impact on the development of European science in the following centuries. Alhazen's book was considered the most important book on optics until Johannes Kepler's "Astronomiae Pars Optica" in 1604. Alhazen's idea about a finite speed of light led to "Fermat's principle" in 1657, the foundation of geometrical optics.

www.astro.ku.dk/~erik/HoegAlhazen.pdf

Contribution to the symposium in Leiden in September 2008:

**400 Years of Astronomical Telescopes: A Review of History, Science and Technology**

# Overview with links to Nos. 10-13

No. 3.2 – 2011.01.27, update from a version of 2008.05.27:

## Miraculous approval of Hipparcos in 1980: (2)

ABSTRACT: The approval of the Hipparcos mission in 1980 was far from being smooth since very serious hurdles were encountered in the ESA committees. This process is illuminated here by means of documents from the time and by recent correspondence. The evidence leads to conclude that in case the approval would have failed, Hipparcos or a similar scanning astrometry mission would never have been realized, neither in Europe nor anywhere else.

www.astro.ku.dk/~erik/HipApproval.pdf

No. 10 - 2011.03.26:

# Astrometry Lost and Regained

## From a modest experiment in Copenhagen in 1925 to the Hipparcos and Gaia space missions

ABSTRACT: Technological and scientific developments during the past century made a new branch of astronomy flourish, i.e. astrophysics, and resulted in our present deep understanding of the whole Universe. But this brought astrometry almost to extinction because it was considered to be dull and old-fashioned, especially by young astronomers. Astrometry is the much older branch of astronomy which performs accurate measurements of positions, motions and distances of stars and other celestial bodies. Astrometric data are of great scientific and practical importance for investigation of celestial phenomena and also for control of telescopes and satellites and for monitoring of Earth rotation. Our main subject is the development during the 20[th] century which finally made astrometry flourish as an integral part of astronomy through the success of the Hipparcos astrometric satellite, soon to be followed by the even more powerful Gaia mission.

www.astro.ku.dk/~erik/AstromRega3.pdf

No. 11 - 2011.04.06:

# Roemer and Gaia

ABSTRACT: During the Hipparcos mission in September 1992, I presented a concept for using direct imaging on CCDs in scanning mode in a new and very powerful astrometric satellite, Roemer. The Roemer concept with larger aperture telescopes for higher accuracy was developed by ESA and a mission was approved in 2000, expected to be a million times better than Hipparcos. The present name Gaia for the mission reminds of an interferometric option also studied in the period 1993-97, and the evolution of optics and detection in this period is the main subject of the present report. The transition from an interferometric GAIA to a large Roemer was made on 15 January 1998. It will be shown that without the interferometric GAIA option, ESA would hardly have selected astrometry for a Cornerstone study in 1997, and consequently we would not have had the Roemer/Gaia mission.
www.astro.ku.dk/~erik/RoemerGaia.pdf

No. 12 - 2011.01.15:   On the website of the Niels Bohr Institute:

# Surveying the sky

"An astrometric experiment in 1925 was the beginning of a development which Erik Høg, Associate Professor Emeritus, took part in for 50 years. A scientific highlight is the star catalogue Tycho-2 from the year 2000, which describes the positions and movements of 2.5 million stars and is now absolutely essential to controlling satellites and for astronomical observations."

In English:  http://www.nbi.ku.dk/english/www/   and   in Danish:  http://www.nbi.ku.dk/hhh/

and

# En landmåler i himlen

In Danish: En artikel i tidsskriftet KVANT, oktober 2010, om 50 års arbejde

Erindringer om 50 år med astrometrien, der begyndte ved en høstak syd for Holbæk og førte til bygning af to satellitter. Et videnskabeligt højdepunkt er stjernekataloget Tycho-2, der nu er helt uundværligt ved styring af satellitter og ved astronomiske observationer.
www.astro.ku.dk/~erik/kv-2010-3-EH-astrometri.pdf

No. 13 - 2011.03.26:

# Lectures on astrometry

Overview, handout, abstracts at:  www.astro.ku.dk/~erik/Lectures2.pdf
# Brief overview :

**Lecture No. 1.**  45 minutes
 Astrometry Lost and Regained
   **From a modest experiment in Copenhagen in 1925**
   **to the Hipparcos and Gaia space missions**
  The lecture has been developed over many years and was held in, e.g., Copenhagen, Vienna, Bonn, Düsseldorf, Vilnius, Oslo, Nikolaev, Poltava, Kiev, Thessaloniki, Ioannina, Athens, Rome, Madrid, Washington, and Charlottesville - since 2007 in PowerPoint.  Revised in 2009 and with the new title

*Astrometry Lost and Regained* it was held in Heidelberg, Sct. Petersburg, Rio de Janeiro, Morelia, Mexico City, Beijing, Montpellier, Groningen, Amsterdam, and Leiden.

---

**Lecture No. 2.**  45 minutes
**Hipparcos - Roemer - Gaia**

   **The lectures briefly outlines the development of photoelectric astrometry culminating with the Hipparcos mission. Development of the Gaia mission beginning in 1992 is followed in detail.**

   The lecture has been held since 2010 in Toulouse and at ESTEC in Holland.

---

**Lecture No. 3.**  45 minutes.  Suited for a broad audience, including non-astronomers
**The Depth of Heavens - Belief and Knowledge during 2500 Years**

   **The lecture outlines the structure of the universe and the development of science during 5000 years, focusing on the distances in the universe and their dramatic change in the developing cultural environment from Babylon and ancient Greece to modern Europe.**

   The lecture was first held in 2002, and since 2007 in PowerPoint. Held in Copenhagen, Vilnius, Nikolaev, Athens, Catania, Madrid, and Paris
       Handouts at:  www.astro.ku.dk/~erik/DepthHeavens2.pdf
   and   www.astro.ku.dk/~erik/DepthHeavens.pdf

   **An article with the same title as the lecture** appeared in Europhysics News (2004) Vol. 35 No.3.
   Here slightly updated, 2004.02.20:  www.astro.ku.dk/~erik/Univ7.5.pdf

---

**Lecture No. 4.**  45 or 30 minutes.
**400 Years of Astrometry: From Tycho Brahe to Hipparcos**

   **The four centuries of techniques and results are reviewed, from the pre-telescopic era until the use of photoelectric astrometry and space technology in the first astrometry satellite, Hipparcos, launched by ESA in 1989.**

   The lecture was presented as invited contribution to the symposium at ESTEC in September 2008: **400 Years of Astronomical Telescopes: A Review of History, Science and Technology.** The report to the proceedings is included as No. 8 among the "Contributions to the history of astrometry ".

++++++++++++++++++++++++++++++++++++++++++++++++++++++++++++++++++++

Further installments in preparation:  On the Hipparcos mission studies 1975-79 and on the Hipparcos archives.

*Best regards      Erik*          *http://www.astro.ku.dk/~erik*

# Reports from 2008 and 2011 on History of Astrometry:

Overview, summary and link to individual reports from 2008 and 2011 are placed in an index file: www.astro.ku.dk/~erik/erik-hoeg-history-of-astrometry-1104-index.pdf .

The two collections of reports are placed in two big files at the following links, including overview and summary pages:

The reports from 2008 are placed at arXiv and in a file printing on 8+94 pages: www.astro.ku.dk/~erik/HistoryAll.pdf   and the title is: "Astrometry and optics during the past 2000 years"

The reports from 2011 are placed at arXiv and in a file printing on 8+46 pages: www.astro.ku.dk/~erik/History2All.pdf   and the title is: "Astrometry during the past 100 years"

# Astrometry and optics during the past 2000 years

*Erik Høg    Niels Bohr Institute, Copenhagen*

2008.11.25:  (update of  version from June 2008)

ABSTRACT: Photoelectric astrometry began with experiments by Bengt Strömgren in 1925 and ended with the Hipparcos satellite mission in the years 1989-93. This period was discussed in reports placed on my website in June 2008, including my proposal in 1992 for CCD astrometry with a scanning satellite called Roemer, which led to the Gaia mission due for launch in 2011. A new lecture and four reports have now been added with overviews of astrometry and optics during the past 2000 years. -  Further installments are planned.

The *present big* file at www.astro.ku.dk/~erik/HistoryAll.pdf of 5 MB will print on 94 pages. It contains a table of contents, an overview with links, and all the reports.

The *short* file at  www.astro.ku.dk/~erik/History.pdf  of 4 pages contains a table of contents and an overview with links to the individual reports.

**Green colour indicates new or newly updated reports or new numbering, compared with June 2008.**

# CONTENTS

# Overview with links

No. 1 -  2008.05.27:
## Bengt Strömgren and modern astrometry:
## Development of photoelectric astrometry
## including the Hipparcos mission

ABSTRACT:  Bengt Strömgren is known as the famous astrophysicist and as a leading figure in many astronomical enterprises. Less well-known, perhaps, is his role in modern astrometry although this is equally significant. There is an unbroken chain of actions from his ideas and experiments with photoelectric astrometry since 1925 over the new meridian circle in Denmark in the 1950s up to the Hipparcos and Tycho Catalogues published in 1997.
www.astro.ku.dk/~erik/Stroemgren.pdf.
Contribution to IAU Symposium No. 254 in Copenhagen, June 2008: The Galaxy Disk in Cosmological Context – Dedicated to Professor Bengt Strömgren (1908-1987).

No. 1A - 2008.06.10:
## Bengt Strömgren and modern astrometry ... (Short version)
www.astro.ku.dk/~erik/StroemgrenShort.pdf.
The same title as No. 1, but containing the short version posted at the symposium.

No. 2 - 2008.03.31:
## Lennart Lindegren's first years with Hipparcos

ABSTRACT: Lennart Lindegren has played a crucial role in the Hipparcos project ever since he entered the scene of space astrometry in September 1976. This is an account of what I saw during Lennart's first years in astrometry after I met him in 1976  when he was a young student in Lund.
www.astro.ku.dk/~erik/Lindegren.pdf.

No. 3 – 2008.05.28:
## Miraculous approval of Hipparcos in 1980

ABSTRACT: The approval of the Hipparcos mission in 1980 was far from being smooth since very serious hurdles were encountered in the ESA committees. This process is illuminated here by means of documents from the time and by recent correspondence. The evidence leads to conclude that in case the approval would have failed, Hipparcos or a similar scanning astrometry mission would never have been realized, neither in Europe nor anywhere else.
www.astro.ku.dk/~erik/HipApproval.pdf

No. 4 -  2007.12.10:
## From the Roemer mission to Gaia

ABSTRACT: At the astrometry symposium in Shanghai 1992 the present author made the first proposal for a specific mission concept post-Hipparcos, the first scanning astrometry mission with CCDs in time-delayed integration mode (TDI). Direct imaging on CCDs in long-focus telescopes was described as later adopted for the Gaia mission. The mission called Roemer was designed to provide accurate astrometry and multi-colour photometry of 400 million stars brighter than 18 mag in a five-year mission. The early years of this mission concept are reviewed. www.astro.ku.dk/~erik/ShanghaiPoster.pdf.
Presented as poster at IAU Symposium No. 248 in Shanghai, October 2007. Only the first three pages appear in the Proceedings.

No. 5 - 2008.05.23, updated 2008.11.25:

# Four lectures on the general history of astrometry

Overview, handout, abstracts at:   www.astro.ku.dk/~erik/Lectures.pdf
**Brief overview :**
Lecture No. 1:

### Astrometry and photometry from space: Hipparcos, Tycho, Gaia

The introduction covers 2000 years of astronomy from Ptolemy to modern times. The Hipparcos mission of the European Space Agency was launched in 1989, including the Tycho experiment. The Hipparcos mission and the even more powerful Gaia mission to be launched in 2011 are described.

Lecture No. 2:

### From punched cards to satellites: Hipparcos, Tycho, Gaia

A personal review of 54 years development of astrometry in which I participated.

Lecture No. 3:

### The Depth of Heavens - Belief and Knowledge during 2500 Years

The lecture outlines the understanding of the structure of the universe and the development of science during 5000 years, focusing on the concept of distances in the universe and its dramatic change in the developing cultural environment from Babylon and ancient Greece to modern Europe.

Lecture No. 4, included on 2008.11.25:

### 400 Years of Astrometry: From Tycho Brahe to Hipparcos

Four centuries of techniques and results are reviewed, from the pre-telescopic era up to the use of photoelectric astrometry and space technology in the first astrometric satellite, Hipparcos, launched by ESA in 1989. The lecture was presented as invited contribution to the symposium at ESTEC in September 2008: **400 Years of Astronomical Telescopes: A Review of History, Science and Technology.** The report submitted to the proceedings is included as No. 8 among "Contributions to the history of astrometry".

No. 6 – 2008.11.25:

## Selected astrometric catalogues

ABSTRACT: A selection of astrometric catalogues are presented in three tables for respectively positions, proper motions and trigonometric parallaxes. The tables contain characteristics of each catalogue to show especially the evolution over the last 400 years in optical astrometry. The number of stars and the accuracy are summarized by the weight of a catalogue, proportional with the number of stars and the statistical weight.
www.astro.ku.dk/~erik/AstrometricCats

No. 7 – 2008.11.25:

## Astrometric accuracy during the past 2000 years

ABSTRACT: The development of astrometric accuracy since the observations by Hipparchus,

about 150 B.C., has been tremendous and the evolution has often been displayed in a diagram of accuracy versus time. Some of these diagrams are shown and the quite significant differences are discussed. A new diagram is recommended and documented.
www.astro.ku.dk/~erik/Accuracy.pdf.
The two diagrams, Fig. 1a and 1b, in black/white and colour :      AccurBasic.pdf
AccuracyBW.jpg,  AccuracyColour.jpg                AccuracyBW.wmf  AccuracyColour.wmf

No. 8 -  2008.11.25:

## 400 Years of Astrometry: From Tycho Brahe to Hipparcos

ABSTRACT: Galileo Galilei's use of the newly invented telescope for astronomical observation resulted immediately in epochal discoveries about the physical nature of celestial bodies, but the advantage for astrometry came much later. The quadrant and sextant were pre-telescopic instruments for measurement of large angles between stars, improved by Tycho Brahe in the years 1570-1590. Fitted with telescopic sights after 1660, such instruments were quite successful, especially in the hands of John Flamsteed. The meridian circle was a new type of astrometric instrument, already invented and used by Ole Rømer in about 1705, but it took a hundred years before it could fully take over. The centuries-long evolution of techniques is reviewed, including the use of photoelectric astrometry and space technology in the first astrometry satellite, Hipparcos, launched by ESA in 1989. Hipparcos made accurate measurement of large angles a million times more efficiently than could be done in about 1950 from the ground, and it will soon be followed by Gaia which is expected to be another one million times more efficient for optical astrometry.
www.astro.ku.dk/~erik/Astrometry400.pdf
Invited contribution to the symposium in Leiden in October 2008:
**400 Years of Astronomical Telescopes: A Review of History, Science and Technology**

No. 9 -  2008.11.25:

## 650 Years of Optics: From Alhazen to Fermat and Rømer

ABSTRACT: Under house arrest in Cairo from 1010 to 1021, Alhazen wrote his Book of Optics in seven volumes. (The caliph al-Hakim had condemned him for madness.) Some parts of the book came to Europe about 1200, were translated into Latin, and had great impact on the development of European science in the following centuries. Alhazen's book was considered the most important book on optics until Johannes Kepler's "Astronomiae Pars Optica" in 1604. Alhazen's idea about a finite speed of light led to "Fermat's principle" in 1657, the foundation of geometrical optics.
www.astro.ku.dk/~erik/HoegAlhazen.pdf.
Contribution to the symposium in Leiden in September 2008:
**400 Years of Astronomical Telescopes: A Review of History, Science and Technology**

Further installments in preparation:  On the Hipparcos mission studies 1975-79 and on the Hipparcos archives.

*Best regards*        *Erik*            *http://www.astro.ku.dk/~erik*

# Bengt Strömgren and modern astrometry
## *Development of photoelectric astrometry including the Hipparcos mission*

*Erik Høg, Niels Bohr Institute, Copenhagen*

***ABSTRACT****:  **Bengt Strömgren is known as the famous astrophysicist and as a leading figure in many astronomical enterprises. Less well-known, perhaps, is his role in modern astrometry although this is equally significant. There is an unbroken chain of actions from his ideas and experiments with photoelectric astrometry since 1925 over the new meridian circle in Denmark in the 1950s up to the Hipparcos and Tycho Catalogues published in 1997.***

**1. Introduction**
This account follows a chain of actions beginning with an experiment on photoelectric astrometry in 1925 and culminating with the Hipparcos mission at the end of the 20[th] century, thus, in 1925 began a new era of positional astronomy comparable in significance to that of Tycho Brahe four centuries earlier. This brief account is far from being a complete history of Hipparcos, nor, of course, of the many other developments of photoelectric astrometry in the same period.

**2. Strömgren's experiments with photoelectric astrometry**

Bengt Strömgren in 1925 at the age of 17 years reported about experiments with photoelectric recording of star transits. In the focal plane of the meridian circle in Copenhagen he had placed a system of slits parallel to the meridian, Fig. 1. Behind the slits was a photo cell which received the light from the star after it had passed the slits. As the star moved across the slits the variations of light intensity gave corresponding variations in the photo current, and these variations of current were amplified and recorded.

At the annual meeting in 1926 of the Astronomische Gesellschaft in Copenhagen, Strömgren reported again. Karl Friedrich Küstner, at the time a great veteran in meridian astronomy, then drew a line from his predecessor in Bonn, Friedrich Argelander, who mastered the "eye-and-ear" method, to the young Strömgren, who wanted to introduce photoelectric recording of transits. Küstner told that Argelander in his old days was once introduced to the then new chronograph. He held the pushbutton in his hand - but then put it down with a shake of the head (Nielsen 1962).

Strömgren, however, found a serious drawback of his initial method: For reasons of statistical noise, it would only allow recording of stars to 6[th] or 7[th] magnitude with a medium size meridian circle. He therefore proposed a method of integration (Strömgren 1933) which should allow observation of much fainter stars. A mirror was placed behind a system of equidistant slits. It was switched quickly at predetermined times between two positions. In one position the light hits one photo cell, in the other another photo cell, both of them able to integrate all the light reaching them. If the switching takes place when the star is near the centre of a slit and midway between them, the cells will integrate almost equal amounts of light, and the ratio between these amounts can be translated to an accurate transit time, provided the dimensions of the system are known.

The mirror and the whole operation of this "second generation" system posed technical problems and no further experiments have been reported. The author of these pages heard about the two proposals as a student and that bore fruit later on as I shall explain, but I probably never discussed them with Strömgren in those early days.

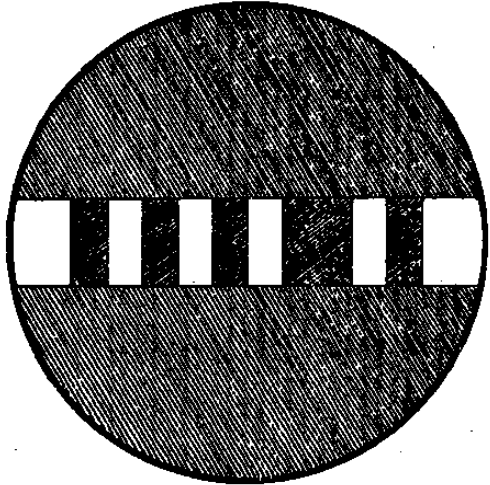

**Figure 1.** Slit system for photoelectric measurement of right ascension with a meridian circle (Strömgren 1925)

## 3. Developments in Denmark and Hamburg

In 1940 Bengt Strömgren succeeded his father, Elis, as director of the observatory, then located in the centre of Copenhagen. The same year he took the initiative to build a new observatory outside Copenhagen. A site in Brorfelde 50 km west of Copenhagen was selected. Ten years later the concrete for the foundation of the main instrument, a new meridian circle, was poured, and finally, in 1953 the instrument itself could be mounted. I then got the task as a fourth year student to test the stability of the new instrument by photographic observations of a star very close to the North pole. Strömgren did not intend to implement a photoelectric method at the new instrument, but preferred a method where a photographic plate is moved along with the star. This method was technically less challenging, and was in fact put into operation in Brorfelde in the 1960s.

Most important for me as a young astronomer, was to grow up in an observatory where a new meridian circle was the main instrument and where this course for the institute had been defined by an outstanding scientist. Bengt Strömgren gave everybody, not only a youngster as me, confidence about the future line of astronomy. How very different at most other places in the world where astrometry, the astronomy of positions, was being discarded as old-fashioned stuff. At such places I would probably have become an astrophysicist, since I certainly did not want to do old stuff.

In 1956 I finished my studies and became a conscript soldier. Most of the time I had the opportunity to work in a laboratory (Niels Finsen Institute) measuring radioactive decay of dust, collected to follow the nuclear weapon testing of the two superpowers. The measurements were obtained by radioactive counting techniques and my experience with this brand new technique I could later apply to photoelectric astrometry.

In 1958 I moved to the Hamburg Observatory where both astrometry and astrophysics were held in high esteem; Otto Heckmann was the powerful director. My interest went in direction of classification of stars by objective prism spectra obtained with the big Schmidt telescope. I built an electronic equipment for digital recording of spectra on punched cards, which was also used for digital recording at the iris photometer, used for photometric measurement of photographic plates (Høg 1959). But soon I got the best idea I ever had (Høg 1960): I realized that Strömgren's method with the switching mirror could be implemented very elegantly by a photon counting technique. I do not think any other astronomers used photon counting at that time; I had the idea from the counting of radioactive decay.

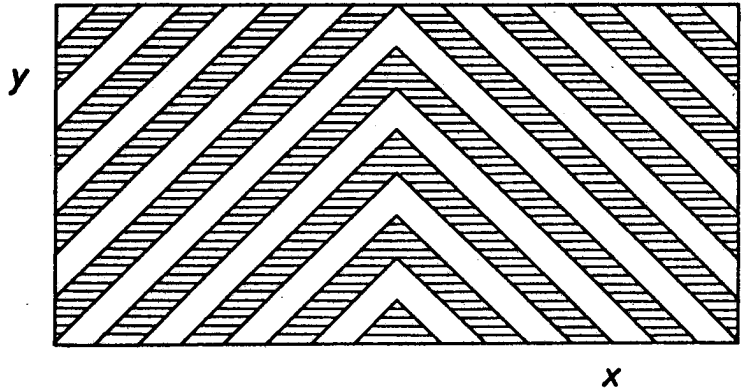

**Figure 2.** Slit system for two-dimensional photoelectric measurement in the focal plane of a meridian circle (Høg 1960)

Briefly, the technique was as follows: A photo multiplier tube is placed behind a slit system, the photo-electrons are counted in short time intervals, e.g. 0.2 seconds, controlled by an accurate clock, and these counts are recorded on punched tape. Later numerical analysis of the counts in a computer gives the transit times across the slits. In principle, the transit time for individual slits could be derived, or the transit time for a group of slits. The latter method would be less sensitive to noise, and in the course of time both method have been widely applied.

The slits should be inclined to the stellar motion by 45 degrees in alternating directions. By such a "fishbone grid" a two-dimensional measurement of the star in the focal plane became possible (Fig. 2), corresponding to right ascension and declination.
Astrometry by means of accurate slits and photon counting was subsequently applied on meridian circles, on long-focus telescopes, and ultimately on the first astrometry satellite, Hipparcos. French astronomers became interested in the method, and I saw reports in the early 1960s from Lille and Besançon where they worked with "une grille de Høg", as they called the system of inclined slits, but I do not recall if they used photon counting. The method with the fishbone grid and photon counting was crucial in the proposal for space astrometry by Pierre Lacroute.

Heckmann was immediately interested in my proposal, and I recall that he helped me write the report in 1960. He wanted the method implemented on the Hamburg meridian circle for the planned expedition to Perth, Western Australia. That kept me busy for the next decade and resulted in the Perth 70 Catalogue of positions for 24,900 stars (Høg & von der Heide 1976).

**4. Space astrometry**

In 1967 I heard the presentation in Prague by Lacroute (1967) about space astrometry. This was the first time that such type of astronomy was proposed for a space mission. The potential advantages were clear, no atmosphere and no gravity, but the technical problems seemed utterly underestimated since a total mission cost of only 10 million French francs was claimed. The proposal did not start any activity outside France, and I was fully occupied with other matters; at that time I did not have any vision of space astrometry. But Lacroute's vision was fortunately shared by other French astronomers, especially Jean Kovalevsky. He supported the idea and finally had the project converted from being a French national project to become European through ESA.

In 1975 I was invited to be member of a small working group of astronomers and ESA engineers set up to make a mission definition study of space astrometry. I felt that I had to join the group in spite of my profound scepticism about Lacroute's proposals and also a lack of interest in space techniques. At the first meeting of the group on 14 October the ESA chairman urged us to be independent of previous ideas and to propose whatever space techniques could most efficiently achieve our scientific goals. With this encouragement in mind I designed in six weeks my vision of a scanning astrometry satellite, called TYCHO. From this design study many (seven) new features were adopted in the final Hipparcos satellite. The name seemed proper for the first satellite especially designed for astrometry, the art of science mastered by Tycho Brahe.

TYCHO used an image dissector tube as detector behind a modulating grid. This detection was 100 times more efficient than the system of slits in front of a photo-multiplier tube as in Lacroute's previous proposals. TYCHO required active attitude control in order to perform an optimal scan along great circles. The spin axis should revolve around the Sun at a constant angle. It included a star mapper so that it would be able to use an input catalogue of 100,000 stars selected for their scientific interest and being observed with a carefully selected observing strategy. The grid measured only one-dimensionally along the scan direction. Therefore the beam combiner in front of the telescope needed only two reflecting surfaces. It would then be easier to manufacture than the beam combiner with three or five surfaces required for the two-dimensional scanning always preferred by Lacroute in his designs called TD-options.

Lacroute adopted only two ideas from TYCHO: an image dissector tube and a modulating grid for a new TD-option of a scanning satellite, which he considered to be technically simpler than TYCHO.

I had called my design TYCHO, but for the study report in the spring of 1976 the chairman suggested that the names TYCHO and TD were not good. They were then changed to respectively Option A and Option B. More details about the two options A and B are given, e.g., by Høg (1997), Kovalevsky (2005) and Turon & Arenou (2008).

Several years later Kovalevsky introduced the name HIPPARCOS as an acronym for the final satellite, based on the TYCHO/Option A. I had preferred TYCHO which I still considered to be a proper name for an ESA astrometric satellite. Luckily, in 1981 I was able to invent the sky mapper experiment and gave it the proper name without hearing any objections. This experiment finally resulted in the Tycho-2 Catalogue with astrometry and two-colour photometry of 2.5 million stars, published by Høg et al. (2000). Tycho-2 is now the preferred astrometric reference catalogue for star brighter than 11[th] magnitude, used to tie the bright 120,000 stars of the Hipparcos system to astrometric observations of fainter stars obtained by ground-based CCD telescopes.

In a large team a few will often stand out; Andrew Murray, my old colleague and member of the Hipparcos science team, once said: "Erik, the best you have ever done for astronomy was to find Lennart!" and I agreed. I "found" Lennart Lindegren in 1973 while a 23 year old student at Lund Observatory (see Høg 2008b), and I have had the privilege to work with him ever since. I brought him into Hipparcos in 1976 and without his unfailing genius in all mathematical, computational and optical matters the project would not have been ripe for approval in 1980, and probably never.

The Hipparcos project won the competition with the EXUV project in ESAs Astronomy Working Group, but only barely so according to Edward van den Heuvel (see Høg 2008c), X-ray astronomer and a member of AWG until the end of 1979, and much in favour of Hipparcos. Several votings took place in AWG before 1980, and at one of the crucial ones Hipparcos stayed for further consideration only because one person had been convinced to change position.

My own attitude then was that if Hipparcos had lost I was ready to quit the project for lack of faith that the astrophysicists would ever let it through.

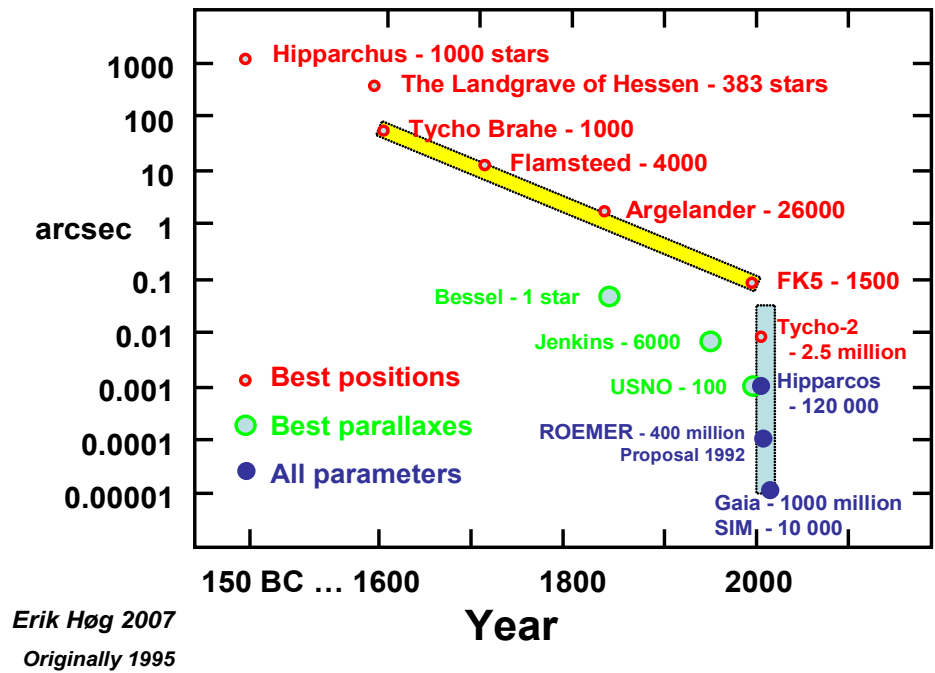

**Figure 3.** Astrometric accuracy through 22 centuries from Hipparchus to Gaia. More explanation will be given in Høg (2008a), presently in preparation.

The final voting in AWG took place on 24 January 1980 (ESA 1980a): Of the 13 members present, 8 voted in favour of Hipparcos and 5 in favour of EXUV, but dangers for Hipparcos laid ahead. At its meeting on 6th and 7th February 1980 the Science Advisory Committee (SAC) discussed six missions and preferred (ESA 1980b) the combined Comet/Geos-3 mission and the Hipparcos mission. The SAC did not make the choice between these two missions which represented the interests of the ESA working groups for respectively the solar system and astronomy. Both missions were therefore recommended, though on certain conditions, and the process ultimately led ESA to do something ESA had never done before: approve two missions at the same time. SAC expressed a preference for Hipparcos over the EXUV mission if the payload is funded *outside* the mandatory budget of ESA. In the end Hipparcos was funded *within* the mandatory budget, so Hipparcos was up against great hurdles all the time, but our mission won in the end, thanks to negotiations of which details are reported especially by Jean Kovalevsky in another report (Høg 2008c).

After approval the project gained great momentum and was carried through by large enthusiastic teams (Perryman et al. 1997) working many years guided by the Hipparcos Science Team whose chairman Michael Perryman personifies this phase of the mission more than anyone.

**5. Conclusion**

Bengt Strömgren appears clearly at the root of my contributions to astrometry, including Hipparcos, and he was directly active before the mission approval in 1980 in order to ensure Danish and Swedish support. He would have seen the Gaia mission (Fig. 3) with astrometry, photometry and radial velocities as the ultimate fulfilment of his quest since the 1930s for comprehensive studies of our Milky Way. It seems from the unbroken chain of actions defined above that there would have been no Hipparcos, no space astrometry with a scanning satellite, if any of the five persons Bengt Strömgren, myself, Pierre Lacroute, Jean Kovalevsky or Lennart Lindegren had been absent from the scene before 1980.

*Acknowledgements:* I am grateful to Catherine Turon, Jean Kovalevsky, and Lennart Lindegren for comments on earlier versions and for their agreement to the present text. I also acknowledge comments from Andrew Murray, Edward van den Heuvel, and colleagues in Copenhagen: Jens Knude and Holger Pedersen.

No. 1A - 2008.06.10: -  Here follows the shorter text of the poster as shown at IAU Symposium 254.

# Bengt Strömgren and modern astrometry

### Development of photoelectric astrometry including the Hipparcos mission

*Erik Høg, Niels Bohr Institute, Copenhagen*

**ABSTRACT**: *Bengt Strömgren is known as the famous astrophysicist and as a leading figure in many astronomical enterprises. Less well-known, perhaps, is his role in modern astrometry although this is equally significant. There is an unbroken chain of actions from his ideas and experiments with photoelectric astrometry since 1925 over the new meridian circle in Denmark in the 1950s up to the Hipparcos and Tycho Catalogues published in 1997.*

## 1. Introduction

This account follows a chain of actions beginning with an experiment on photoelectric astrometry in 1925 and culminating with the Hipparcos mission at the end of the 20[th] century, thus, in 1925 began a new era of positional astronomy comparable in significance to that of Tycho Brahe four centuries earlier. This brief account is far from being a complete history of Hipparcos, nor, of course, of the many other developments of photoelectric astrometry in the same period.

## 5. Conclusion

Bengt Strömgren appears clearly at the root of my contributions to astrometry, including Hipparcos, and he was directly active before the mission approval in 1980 in order to ensure Danish and Swedish support. He would have seen the Gaia mission (to be launched in 2011) with astrometry, photometry and radial velocities as the ultimate fulfilment of his quest since the 1930s for comprehensive studies of our Milky Way. It seems from the unbroken chain of actions defined here that there would have been no Hipparcos, no space astrometry with a scanning satellite, if any of the five persons Bengt Strömgren, myself, Pierre Lacroute, Jean Kovalevsky or Lennart Lindegren had been absent from the scene before 1980.

## 2. Strömgren's experiments with photoelectric astrometry

Bengt Strömgren in 1925 at the age of 17 years reported about experiments with photoelectric recording of star transits. In the focal plane of the meridian circle in Copenhagen he had placed a system of slits parallel to the meridian, Fig. 1. Behind the slits was a photo cell which received the light from the star after it had passed the slits. As the star moved across the slits the variations of light intensity gave corresponding variations in the photo current, and these variations of current were amplified and recorded.

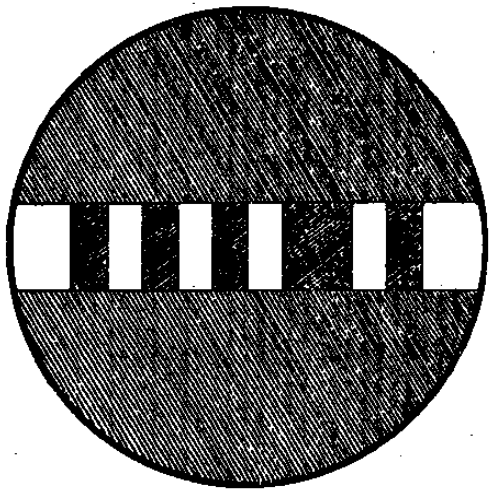

**Figure 1.** Slit system for photoelectric measurement of right ascension with a meridian circle (Strömgren 1925)

Strömgren, however, found a serious drawback of his initial method: For reasons of statistical noise, it would only allow recording of stars to 6[th] or 7[th] magnitude with a medium size meridian circle. He therefore proposed a method of integration (Strömgren 1933) which should allow observation of much fainter stars. A mirror was placed behind a system of equidistant slits. It was switched quickly at predetermined times between two positions. In one position the light hits one photo cell, in the other another photo cell, both of them able to integrate all the light reaching them.

The mirror and the whole operation of this "second generation" system posed technical problems and no further experiments have been reported. The author of these pages heard about the two proposals as a student and that bore fruit later on as I shall explain, but I probably never discussed them with Strömgren in those early days.

## 3. Developments in Denmark and Hamburg

In 1940 Bengt Strömgren succeeded his father, Elis, as director of the observatory, then located in the centre of Copenhagen. The same year he took the initiative to build a new observatory outside Copenhagen. A site in Brorfelde 50 km west of Copenhagen was selected. Ten years later the concrete for the foundation of the main instrument, a new meridian circle, was poured, and finally, in 1953 the instrument itself could be mounted. I then got the task as a fourth year student to test the stability of the new instrument by photographic observations of a star very close to the North pole. Strömgren did not intend to implement a photoelectric method at the new instrument, but preferred a method where a photographic plate is moved along with the star. This method was technically less challenging, and was in fact put into operation in Brorfelde in the 1960s.

Most important for me as a young astronomer, was to grow up in an observatory where a new meridian circle was the main instrument and where this course for the institute had been defined by an outstanding scientist. Bengt Strömgren gave everybody, not only a youngster as me, confidence about the future line of astronomy. How very different at most other places in the world where astrometry, the astronomy of positions, was being discarded as old-fashioned stuff. At such places I would probably have become an astrophysicist, since I certainly did not want to do old stuff.

In 1956 I finished my studies and became a conscript soldier. Most of the time I had the opportunity to work in a laboratory (Niels Finsen Institute) measuring radioactive decay of dust, collected to follow the nuclear weapon testing of the two superpowers. The measurements were obtained by radioactive counting techniques and my experience with this brand new technique I could later apply to photoelectric astrometry.

In 1958 I moved to the Hamburg Observatory where both astrometry and astrophysics were held in high esteem; Otto Heckmann was the powerful director. My interest went towards observational astrophysics, but soon I got the best idea I ever had (Høg 1960): I realized that Strömgren's method with the switching mirror could be implemented very elegantly by a photon counting technique. I do not think any other astronomers used photon counting at that time; I had the idea from the counting of radioactive decay.

Briefly, the technique was as follows: A photo multiplier tube is placed behind a slit system, the photo-electrons are counted in short time intervals, e.g. 0.2 seconds, controlled by an accurate clock, and these counts are recorded on punched tape. Later numerical analysis of the counts in a computer gives the transit times across the slits. In principle, the transit time for individual slits could be derived, or the transit time for a group of slits. The latter method would be less sensitive to noise, and in the course of time both method have been widely applied.

The slits should be inclined to the stellar motion by 45 degrees in alternating directions. By such a "fishbone grid" a two-dimensional measurement of the star in the focal plane became possible (Fig. 2), corresponding to right ascension and declination.

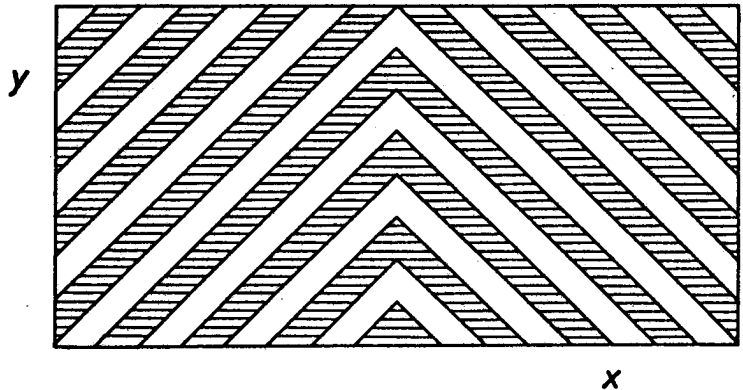

**Figure 2.** Slit system for two-dimensional photoelectric measurement
in the focal plane of a meridian circle (Høg 1960)

Astrometry by means of accurate slits and photon counting was subsequently applied on meridian circles, on long-focus telescopes, and ultimately on the first astrometry satellite, Hipparcos. French astronomers became interested in the method, and I saw reports in the early 1960s from Lille and Besançon where they worked with "une grille de Høg", as they called the system of inclined slits, but I do not recall if they used photon counting. The method with the fishbone grid and photon counting was crucial in the proposal for space astrometry by Pierre Lacroute.

Heckmann was immediately interested in my proposal, and I recall that he helped me write the report in 1960. He wanted the method implemented on the Hamburg meridian circle for the planned expedition to Perth, Western Australia. That kept me busy for the next decade and resulted in the Perth 70

Catalogue of positions for 24,900 stars (Høg & von der Heide 1976).

## 4. Space astrometry

In 1967 I heard the presentation in Prague by Lacroute (1967) about space astrometry. This was the first time that such type of astronomy was proposed for a space mission. The potential advantages were clear, no atmosphere and no gravity, but the technical problems seemed utterly underestimated since a total mission cost of only 10 million French francs was claimed. The proposal did not start any activity outside France, and I was fully occupied with other matters; at that time I did not have any vision of space astrometry. But Lacroute's vision was fortunately shared by other French astronomers, especially Jean Kovalevsky. He supported the idea and finally had the project converted from being a French national project to become European through ESA.

In 1975 I was invited to be member of a small working group of astronomers and ESA engineers set up to make a mission definition study of space astrometry. I felt that I had to join the group in spite of my profound scepticism about Lacroute's proposals and also a lack of interest in space techniques. At the first meeting of the group on 14 October the ESA chairman urged us to be independent of previous ideas and to propose whatever space techniques could most efficiently achieve our scientific goals. With this encouragement in mind I designed in six weeks my vision of a scanning astrometry satellite, called TYCHO. From this design study many (seven) new features were adopted in the final Hipparcos satellite. The name seemed proper for the first satellite especially designed for astrometry. Several years later Kovalevsky introduced the name HIPPARCOS as an acronym for the final satellite, based on the TYCHO/Option A.

In a large team a few will often stand out; Andrew Murray, my old colleague and member of the Hipparcos science team, once said: "Erik, the best you have ever done for astronomy was to find Lennart!" and I agreed. I "found" Lennart Lindegren in 1973 while a 23 year old student at Lund Observatory (see Høg 2008b), and I have had the privilege to work with him ever since. I brought him into Hipparcos in 1976 and without his unfailing genius in all mathematical, computational and optical matters the project would not have been ripe for approval in 1980, and probably never.

The Hipparcos project won the competition with the EXUV project in ESAs Astronomy Working Group, but only barely so. - If Hipparcos had lost I was ready to quit the project for lack of faith that the astrophysicists would ever let it through.  -  *See the conclusion on the first page!*

## Further illustrations to the poster

The poster was shown with four large pictures (A4):

**The Copenhagen meridian circle used by Bengt Strömgren**
Courtesy Steno Museum, Aarhus

**The Carlsberg Meridian Circle from Brorfelde,  here on La Palma**

**The Hipparcos astrometry satellite launched by ESA in 1989**

**Astrometric accuracy through 22 centuries from Hipparchus to Gaia.**

# Lennart Lindegren's
# first years with Hipparcos

*Erik Høg, Niels Bohr Institute, Copenhagen*

**ABSTRACT**: *Lennart Lindegren has played a crucial role in the Hipparcos project ever since he entered the scene of space astrometry in September 1976. This is an account of what I saw during Lennart's first years in astrometry after I met him in 1976 when he was a young student in Lund.*

A new era of my life began on 1 September 1973 when I returned to Denmark with my family of five, after 15 years in Hamburg. I had obtained a tenure at the Copenhagen University where I was going to work on the construction of automatic control of the meridian circle in Brorfelde. Very soon, however, I heard of a young student at Lund Observatory who worked alone on modernizing the old meridian circle there. I went to Lund and "found" Lennart. A few years later, Andrew Murray, my old colleague and member of the Hipparcos science team, would say: "Erik, the best you have ever done for astronomy was to find Lennart!" and I agreed.

At my visit in the autumn of 1973, Lennart showed all he had done on the mechanics, the chronograph etc. I was impressed, but very curious about his intentions. He explained that he had been attracted by the fine mechanics of the old instrument, but he knew well that he could not make it operational without a large funding, which he did not aim for.

I offered two subjects for a thesis which he wanted to think about. He visited me in Brorfelde some time later and had made his choice: The Glass Meridian Circle, a new type of horizontal meridian circle which I had proposed several years earlier. But that subject had just been taken by a Danish student. Lennart then immediately chose the second proposal: Data analysis of astrometric observations of the major planets made with the photoelectric meridian circle in Perth. He made a brilliant analysis where he, e.g., took physical limb darkening into account, not only the classical geometric darkening. Nobody had done that before, and he used available satellite observations of the planets to get the most correct darkening.

By the time he had finished that work, time was ripe to introduce him to the astrometric satellite project which then consisted of two concepts or proposals: Option A and Option B (see Høg 2008). I called a meeting in Copenhagen on 22 September 1976 with Lennart, a Danish student, and a colleague where I explained the project and especially the challenging task of data analysis. I wanted a student to propose methods especially related to Option A, to derive the astrometric parameters from one-dimensional observations with the satellite operating in revolving scanning mode, i.e. with the spin axis at a constant angle from the sun direction and moving around the sun. I had proposed such scanning in December 1975, but I did not know whether the observations would at all produce a rigid coordinate system nor whether the computations could be accomplished with computers of the time.

In the two hour meeting the Danish student repeatedly said that it was a too big task for him. Lennart only asked two or three questions and made no comments, but the meeting started an intensive correspondence between us two. Four weeks later I received a letter and a nicely typed report. On 9 pages (Lindegren 1976a) he gives the mathematical description of the "three-step procedure" for Option A which later became the fundamental method used by both Hipparcos data reduction consortia. It broke down the enormous system of least-squares equations to smaller systems that could be solved with an acceptable computational effort. He gives the variances of the five parameters and the coefficients of correlation between them. He finds much higher variances for some of the parameters than in the previous ESA study report. He has found a very good resolution for all five parameters with revolving scanning for the whole sky in this mode.

Two weeks thereafter came a report (Lindegren 1976b) of 8 pages with more complete results of the first simulated observations with one-dimensional scanning and a revolving spin axis. It included tables and plots showing the accuracies of the five parameters as function of ecliptic latitude and longitude.

This pace of meticulous and crucial reports coming from Lennart's hand has been maintained ever since. He once said: "If a problem can be stated mathematically it is simple to solve." This could be construed as immodesty, but everybody knowing Lennart, will say that for him it is simply true. I do not recall any error of mathematics has been found in his reports, or for that matter any other kind of error; nor any other sign of immodesty.

ESA has a system of advisory groups, one of them is Astronomy Working Group (AWG). I had become a member of this group in December 1975, just after I had made my TYCHO mission proposal, later named Option A, and the working group had altered its name from *Astrophysical* Working Group in order to accommodate, for the first time, an astrometrist in this group otherwise solely astrophysicists. On 9 December the AWG had to select members for a Space Astrometry Team (SPT) to follow the coming feasibility study. Many had applied to become member of this astrometry team, and that was of course very good so that a real selection could be made.

At some moment there were still too many candidates for the team, and the discussion was about including Lennart or not. I had urged Lennart to apply and in AWG I argued strongly for his membership pointing, e.g., at his recent reports about the data analysis and saying that it might be crucial for the success of the entire study to include him. A person (probably Niels Lund from Copenhagen) then injected: "Lund Observatory is so close to Copenhagen that you might be able to collaborate with Lindegren even if he is not in the astrometry team." I answered: "OK, if there is only room for one of us in the team, it must be Lennart Lindegren."

A team of seven astronomers was thus selected. It included Pierre Lacroute, although a leading member of the AWG had previously questioned the usefulness hereof, but this matter was not brought up at the meeting. Of course, Lacroute deserved to be able to follow the project study closely. We owe him so much as the originator of space astrometry and as the never doubting believer in its future.

When the meetings of the Space Astrometry Team began we noticed that Lennart never said anything, except when asked, and that he then gave his opinion, often quite brief, but always clear and well spoken, while everyone would listen. He took notes in a protocol, and does so even today although he now brings also a laptop. Historians of astronomy will once appreciate his clear hand when they have to read these protocols; there must be dozens.

Of his numerous papers I will only mention two. He wrote a paper on "Photoelectric astrometry" (Lindegren 1978), a subject I had proposed,

where he systematically discussed the performance of methods for precise image location from observations. It remains a classical paper. The second paper to mention is about the rigidity of the celestial coordinate system obtained by the one-dimensional observations in a scanning satellite as TYCHO/Option A/Hipparcos. The question was asked in 1976 as mentioned above, but it took years before we had the answer which was affirmative as given by Høyer et al. (1981). The study was lead by Lennart and contains his brilliant mathematical analysis of the simulations, but he modestly left the position as first author to another person.

More information about the scientific environment in which Lennart played a crucial role is given in a recent report (Høg 2008) on the development of photoelectric astrometry including the Hipparcos mission.

*Acknowledgements:* I am grateful to Andrew Murray for comments to earlier versions of the text and to Lennart for verifying that it contains no factual errors. Lennart agreed, though with great reluctance, that I distribute this report.

# Miraculous approval of Hipparcos in 1980

## *Erik Høg, Niels Bohr Institute, Copenhagen*

**ABSTRACT: The approval of the Hipparcos mission in 1980 was far from being smooth since very serious hurdles were encountered in the ESA committees. This process is illuminated here by means of documents from the time and by recent correspondence. The evidence leads to conclude that in case the approval would have failed, Hipparcos or a similar scanning astrometry mission would never have been realized, neither in Europe nor anywhere else.**

## 1. Introduction

The discussions in ESAs Astronomy Working Group (AWG) and the Science Advisory Committee (SAC) in 1979-80 have been summarised in a previous report (Høg 2008) as repeated here in section 2. I have in the present report chosen to let documents and witnesses speak separately, through quotations and recent correspondence. It may look a bit complicated, but I hope at least some readers will appreciate to get in closer touch with history in this manner.

Correspondence with Ed van den Heuvel is collected in section 3, and I am quoting in extenso because I think the drama is of some interest for a wider audience. Section 4 brings further quotations from the meetings in AWG, SAC, and the Scientific Programme Committee (SPC) and from recent correspondence with Jean Kovalevsky and Catherine Turon. I conclude that Hipparcos prevailed thanks to a kind of miracle. In section 5 I argue that in case the approval would have failed, Hipparcos would never have been realized.

Lennart Lindegren just wrote that he intends to write down the developments up to 1980 from his own perspective, but he cannot promiss a certain date. Jean Kovalevsky will try to write before summer on the 1965-1975 period. I will update the present report if further evidence of sufficient interest should become available.

## 2. Summary of discussions in AWG, SAC, and SPC

The Hipparcos project won the competition with the EXUV project in ESAs Astronomy Working Group, but only barely so according to Edward van den Heuvel (2008, priv. comm.), X-ray astronomer and a member of AWG until the end of 1979, and much in favour of Hipparcos. Several votings took place in AWG before 1980, and at one of the crucial ones Hipparcos stayed for further consideration only because one person had been convinced to change position.

My own attitude then was that if Hipparcos had lost I was ready to quit the project for lack of faith that the astrophysicists would ever let it through.

The final voting in AWG took place on 24 January 1980 (ESA 1980a): Of the 13 members present, 8 voted in favour of Hipparcos and 5 in favour of EXUV, but dangers for Hipparcos laid ahead. At its meeting on 6[th] and 7[th] February 1980 the Science Advisory Committee (SAC) discussed six missions and preferred (ESA 1980b) the combined Comet/Geos-3 mission and the Hipparcos mission. The SAC did not make the choice between these two missions which represented the interests of the ESA working groups for respectively the solar system and astronomy. Both missions were therefore recommended, though on certain conditions, and the process ultimately led ESA to do something ESA had never done before: approve two missions at the same time. SAC expressed a preference for Hipparcos over the EXUV mission if the payload is funded *outside* the mandatory budget of ESA. In the end Hipparcos was funded *within* the mandatory budget, so Hipparcos was up against great hurdles all the time, but our mission won in the end, thanks to negotiations of which details are reported by Jean Kovalevsky in section 4. This leads to a summary of the ESA committee meetings in January to July of 1980:

**24 Jan. AWG:** Hipparcos is recommended.

**6/7 Feb. SAC:** Comet/Geos3 and Hipparcos are recommended, no choice is made within SAC, but there are conditions on both.

**4/5 Mar. SPC:** Hipparcos is selected as the next scientific project of ESA. The Hipparcos instrumental payload is included on certain conditions. The mission to Halley comet shall be pursued on certain conditions, and if these conditions are met SPC will in fact have approved two missions simultaneously, resulting in consequences for the schedules.

**8/9 July SPC:** Giotto is included for a flyby in 1986 of Comet Halley as a purely

European project since NASA could not make a firm commitment. The schedule of Hipparcos is accordingly stretched by six months.

## 3. Edward van den Heuvel (2008, priv. comm.)

The summary in the first paragraph of section 2 was based on the following mails, here slightly shortened and quoted with permission from Ed van den Heuvel. I asked Ed on 17 March 2008 how close the vote in AWG was. He answered at 6:07 PM our time, the same day:

Dear Erik,
The vote was indeed very close. I was able to convince one of the X-ray astronomers (Spada) not to vote for the EUV/Soft X-ray mission which was then the competitor of Hipparcos, and his vote was just the one that made the difference ....

Spada, although director of the X-ray astronomy lab in Bologna, casted the vote that made the difference

very sadly, Spada has completely disappeared from the scene in Italy. …

I am at the moment working at the Institute for Theoretical Physics, University of California Santa Barbara. If you wish to call me …

Best wishes,
Ed van den Heuvel

## An hour later, at 7:21 he added:

Dear Erik,

It is a long time ago, and there have perhaps been various stages of voting in the AWG. I do not have any of my papers here in California, so I cannot check. I know I kept my papers from that time in the AWG in my archive in Amsterdam, so when I am back I can check.

What I remember is that we first had Setti as the AWG chair (I thought you were in the AWG at that time), and under his chairmanship we had many discussions of the projects but not a final vote. When the vote had to be taken, Setti had been replaced by De Jager from my country, who had a big stake in the EUV/X-ray mission. .... It was under his guidance that the vote which I mentioned in my last e-mail to you was taken and in which Spada and I (as X-ray astronomers) voted in favour of Hipparcos ...
Now that you say that I was no longer in the AWG in 1980 when apparently a final vote was taken, I am getting a bit confused, about whether there may have been a still later (definitive?) round of votes and whether the votes which I mentioned was perhaps an earlier round.
I presume that it must be possible to trace that back in the minutes of the
AWG from 1979 and 1980.

As you know, memory is not fully reliable, and this was almost 30 years ago.
But I vividly remember that there was this one voting round where Spada's vote
made the difference. I thought that what I remembered is that if in that
voting round Hipparcos would have lost, then the AWG from that moment would
have gone further with the EUV/X mission. But I hope this can be traced back
in the AWG minutes.

There you also could trace back whether Spada was still in the AWG when the
final vote was made.
I do not know whether the minutes tell whom voted in favour and whom voted
against? *(No, the minutes do not give such details, EH)*

Since I am just saying this all from the top of my head, without any papers
here that may support it, and since- as said- memory may be unreliable, please
consider all this as confidential, and not for circulation. *(Permission has later been
given, EH)*

Best wishes,
Ed

*Note by EH:  It seems that Ed has been member of AWG with his period of three years
1976-79 overlapping my years 1976-78. But I do not remember him from that time in
spite of his great sympathy for the space astrometry project and the important role he has
played in the mission approval. About twenty years ago, however, he told me what I just
reported, and he has recalled it ever since when we happened to meet with years
between. Therefore I contacted him when I was writing (Høg 2008) and got immediate
reply.*

## 4.  From the committee meetings in 1980

Some further quotations from AWG and SAC meetings (ESA 1980a and
1980b) illustrate the difficulties Hipparcos encountered. At a meeting on 24
January 1980 the AWG considered the Astrometry and EXUV missions,
concluding that both missions will give excellent scientific return. This is
elaborated for the two missions. On astrometry for instance this: *"The
Astrometry mission, HIPPARCOS, will give fundamental quantitative results
to all branches of Astronomy. It emphasises typical European know how and
will serve a community never before involved in space research";*  on the
EXUV mission for instance this: *"The fact that the scientific objectives of
this mission are being covered by two different missions proposed by other*

*agencies (EUVE by NASA and ROBISAT by Germany) emphasises its timeliness."*

It is somewhat surprising then that 5 members were still in favour of EXUV and only 8 in favour of HIPPARCOS. One could have thought that a unique mission as Hipparcos would come above anything else in everybody's mind.

SAC discussed the missions on 6[th] and 7[th] February 1980 and unanimously recommended that the combined Comet/Geos-3 mission be selected as proposed by the Solar System Working Group (SSWG) on certain conditions. Strong advocates for EXUV were also present at the SAC meeting: *"in the event that the Hipparcos payload would need to be funded within the mandatory programme, the SAC was divided as to whether Hipparcos should then remain the Agency's choice or <u>EXUV</u> should be carried out because this mission was considered by some members to be just as interesting."* (The quotation is literal, including spellings and emphasis.) In the end, Hipparcos was in fact financed *within* the mandatory programme.

In view of all these hurdles it seems a kind of miracle that Hipparcos could prevail, but it was of course because the right people worked hard to make it happen. The final solution was that SPC approved two missions: Giotto, the mission to comet Halley, to be launched first and to be followed by Hipparcos, and that SPC decided to finance the Hipparcos scientific payload out of the mandatory programme. ESA otherwise always assumes that payloads are financed by the member states.

Where were the competing EXUV people in all this? An answer may be found in the following letters from Jean Kovalevsky.

**Jean Kovalevsky** *wrote on 2008.05.11:*
*I was invited to the AWG for the Hipparcos presentation, but did not attend the discussions.*

*I was member of SAC and I remember very well that, at some point, there was a vote between Hipparcos and EXUV: Hipparcos had 5 votes out of 6, the only tenant of EXUV was H Elliot from the UK. The other members were: Egidi (Frascati), Tammann (Basel), Weiss (Erlangen) and Pinkau (Chairman). The fact that SAC proposed that Hipparcos payload was to be paid nationally was simply repeating the SSWG statement.*

*It was evident for me and (at least as far as I remember) Tammann,*

*that the responsibility of the payload had to be taken over by ESA, but I felt that insisting on this point would have been counter-productive, because the announced costs of the two proposals without the payload were identical while adding 50 MAU to the cost of Hipparcos would have killed it.*

*So I decided, in order to save the mission, to accept this point. After all, SAC was only an advisory group and had no financial responsibility. The only ESA body that could overrule the normal procedure (following which nations should fund and prepare the payload) was the SPC. An additional problem was that the laboratories involved in space hardware had experience in receivers and in conventional optics, but no one was reasonably able to built the delicate parts of Hipparcos. I knew that at least the French delegation at SPC, and possibly others will lobby in favour of an indoor payload funding. The March decision by SPC proved that I was right.*

*Pinkau had reported to the March SPC meeting of the views of SAC. I prepared, as an attachment for you, the part which concerns Hipparcos and EXUV.*

*From the part on Hipparcos: "The SAC realized the extremely fundamental nature of the mission, and the impact it will have on many branches of science and our conception of the world we live in. The SAC also noted the strong support for this mission within the AWG." Then the three areas of concern to the SAC are outlined: Technical difficulties, the data analysis problem, and the cost of the mission.*

**Catherine Turon** *wrote on 2008.05.13:*
*Hipparcos was approved in March 1980, and Giotto later, after still another meeting of the SPC (exceptional ???), in July 1980. I do not have the minutes of these SPCs neither their decisions, but the letter of information sent to "the wide scientific community" by E.A. Tredelenburg, then Director of the Scientific Programme.*
*I'll send these to you.*

**EH** *wrote on 2008.05.15:*

*I was the only astrometrist in the AWG about 1977 and I remember saying to Malcolm Longair in a coffee break: "You astrophysicists will decide about the astrometry project and you should be aware that you have only one opportunity to approve such a mission. It you reject it this time it cannot be revived because the astrometrists would never again believe astrophysicists could ever let it pass. We would believe that no matter how much you are impressed by space astrometry, in the end the majority would always put their own project higher." He said that I should not use this as an argument, but only argue with the qualities of the project. That was all he said, a wise advice, I think, which I*

*followed. But the insight I believed to have then has become certainty after seeing the evidence presented here.*

**Jean Kovalevsky** *wrote on 2008.05.23:*

*Dear Erik.*

*Let me make some further remarks that could enrich your text, a text which I fully appreciate.*

*Coming back to the February 1980 SAC meeting, there was really NO competition between the Comet/Geos3 mission and the astronomical missions. From the very beginning of its session, SAC did not like the idea of choosing between an astronomical and a Solar system mission. It considered that it would be more fair to give a chance to both working groups' proposals, and that ESA, rather than deciding missions one by one every year or so, must have a broader and more prospective policy.*

*So, indeed, the choice was only between EXUV and Hipparcos. I think that the key sentence in the pages I sent you is the following:*
*"It was thought that then a new proposal for an EUV-mission would be very worthwhile". This was really killing EXUV.*

*Now, there were two conditions:*
*-For Hipparcos, it was the funding of the payload*
*-For the Comet/Geos3 mission, it was the necessary re-assesment to transform it into a really cometary mission.*

*In March, SPC solved the first problem (and this is probably the most miraculous part of the adventure) and, letting time for the re-assesment of the cometary mission, Hipparcos found itself as the ONLY approved mission!*

*What followed is interesting. The re-assesment of the cometary mission, becoming Giotto, put ESA in an awkward situation: the non-approved mission was evidently more urgent because of Halley's orbit. We had an additional SAC meeting end of June or July. I do not have documentation on it, but I remember well how insistently Trendelenbourg (Director of Science) tried to convince me (as he assumed I was the toughest proponent of Hipparcos), that I should accept that Hipparcos be delayed by a year or so, to allow the maximum money to be spent on Giotto. Of course, SAC unanimously agreed and the next SPC followed the recommendation.*

*The decision of the SPC that the payload should be the responsibility of ESA was taken very seriously and ESA started to study how to manage it. In the October 1980 meeting of SAC, the Executive presented a document which described the management as we have known it, and SAC approved it.*

*Best regards,*

*Jean*

**Catherine Turon** *agreed to this later the same day, and did not want to add anything.*

**EH** *wrote on 2008.05.26:*

*The reports mentioned by Catherine have been received (ESA 1980c and d). They spell out in detail what Jean has said in his two letters. Finally, therefore, the summary of the ESA committee meetings in January to July of 1980 can be written  and is placed at the end of section 2.*

## 5.  In case the approval had failed

It appears that the approval could well have failed in which case I am sure Hipparcos would never have been realized. This proposition has been countered by a colleague: *"You can never know that, something could have happened."* But please consider the situation of astrometry at that time. For decades up to 1980 the astrometry community was becoming ever weaker, the older generation retired and very few young scientists entered the field. I myself would have lost the faith that the astrophysicists would ever let such a mission through, and others would also have left the field of space astrometry.

If someone would have tried a Hipparcos revival one or two decades later the available astrometric competence would have been weaker, and where should the faith in space astrometry have come from? When Hipparcos became a European project in 1975 and the hopes were high for a realization, the competence from many European countries gathered and eventually was able to carry the mission. This could not have been repeated after a rejection of the mission.

But NASA could have realized a Hipparcos-like mission? No, for two reasons: The American astrometric community had much less resources of competence to draw from than there were in Europe, and secondly, as an American colleague said: *"You can convince a US Congressman that it is important to find life on other planets, but not that it is important to measure a hundred thousand stars."*

Thanks to the completion of the Hipparcos mission a strong astrometric community now exists in Europe which has been able to propose and develop the Gaia mission and which will carry it to a successful completion. Without Hipparcos the faith in the much more difficult CCD technology of Gaia would have been missing.

**Acknowledgements:** I am grateful to Catherine Turon for providing the reports ESA 1980a-d, to Edward van den Heuvel for permitting his letters to be included here, and to Jean Kovalevsky for providing more information from the ESA meetings. I also thank all of them and Holger Pedersen for comments to earlier versions of this report.

Intentionally empty page

Contribution to the history of astrometry No. 4                  2007.12.10

The following was posted at IAU Symposium No. 248 15-19 October 2007, Shanghai, China: *A Giant Step: From Milli- to Micro-Arcsecond Astrometry*. Sections 1, 2, and the references appeared on three pages (300-302) in the proceedings edited by Wenjing Jin, Imants Platais & Michael A.C. Perryman.

# From the Roemer mission to Gaia

Erik Høg, Niels Bohr Institute, Copenhagen University
Juliane Maries Vej 30, 2100 Copenhagen Ø, Denmark

**Abstract.** At the IAU symposium in Shanghai September 1992 the present author made the first proposal for a specific mission concept post-Hipparcos, the first scanning astrometry mission with CCDs in time-delayed integration mode (TDI). Direct imaging on CCDs in long-focus telescopes was described as later adopted for the Gaia mission. The mission called Roemer was designed to provide accurate astrometry and multi-colour photometry of 400 million stars brighter than 18 mag in a five-year mission. The early years of this mission concept are reviewed.

When Hipparcos was launched in August 1989 the Hipparcos Science Team (HST) was present in Kourou and we were greatly relieved seeing the take-off after the many years of preparation. But that changed to grim disappointment the next day when we learned that the apogee boost motor had not started so that the satellite was stuck in an elliptical transfer orbit instead of the intended geostationary. This endangered the whole mission and we would possibly only get a much shorter set of poor observations, perhaps only months and not the planned three years. Passing through the radiation belts every few hours could soon destroy the electronics and solar cells.

In this mood, but optimistic as always, I presented the Hipparcos mission on behalf of Michael Perryman who could not be present, and the Tycho project at the IAU Symposium No.141 October 17-21, 1989 in Leningrad (now St. Petersburg). The audience was full of high hopes for Hipparcos - hopes which were in fact justified as we should later see. We noticed that Soviet (later Russian) colleagues presented ideas at the Symposium for a successor to Hipparcos. They themselves had three projects on the drawing boards: AIST/ STRUVE, LOMONOSOV, and REGATTA-ASTRO. The basic idea was to re-observe the 120,000 Hipparcos stars and utilize the positions from Hipparcos and those from a new epoch to get much better proper motions than Hipparcos alone would achieve, even if its severe problems would be cured.

Such ideas were far beyond the horizon of anyone in the Hipparcos team, busy as we were to get our mission to work and to perform the very complex data analysis. I was myself leader of one of the two Hipparcos data analysis teams and of the Tycho team and thus had more than enough to look after.

Shortly later I was invited to lecture about Hipparcos at the Pulkovo Observatory in Leningrad, the Mission Control Center in Moscow, and the Kislovodsk Observatory in Caucasus. I was accompanied on the journey in August 1990 by M.S. Chubey, V.V. Makarov, and V.N. Yershov so we had plenty of time for discussions. I wanted to understand how their AIST project functioned, but unexpectedly, after a day I was more occupied by designing a second Hipparcos myself, realizing that it could easily be made ten times more efficient in utilizing the star light, mainly by employing more detectors, while keeping the same telescope aperture of 0.29 m.

In June 1991 an International Symposium "Etalon" Satellites was held in Moscow where I presented a paper with Mark Chubey "Proposal for a second Hipparcos", but the proceedings were not published. If launched ten years after Hipparcos the mission could obtain proper motions for the 120,000 Hipparcos stars with an accuracy 10 times better than expected from Hipparcos as well as 1 mas accuracy for all astrometric parameters of some 400,000 stars and four-colour photometry for two million stars. This proposal was considered by the Mission Control Centre in Moscow.

During 1990-91 we met many times for discussion of our ideas as they developed, and Lennart Lindegren joined us. At the HST meetings I only got a few minutes to present the progress: HST was a body put in place to supervise the scientific development and optimisation of Hipparcos, not as a body to develop ideas for a new mission concept.

In 1991 I had left the study of photon counting techniques as in Hipparcos and tried to use CCDs, a completely new technique for me. I learnt it from our engineer, R. Florentin Nielsen, and designed a detector system using a modulating grid as in Hipparcos. The result was 1000 times better light efficiency than Hipparcos (see Høg & Lindegren in IAU Symposium 156, 1993).

Having done that I dropped the modulating grid and tried direct imaging on the CCDs imployed in drift-scan mode or time-delayed integration (TDI). That design was called Roemer and gave 100,000 times better light efficiency with the same telescope aperture (0.29 m), but a very long focal length was needed, 5 m instead of the 1.4 m in Hipparcos (see Høg 1993). Both systems were presented at the IAU Symposium 156 in Shanghai September 15-19, 1992.

The Roemer design was proposed in June 1993 for the Third Medium Size ESA Mission (M3) by a team mainly from the HST. The proposal got a high rating in the ESA selection committee, but was not finally selected because it was considered to come too early after Hipparcos. This view was not shared by the proposers, but in hindsight it was a wise decision because it gave us time for much development in the subsequent years.

Interferometry was proposed at the IAU Symposium No. 166 in August 1994 by Lindegren & Perryman "A small interferometer in space for global astrometry: The GAIA concept", stating the "very strong scientific case for global optical astrometry at the 20 microarcsec accuracy level." The satellite should contain three Fizeau-type interferometers with 2.5 m baselines.

At the same IAU Symposium a 10 microarcsec mission (Roemer+) with 9-colour intermediate- and wide-band filter photometry was proposed by the present author. The better performance was obtained with two telescopes of larger apertures of 70 cm instead of 29 cm. Picometer gauges were adopted to monitor the alignment of the telescopes.

The development of instrument ideas had mainly three scientific goals: higher astrometric accuracy of 10 microarcsec instead of the 100 microarcsec envisaged in Roemer, measurement of radial velocities for the brighter stars with the satellite, and better multi-colour photometry. These improvements were considered crucial for an ambitious ESA mission aiming for understanding the details of our Galaxy. Thorough assessment of the scientific goals and the data analysis was also made. - Finally, Gaia is now scheduled for launch in 2011 on a 5 year mission to measure 1000 million stars brighter than 20 mag.

## 2. Why interferometry - and why not?

There was a widespread belief at the time of the Roemer proposal, Lindegren et al. (1993), that interferometry could give better astrometry from space, and a section was included: "Towards 10 microarcsec astrometry: The FIZEAU option". It was not part of the baseline Roemer proposal, but was meant "to point out a possible development towards a scanning satellite with ten times the angular accuracy of Roemer", and the enourmous scientific benefits of such an accuracy for millions of stars were outlined. A Fizeau principle was subsequently used in several proposals for scanning astrometric satellites, e.g., the GAIA concept of 1994 mentioned above, FAME and DIVA.
.
I agreed that interferometric options should be deeply studied as they in fact were during the following years. Perhaps the complications of interferometry could be alleviated, or at least the fallout from studies could bear fruit in other (unforeseen) contexts. These studies always focussed on a scanning astrometry satellite similar to Hipparcos because a systematic scanning of the sky was considered the only way to measure the millions of stars required for our scientific goal. A pointing satellite could never do that, but would of course have the advantage of allowing longer integration time on any selected area.

My own preference in instrument design has always been to identify and focus on difficulties and try to solve or circumvent them. So I believed more in direct imaging on CCDs from full-aperture telescopes than in the diluted apertures required for interferometry. The Roemer+ design of 1994 used full apertures and obtained 10 microsec, but it required picometer gauges to monitor the alignment of telescopes, a technique nearly always required in interferometric options.

In 1995 we designed an interferometric option later published by Høg et al. (1997). It used a beam-combiner of 150 cm aperture and a simple telescope, basically an aplanatic Gregorian system. A prism provided a low dispersion perpendicular to the scanning direction so that spectrophotometry could be obtained. This new option of Gaia was adopted in ten times smaller size for the proposal by Röser et al. (1997). This was a small German astronomy satellite, DIVA, planned for launch in 2003 to measure about 40 million stars as a fore-runner for Gaia. But funding did not follow suit.

The ESA studies of the interferometric option are described at length in a section (pp.331-338) of ESA (2000) and complete references are given. The history of the development of Gaia is briefly summarized in Perryman et al. (2001). One of the problems was that the split pupil of an interferometer did not allow accurate measurement of the stars about 20 mag required for the ambitious scientific goal, but only about 17 mag. Another problem came from the required data rate to be transmitted from the satellite. An interferometric image requires a lot more data points to cover the fringes of a star than a direct stellar image from a full aperture. The higher data rate could well be accepted from a geostationary orbit, but the thermal control during eclipses would jeopardize the instrument stability, so the orbit around L2 was required for thermal stability. Here the data rate of one Megabit per second for the full aperture option was acceptable, but not the higher rate for interferometry. Other problems of interferometry were identified and in the end the full aperture could be selected and we were sure that all had been done to investigate both options, based on industrial studies by Matra Marconi Space for the baseline design and Alenia Aerospazio for the interferometric.

# 3. Roemer instead of Gaia

The name GAIA was introduced as an acronym for Global Astrometric Interferometry for Astrophysics. The name was retained after interferometry had been dropped, some said it was too late to change the name. But changes of name for great satellite missions have been made before, even close to launch or after, e.g. Hubble and Chandra. Gaia gives association with the Greek word for Earth and is always associated with the Gaia whole-earth hypothesis of James Lovelock, a source of confusion.

**It should be considered to give Gaia a new name. A good choice would be Roemer, the original name given by the thirteen proposers to the mission concept for the ESA M3 mission.** It is the name of a scientist who deserves a satellite to be called after him. An astrometric satellite matches especially well with Roemer since he invented the meridian circle, the main instrument for fundamental astrometry during several centuries, and his observations were used by Tobias Mayer to derive the first proper motions of stars from modern observations. - The following details are extracted from the M3 proposal.

The Danish astronomer OLE RØMER (1644—1710) is best known as the discoverer of the speed of light. Around 1675, while working at the newly created Royal Observatory in Paris, he noticed that the intervals between successive eclipses of Jupiter's moons were not always in agreement with the ephemerides that had recently been calculated by Cassini. Depending on the relative motion of Earth and Jupiter the intervals were sometimes larger, sometimes smaller. Rømer correctly inferred that these discrepancies were due to the finite time it took for light to travel from Jupiter to Earth. He computed a value of 22 minutes for the time it takes light to travel one diameter of the Earth orbit. Not knowing this diameter with any reliability he did not calculate the speed of light in terrestrial units.

After his return to Copenhagen in 1681, Rømer constructed a series of instruments for measuring the positions of celestial bodies. His instruments gradually incorporated several new and ingenious concepts which were perfected during the next two centuries: the use of a long axis resting on two bearings for better definition of the viewing plane; microscopic reading of a graduated full circle; the use of counterweights to reduce flexure; and an emphasis on symmetrical design and measuring principles to eliminate otherwise uncontrollable errors. His *rota meridiana* constructed in 1704 is the prototype for the modern meridian circle, one of the most efficient and accurate instruments for ground-based positional measurements. Rømer's strive for ever improved accuracy may have been motivated by the search for stellar parallax. This phenomenon, the ultimate proof of the Copernican theory, would however elude astronomers for yet another century.

All Rømer's instruments and all the observations except those from three nights called *triduum* were destroyed in the fire of 1728. The *triduum* observations of 88 stars were used in 1756 by Tobias Mayer together with his own observations to discover that a fourth of the stars showed a significant change of position, thus deriving the first proper motions of stars from modern observations.

# 4. Appendix: Design and performance of Roemer 1992

Figures of the optical and mechanical design are included, and a table of the predicted astrometric and photometric performance, all with the original captions from 1992. Please note therefore, that milliarsec and microarcsec should be written without hyphen, thus not as in Table 1. This notation was used in the published Hipparcos and Tycho Catalogues as agreed in 1995 in connection with IAU Symposium No. 166, in analogy with millimeter, kilogram etc. But a hyphen is required in sub-milliarcsec.

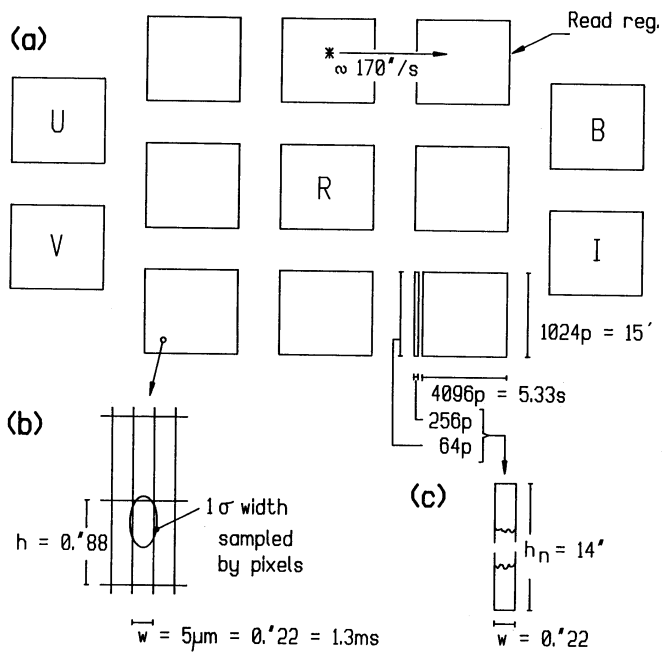

**Figure 1.** Focal plane arrangement of Roemer. (**a**) The stars drift across 13~chips, each containing a narrow and a wide CCD. Reading of the number of accumulated electrons (counts) takes place in a special register at the right edge of each CCD. Eight of the chips measure in a wide spectral band *W* and five in the photometric bands *UBVRI*. (**b**) The 1σ contour of the sampled diffraction image is shown superposed on the pixels of a wide CCD. (**c**) A pixel of a narrow CCD.

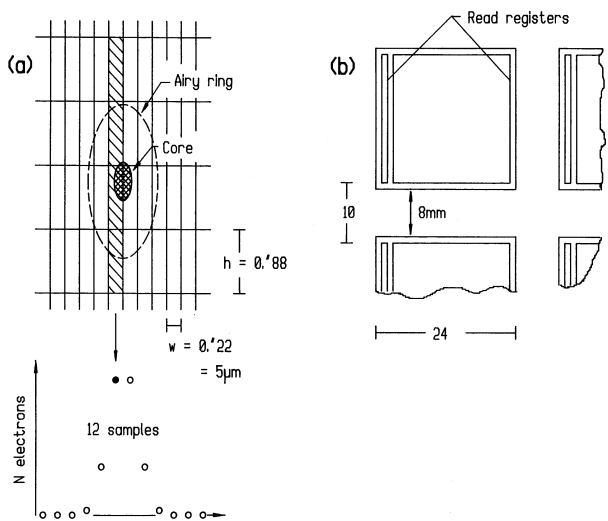

**Figure 2.** The sampling and the slits. (**a**) Sampling of the star image. Single pixels may be read but usually the number of electrons in four pixels (hatched) are summed into one sample (filled dot). Any star in the input catalogue will be covered by about 16 samples and each of these corresponds to 5.3 s integration over the crossing of a wide CCD. (**b**) The chips, each containing a narrow and a wide CCD, are mounted on a frame with the required edge-to-edge spacing.

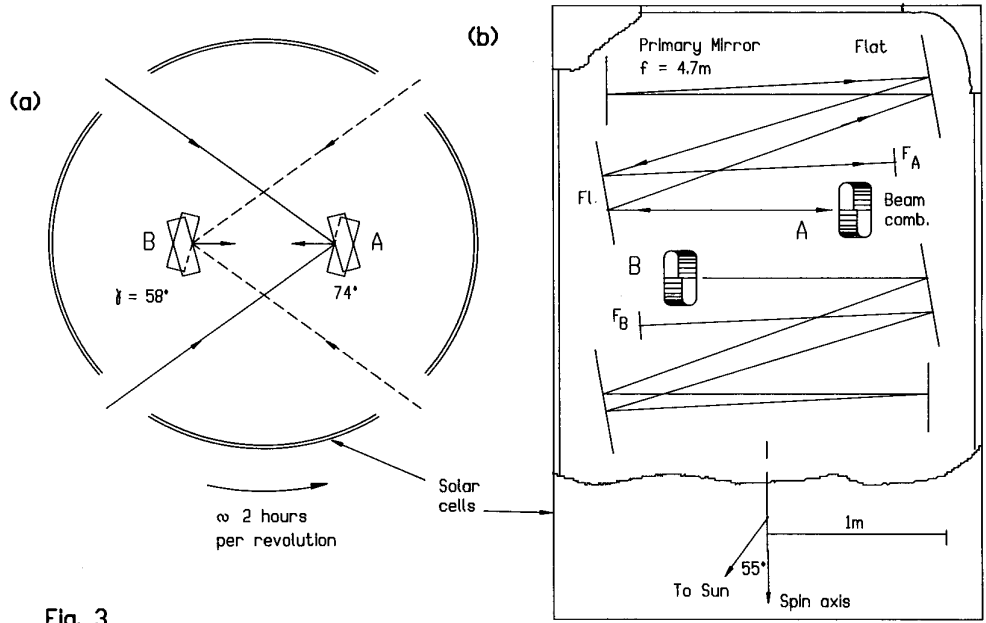

**Fig. 3**

**Figure 3.** Optical system in case *two* telescopes are required, cf. the text. **(a)** Two beam combiner mirrors A and B with basic angles e.g. 58 and 74 degrees, shown in a projection perpendicular to the spin axis. **(b)** Two folded off-axis telescope systems with focal length about 5 m in a cylindrical spacecraft.

Table 1. Predicted mean errors in astrometry and photometry for a 5 year Roemer mission. Columns 3 and 4 give errors for parallax and annual proper motions in milli-arcsec (mas), but asymptotic errors should be added as discussed in the text. Photometric errors are given for the $W$ band from 300-950 nm, and for the five standard colours for stars of spectral type G0. Assumptions: Two beam combiner telescopes of 0.29 m aperture using 13 CCD chips in each focal plane.

| $V$ mag | Astrometry | | Photometry ($W$ = Wide band) | | | | | |
|---|---|---|---|---|---|---|---|---|
| | par. mas | p.m. mas | $W$ mag | $U$ mag | $B$ mag | $V$ mag | $R$ mag | $I$ mag |
| 10 | 0.02 | 0.01 | 0.000 | 0.003 | 0.001 | 0.001 | 0.001 | 0.001 |
| 11 | 0.03 | 0.01 | 0.000 | 0.005 | 0.002 | 0.001 | 0.001 | 0.001 |
| 12 | 0.05 | 0.02 | 0.000 | 0.008 | 0.003 | 0.002 | 0.002 | 0.002 |
| 13 | 0.07 | 0.04 | 0.001 | 0.013 | 0.004 | 0.004 | 0.004 | 0.004 |
| 14 | 0.12 | 0.06 | 0.001 | 0.021 | 0.007 | 0.006 | 0.006 | 0.006 |
| 15 | 0.19 | 0.10 | 0.002 | 0.036 | 0.011 | 0.010 | 0.009 | 0.010 |
| 16 | 0.31 | 0.16 | 0.003 | 0.067 | 0.018 | 0.016 | 0.015 | 0.016 |
| 17 | 0.54 | 0.27 | 0.004 | 0.135 | 0.032 | 0.029 | 0.027 | 0.028 |
| 18 | 0.99 | 0.50 | 0.009 | 0.302 | 0.064 | 0.056 | 0.052 | 0.055 |
| 19 | 1.96 | 0.98 | 0.018 | — | 0.137 | 0.121 | 0.112 | 0.119 |

No. 5 - 2008.05.23, updated 2008.11.25:

# Four lectures on the general history of astrometry
### *overview, handouts and abstracts - 2008-11-25*

Erik Høg - Niels Bohr Institute, Copenhagen University - **email:** erik.hoeg@get2net.dk

**Lecture No. 1.**   45 minutes
## Astrometry and photometry from space: Hipparcos, Tycho, Gaia
   **The introduction covers 2000 years of astronomy from Ptolemy to modern times. The Hipparcos mission of the European Space Agency was launched in 1989, including the Tycho experiment. The Hipparcos mission and the even more powerful Gaia mission to be launched in 2011 are described.**

   The lecture has been developed over many years and was held in, e.g., Copenhagen, Vienna, Bonn, Düsseldorf, Vilnius, Oslo, Nikolajev, Poltava, Kiev, Thessaloniki, Ioannina, Athens, Rome, Madrid, Washington, and Charlottesville - since 2007 in PowerPoint.
   Handouts with 2 and 6 slides per page:
www.astro.ku.dk/~erik/AstrometrySpace2.pdf.  and  AstrometrySpace.pdf.

---

**Lecture No. 2**:    30 minutes. An alternative to No. 1, for astronomers and data engineers. It may be expanded to 45 minutes by including more on Gaia.
## From punched cards to satellites: Hipparcos, Tycho, Gaia
   **A personal review of 54 years development of astrometry in which I took active part.**
   The lecture was developed in 2007 in PowerPoint and was held in Catania and Madrid.
   Handouts at:  /PunchedCards2.pdf.   and   /PunchedCards.pdf.

---

**Lecture No. 3:**  45 minutes. Suited for a broad audience, including non-astronomers
## The Depth of Heavens - Belief and Knowledge during 2500 Years
   **The lecture outlines the structure of the universe and the development of science during 5000 years, focusing on the distances in the universe and their dramatic change in the developing cultural environment from Babylon and ancient Greece to modern Europe.**
   The lecture was first held in 2002, and since 2007 in PowerPoint. Held in Copenhagen, Vilnius, Nikolajev, Athens, Catania and Madrid.
   Handouts at:  /DepthHeavens2.pdf   and   /DepthHeavens.pdf

   **An article with the same title as the lecture** appeared in Europhysics News (2004) Vol. 35 No.3. Here slightly updated, 2004.02.20:  www.astro.ku.dk/~erik/Univ7.5.pdf

---

**Lecture No. 4:**    included on 2008.11.25,   45 or 30 minutes.

### 400 Years of Astrometry: From Tycho Brahe to Hipparcos

   The four centuries of techniques and results are reviewed, from the pre-telescopic era until the use of photoelectric astrometry and space technology in the first astrometry satellite, Hipparcos, launched by ESA in 1989.

   The lecture was presented as invited contribution to the symposium at ESTEC in September 2008: **400 Years of Astronomical Telescopes: A Review of History, Science and Technology.** The report to the proceedings is included as No. 8 among the "Contribution to the history of astrometry ".

## Organization:

Each lecture may stand alone, depending on the audience. The combination of lecture No. 3 (The depth of heavens) and No. 2 (From punched cards...) may however be recommended. That was the arrangement in April 2008 at ESAC, the place near Madrid where Gaia data reduction software is being developed. No. 3 was given before noon where everyone in the ESAC community was invited, and No. 2 was held in the afternoon for a more specialised astronomer audience, in both cases the attendance was very satisfactory.

# Abstracts of the four lectures

**Lecture No. 1.**   45 minutes

## Astrometry and photometry from space: Hipparcos, Tycho, Gaia
   With an historical introduction

The introduction covers 2000 years of astronomy from Ptolemy to modern times.

   The Hipparcos satellite of the European Space Agency was the first satellite specifically designed for astrometry. It obtained high-precision astrometry for 120 000 stars in 3 years of observations (1989-1993), published 1997 in the Hipparcos Catalogue. For 21000 stars the precision of distances is better than 10 per cent. Photometry in a broad spectral band was obtained, with a median precision of 0.0015 mag. The Tycho experiment onboard the satellite gave astrometry and two-colour photometry for 2.5 million stars, published in 2000 in the Tycho-2 Catalogue, including proper motions.

   The Gaia satellite is also an ESA project and will be launched in 2011. It will obtain high-precision astrometry and multi-colour photometry for all the one billion stars brighter than V=20 mag. Astrometric precision for bright stars: 10 microarcsec. Gaia data will have the precision necessary to quantify the early star formation, and subsequent dynamical, chemical and star formation evolution of the Milky Way Galaxy.

   Since all point sources brighter than V=20 mag will be detected and measured astrometrically and photometrically, GAIA will make a deep survey of about one million small objects in our Solar System.

**Lecture No. 2**:   30 minutes. An alternative to No. 1, for astronomers and data engineers. It may be expanded to 45 minutes by including more on Gaia.

## From punched cards to satellites: Hipparcos, Tycho, Gaia

A personal review of 54 years

The Hipparcos satellite of the European Space Agency was the first space mission to perform astrometry, the art of measuring positions, motions and distances to stars. Hipparcos succeeded 1989-93 to measure a million times more efficiently than ground-based instruments in the 1950s when I studied at the Copenhagen University. A personal review is presented of this development in which I took active part, for instance by proposing in 1960 the principle of astrometric measuring with a slit system and photon counting, used for 40 years on meridian circles and for Hipparcos/Tycho, until CCDs became mature. This led to the Gaia mission to be launched in 2011 and it will improve astrometry by yet another million times. The scientific impact of the missions is illustrated.

**Lecture No. 3**   45 minutes.  Suited for a broad audience, including non-astronomers

## The Depth of Heavens - Belief and Knowledge during 2500 Years

The lecture outlines the structure of the universe and the development of science during 5000 years, focusing on the distances in the universe and their dramatic change in the developing cultural environment from Babylon and ancient Greece to modern Europe.

For Dante Alighieri (1265-1321) the spiritual cosmos contained the Heavens, Earth, and Hell, and it was compatible with the physical cosmos known from Aristotle (384-322 B.C.). Dante's many references in his Divine Comedy to physical and astronomical subjects show that he wanted to treat these issues absolutely correct. Tycho Brahe proves three hundred years later by his observations of the Stella Nova in 1572 and of comets that the spheres of heavens do not really exist. It has ever since become more and more difficult to reconcile the ancient ideas of a unified cosmos with the increasing knowledge about the physical universe.

Ptolemy derived a radius of 20 000 Earth radii for the sphere of fixed stars. This radius of the visible cosmos at that time happens to be nearly equal to the true distance of the Sun, or 14 micro-light-years. Today the radius of the visible universe is a million billion (10 to the power 15) times larger than Ptolemy and Tycho Brahe believed.

**Lecture No. 4**  included on 2008.11.25:   45 or 30 minutes

## 400 Years of Astrometry: From Tycho Brahe to Hipparcos

Galileo Galilei's use of the newly invented telescope for astronomical observation resulted immediately in epochal discoveries about the physical nature of celestial bodies, but the advantage for astrometry came much later. The quadrant and sextant were pre-telescopic instruments for measurement of large angles between stars, improved by Tycho Brahe in the years 1570-1590. Fitted with telescopic sights after 1660, such instruments were quite successful, especially in the hands of John Flamsteed. The meridian circle was a new type of astrometric instrument, already invented and used by Ole Rømer in about 1705, but it took a hundred years before it could fully take over. The centuries-long evolution of techniques is reviewed, including the use of photoelectric astrometry and space technology in the first astrometry satellite, Hipparcos, launched by ESA in 1989. Hipparcos made accurate measurement of large angles a million times more efficiently than could be done in about 1950 from the ground, and it will soon be followed by Gaia which is expected to be another one million times more efficient for optical astrometry.

The lecture was presented as invited contribution to the symposium at ESTEC in September 2008: **400 Years of Astronomical Telescopes: A Review of History, Science and Technology.** The report submitted to the proceedings is included as No. 8 among my "Contributions to the history of astrometry".

Contribution to the history of astrometry No. 6            25 November 2008

# Selected astrometric catalogues

*Erik Høg, Niels Bohr Institute, Copenhagen*

ABSTRACT: A selection of astrometric catalogues are presented in three tables for respectively positions, proper motions and trigonometric parallaxes. The tables contain characteristics of each catalogue showing the evolution over the past 400 years in optical astrometry. The number of stars and the accuracy are summarized by the weight of a catalogue, proportional with the number of stars and the statistical weight.

## Introduction

The 400 years of astrometry from Tycho Brahe to the Hipparcos mission have been reviewed (Høg 2008d) for the symposium held at ESTEC in September 2008 to celebrate the 400 years of astronomical telescopes. For this purpose the Tables 1 to 3 were elaborated, containing data for *selected astrometric catalogues* for positions, proper motions and trigonometric parallaxes, respectively. The tables give characteristics of each catalogue showing the evolution over the past 400 years in optical astrometry. The number of stars, $N$, and the accuracy, i.e. the standard error, $s$, are summarized by the weight of a catalogue, $W$, defined in all tables as $W = N s^{-2} 10^{-6}$, proportional with the number of stars and the statistical weight. The table entries are documented in a separate paper (Høg 2008c).

## Position catalogues, Table 1

An approximate standard error of individual mean position coordinates at the mean epoch of the catalogue is given in Table 1, preferably the median value as representative for the bulk of stars in a catalogue. The value $s = 0.04$ arcsec is adopted for FK5 at the mean epoch as we have derived from the comparison with Hipparcos by Mignard & Froeschlé (2000), and we assume that PPM is the only catalogue before Hipparcos which has been more accurate than 0.1 arcsec. The weight is only derived for observation catalogues, not for compiled ones. Note that position catalogues after 1990 already appear a few years after ending the observations thanks to modern computing facilities, to the available reference systems provided by Hipparcos, and to publication on the web. Previously, the reduction and publication on paper could take decades.

Some important effects are not expressed in the tables such as the serious difference in accuracy between the northern and southern celestial spheres for ground-based catalogues before 1997, after which time a N-S effect is absent due to the use of Hipparcos all-sky results. The variation within a catalogue of the accuracy with magnitude of the star, and with the observational history also does not appear from the tables. The tables give only an imperfect impression of the enormous efforts by astrometrists during the centuries. The following information is obtained from Knobel (1877) and Eichhorn (1974), from correspondence with colleagues and the web. No attempt for completeness is made and more details can be found in Høg (2008c).

**Positions from sextants and quadrants:** The catalogues with 1000 stars from Ptolemy, Ulugh Beg and Tycho Brahe were the largest up to Hevelius' catalogue of 1690. Ptolemy's is the only extant catalogue from antiquity, and perhaps partly copied from the 300 years older Hipparchus. Ulugh Beg published his catalogue in 1437 in Samarkand, but it became known in Europe only when published again in 1665. Hevelius' catalogue from 1690 had three times smaller errors than Tycho's. Flamsteed's *Historia Coelestis Britannica* with 2935 stars, published in 1725, remained the largest until it was surpassed by N.L. de Lacaille's catalogue of 10,000 southern stars, observed from the Cape of Good Hope about 1752. Lalande's much larger *Histoire Céleste Francaise* with 50,000 stars was published in 1801. This catalogue was so important that it was republished in England by Baily half a century later, compiled with other observations. Bradley's observations beginning in 1743 were processed by the young Bessel, and again by Auwers, more than a hundred years later, resulting in positions with an accuracy of 1.1 arcsec as mentioned by Høg (2008d).

**Positions from visual meridian circles:** The Geschichte des Fixsternhimmels (GFH) contains the observations of 365,000 stars from 492 catalogues before 1900 compiled after 1899 by F. Ristenpart and many others, and published in 48 volumes between 1922 and 1964. The GFH was supplemented by Index der Sternörter I and II, initiated by R. Schorr in 1924. It gives reference to 401 catalogues of the observations after 1900 of 365,000 stars, half of them from the southern hemisphere(!), and was published 1927-1966.

**Photographic position catalogues:** The Astrographic Catalogue (AC) was obtained from photographic plates taken between 1892 and 1950, for details see Eichhorn (1974). It is the biggest astronomical enterprise ever undertaken by international cooperation and it began with a meeting in Paris in April 1887 invited by the French Academy of Science. It was a revolutionary idea at that time. Up to then, most star positions were obtained from observations on meridian circles. This imposed a limit of about 9[th] magnitude on stars accessible to observation; and (at that time) the typical standard error of a meridian position was about 0.5 arcsec in either coordinate. The photographic method then under consideration made it practical (without too much effort) to derive positions with a standard error about 0.3 arcsec for stars as faint as 13[th] or even 14[th] magnitude.

The plates were taken with identical telescopes at 20 observatories distributed at all geographic latitudes. The telescope, called a Normal Astrograph, had a lens of 33 cm aperture and 3.4 m focal length and a useful field of 2.1 x 2.1 square degrees. The plates were measured and star coordinates were published in about 150 volumes, the last ones in 1971. These books were inconvenient to use and the given coordinates suffered from the lack of an accurate reference system when the reductions were made. Due primarily to the enormous job of getting the data into machine readable form, attempts to attain a usable whole-sky catalogue failed until the 1990s. The USNO made a new reduction of the AC, containing over 4.5 million star positions, and published it in 2001 as AC 2000.2. The positions in AC 2000.2 and more than 140 other ground-based catalogues were used with the Tycho-2 positions to derive proper motions of the 2.5 million stars in the Tycho-2 Catalogue, as well as the UCAC2 (Zacharias 2008).

**The instruments in Col. 2 of Table 1 and 2:** Observations were visual before 1880 and with all meridian circles in the list, except where noted. Abbreviations: *Quad* = quadrant, which e.g. Tycho Brahe supplemented with the sextant, *MC* = meridian circle, *AAC* = alt-azimuth circle, *TI* = transit instrument, *Pgr.* = photographic, *p.e.* = photoelectric, and *sat.* = satellite.

**Table 1** Position catalogues. The standard error of the mean positions in a catalogue is a median value, if available, representing the bulk of stars, mostly faint ones. The standard errors in the three tables are often internal errors only, e.g. those in brackets (...); the external errors may be much larger, cf. Høg (2008c). - Argelander's accurate meridian circle catalogue in the list is *not* the same author's BD survey catalogue with approximate positions of 325,000 stars

| Catalogue | Instrument | Publ. year | Mean epoch year | Obs. period years | N entries | n per star | $s_{star}$ s.e. of star arcsec | W weight |
|---|---|---|---|---|---|---|---|---|
| Ptolemy | Sextant | 150 | 138 | | 1025 | | 1 deg. | |
| Ulugh Beg | Sextant | 1665 | 1437 | 17 | 1018 | | 1 deg. | |
| Wilhelm of Hesse | Quad | 1594 | | | 1004 | | 360 | |
| Tycho Brahe | Many | 1598 | 1586 | 20 | 1005 | | 60 | 0.000,000,3 |
| Hevelius | Quad+Sext | 1690 | 1670 | | 1564 | | 20 | 0.000,003,9 |
| Rømer | MC | 1735 | 1706 | 0.01 | 88 | 2.6 | 4 | 0.000,013 |
| Flamsteed | Quad | 1725 | 1700 | | 2934 | | 20 | 0.000,007 |
| Lacaille | Quad | 1763 | 1752 | 2 | 9766 | | 6? | 0.000,27 |
| Bradley/Auwers | TI+Quad | 1888 | 1760 | 12 | 3222 | | 1.1 | 0.002,7 |
| Lalande | Quad | 1801 | 1795 | 40 | 50,000 | | 3 | 0.006 |
| Piazzi | AAC | 1814 | 1802 | 21 | 7646 | | 1.5 | 0.003,4 |
| Lalande/Baily | compiled | 1847 | | | 47,390 | | | |
| Argelander | MC | 1867 | 1856 | 22 | 33,811 | 2 | 0.9 | 0.042 |
| Küstner | MC | 1908 | 1899 | 10 | 10,663 | 2.4 | 0.34 | 0.092 |
| USNO | MC | 1920 | 1907 | 8 | 4526 | 10 | (0.15) | 0.20 |
| USNO | MC | 1952 | 1945 | 8 | 5216 | 6 | (0.15) | 0.23 |
| Astrographic Cat. | Pgr. | | 1900 | 60 | 4,500,000 | 2 | 0.2 | 110 |
| Stoy | MC | 1968 | 1948 | 13 | 6800 | 2 | 0.43 | 0.037 |
| GC | MC | 1937 | 1900 | 175 | 33,342 | | 0.15 | |
| SAOC | Pgr.+MC | 1965 | 1930 | 50 | 259,000 | | 0.2 | |
| Perth70 | p.e. MC | 1976 | 1970 | 5 | 24,900 | 4 | 0.15 | 1.1 |
| FK5 | MC | 1988 | 1950 | 242 | 1535 | | 0.04 | |
| PPM N+S | Pgr.+MC | 1993 | 1945 | 90 | 379,000 | 6 | 0.04 | |
| CMC1-11 | p.e. MC | 1999 | 1991 | 14 | 176,591 | 6 | 0.07 | 36 |
| Hipparcos | p.e.sat. | 1997 | 1991 | 3 | 118,218 | 110 | 0.001 | 120,000 |
| Tycho-2 | p.e.sat. | 2000 | 1991 | 3 | 2,539,913 | 130 | 0.06 | 700 |
| USNO-B1.0 | Pgr. | 2002 | | 50 | 1,000,000,000 | | | |
| UCAC2 | CCD | 2003 | 2000 | 4 | 48,000,000 | 2 | 0.06 | 13,000 |
| 2MASS | HgCdTe | 2003 | 2000 | 3 | 400,000,000 | | 0.08 | 62,000 |
| CMC14 | CCD MC | 2005 | 2002 | 6 | 95,000,000 | 2 | 0.07 | 19,000 |
| GSC-II | Pgr. | 2005 | | 50 | 945,000,000 | | | |

Other photographic astrometric enterprises with ever better lens astrographs giving a larger usable field were undertaken. They have resulted in hundreds of thousands of star positions but only some of the names can be mentioned: the AGK2 observed about 1930, published 1952, the AGK3 observed 1959-1961, published 1973, including proper motions, the Yale Catalogues observed 1914-1956, published in the 1950s, and the Cape Photographic Catalogue observed 1930-1953, published 1968. Much larger catalogues of a 1000 million stars were derived from Schmidt plates taken in the 1950s and later, going to much fainter magnitudes about 20[th]. From the USNO came the USNO A1.0, and B1.0 catalogues. The most recent Hubble Guide Star Catalogue is the GSC-II, obtained with the Palomar and UK Schmidt telescopes at two epochs and three photometric band passes. These catalogues and USNO B1.0 contain positions, proper motions, and photometry.

**CCD position catalogues:** After 1980 photographic plates were gradually replaced by electronic detectors, especially CCDs because of the higher sensitivity and the immediate digitization of the observation, and because astrometric quality plates were no longer manufactured. CCDs are primarily used in pointing mode. Scanning mode was introduced by Stone & Monet (1990) on a meridian circle. Scanning mode with CCDs on an astrometric satellite, Roemer, was proposed in 1992 by Høg (1993). This proposal initiated other similar projects, FAME and DIVA, and Roemer itself developed into the Gaia mission. From the web: The UCAC2 catalogue used the U.S. Naval Observatory Twin Astrograph of 20 cm aperture and a 4k by 4k CCD camera and it covers the declinations -90 to +40 deg. The catalogue positions have a standard error of 70 mas at the limiting magnitude of R=16, and an error of 20 mas at 11-14[th] mag. Such a smaller error for brighter stars is typical for many catalogues, but it is not expressed in the present tables. The 2MASS all-sky catalogue was obtained by two highly automatic telescopes with 1.3 m aperture equipped with HgCdTe detectors sensitive in the J,H,K bands (1-2 microns) with a limit of 17 mag in J. The CMC14 was obtained with the Carlsberg Meridian Circle with 18 cm aperture using CCDs in scanning mode giving a magnitude range of 9-17 in the red band, r'. The Tycho-2 Catalogue supplied the astrometric reference stars for UCAC2, CMC14 and GSC-II. Further projects for astrometry and photometry with enormous CCD arrays are: the ongoing Sloan Digital Sky Survey (SDSS), the coming Pan-STARRS, and the Gaia satellite.

## Proper motion catalogues, Table 2

This small selection of proper motion catalogues shows especially the increasing number of stars and the improvement of accuracy over three centuries. All catalogues, apart from the first entries, contain both positions and proper motions, the motions being derived from positions observed over long periods of time and sometimes with different types of instruments. The number of catalogues, *n*, used for the proper motions is given, but this number should be regarded with caution since a single meridian circle catalogue and the entire AC are both counted as one catalogue although AC contributes much more weight to the derived proper motions.

Some of the catalogues in Table 2 are the especially accurate fundamental catalogues, Auwers' FC, NFK, N30, FK3, FK4 and FK5. They were compiled in order to provide reference stars for meridian circle observations and very accurate proper motions for the study of kinematics and dynamics of the Galactic stellar system. Scott (1963) gives an overview, including the proper motion errors for FK3 and N30. The FK5 states an error of 0.75 mas/yr, but from Tables 1 to 4 by Mignard & Froeschlé (2000) we have derived that the error is 1.6 times higher i.e. 1.2 mas/yr. A slightly larger value is given for FK4. The larger reference catalogues IRS and ACRS from the 1990s of respectively 36,027 and 320,211 stars for the reduction of photographic plates are described by T. Corbin in Høg (2008c).

**Table 2** Proper motion catalogues. The standard error, *s*, is a median value, if available. Mean epoch, observation period, number of catalogues, and the standard error are sometimes given as round numbers. The observations used for the Tycho-2 proper motions were obtained by Hipparcos, MCs and photography, and the catalogues after 2000 benefit greatly from Tycho-2 as reference catalogue. - Abbr.: mas/yr = milliarcsecond/year

| Catalogue | Instrument | Publ. year | Mean epoch year | Obs. period years | N entries | n catalogues | s s.e. of star mas/yr | W weight | $s_{pos.publ.}$ mas |
|---|---|---|---|---|---|---|---|---|---|
| Halley | | 1718 | | | 3 | | | | |
| Mayer | MC+Quad | 1775 | | | 80 | | | | |
| Mädler | Quad+MC | 1856 | | | 3222 | | | | |
| Auwers' FC | Quad+MC | 1879 | | 117 | 539 Dec > 10 | 9 | 8? | 8 | |
| NFK | Quad+MC | 1907 | | 155 | 925 | | 5? | 40 | |
| FK3 | MC | 1937 | 1900 | 190 | 1535 | 70 | 3 | 170 | 150 |
| GC | Quad+MC | 1937 | 1900 | 175 | 33,342 | 238 | 10 | 330 | 400 |
| N30 | MC | 1952 | 1930 | 100 | 5268 | 60 | 5 | 210 | 150 |
| SAOC | Pgr.+MC | 1965 | 1930 | 50 | 259,000 | 10 | 15 | 1200 | 560 |
| FK4 | MC | 1963 | 1920 | 213 | 1535 | 250 | 2 | 380 | 380 |
| FK5 | MC | 1988 | 1950 | 242 | 1535 | 350 | 1.2 | 1100 | 62 |
| PPM North | Pgr.+MC | 1991 | 1931 | 90 | 182,000 | 12 | 4.2 | 10,300 | 270 |
| PPM South | Pgr.+MC | 1993 | 1962 | 100 | 197,000 | 14 | 3.0 | 22,000 | 110 |
| PPM N+S | | | | | 379,000 | | 3.4 | 32,000 | 144 |
| Hipparcos | p.e.sat. | 1997 | 1991 | 3 | 118,218 | 1 | 0.9 | 120,000 | 6 |
| Tycho-2 | p.e.sat.++ | 2000 | 1991 | 100 | 2,539,913 | 145 | 2.5 | 400,000 | 64 |
| SPM3 | Pgr. | 2004 | 1980 | 23 | 10,700,000 | 2 | 4.0 | 670,000 | 100 |
| UCAC2 | CCD++ | 2003 | 1990 | 100 | 48,000,000 | 146 | 6.0 | 1,300,000 | 80 |
| USNO-B | Pgr. | 2002 | 1975 | 50 | 1,000,000,000 | 2 | 7.0 | 20,000,000 | 275 |

Table 2 includes a standard error of the positions in the year of publication, $s_{pos.publ}$. This positional error is a measure for the ability of the catalogue to provide good positions in the years following the publication. The value is calculated by quadratic addition of the error due to proper motion and the position error at the mean epoch. The resulting error is mainly due to the proper motion error in all catalogues, except Tycho-2 where the 0.06 arcsec error at the mean epoch dominates by far. Tycho-2 will therefore keep its value as reference catalogue for many years.

The large all-sky position and proper motion catalogues in the table of stars brighter than 11 mag are GC, SAOC, PPM, Hipparcos, and Tycho-2, published from 1937 to 2000. The progress from GC to Tycho-2 for the practical user of catalogues appears in the increase by a factor of almost 100 in the number of stars and more than 1000 in the weight. Furthermore, the errors of star positions at the time of publication of the catalogue decreased by a factor 6. After the year 2000 much larger catalogues with up to 1000 million stars cover also the fainter stars between 11 and 20 mag with positions, proper motions and multi-colour photometry.

# Catalogues of trigonometric parallaxes, Table 3

The first three reliable annual parallaxes of stars were published in 1838-40, their standard errors are given according to Høg (2008c). The technology and methodology of parallax measurement (see the table and Høg 2008c,d) remained basically unchanged for about 60 years (ca. 1840-1900). Despite the rapid initial success, the number of stars with reliable parallaxes grew slowly, and is hard to calculate because of disputes about which were reliable. The uncertainties and systematic errors are strikingly illuminated when we see that the review paper with catalogue by Oudemans from 1889 lists 55 observations of 61 Cyg and gives the mean parallax as 0.40". This is 0.11" larger than the true value, a deviation 8 (eight) times larger than the formal error of 0.014" claimed by Bessel in 1840, and Bessel's own parallax deviates four times his own error from the true value. In 1899 Ch. Andre gives the parallax as 0.44" from the same 55 observations, see references and discussion in Høg (2008c).

**Table 3** Catalogues of trigonometric parallaxes.

| Catalogue | Instrument | Publ. year | Obs. period years | N entries | s s.e. of star mas | W weight | Notes |
|---|---|---|---|---|---|---|---|
| Bessel | Heliometer | 1838 | | 1 | 60 | | |
| Henderson | Quad | 1839 | | 1 | 500 | | |
| Struve | Wire micr. | 1840 | | 1 | 100 | | |
| Peters | Visual | 1850 | | 20 | ? | | |
| Oudemans | Visual+Pgr. | 1889 | 60 | 50 | ? | | |
| Bigourdan | Visual+Pgr. | 1909 | | 100 | (50) | 0.04 | With one observation per star |
| - same - | | | | 200 | (30) | 0.2 | With two or more obs. per star |
| Russell | Pgr. | 1910 | | 52 | (40) | 0.03 | |
| Schlesinger | Pgr. | 1935 | 35 | 7534 | 15 | | Includes spectroscopic par. |
| Jenkins | Pgr. | 1952 | 50 | 5822 | 15 | 26 | |
| Van Altena | Pgr. | 1995 | 95 | 8112 | 10 | 81 | |
| - same - | | | | 1649 | | | Error of parallax < 17.5 % |
| - same - | | | | 940 | | | Error of parallax < 10 % |
| Hipparcos | p.e.sat. | 1997 | 3 | 118,218 | 1.0 | 120,000 | |
| - same - | | | | 20,853 | | | Error of parallax < 10 % |
| USNO | Pgr.+CCD | -2008 | 20 | 357 | 0.6 | 1000 | C. Dahn, priv. comm. 2008 |
| HST | CCD, satellite | -2008 | 18 | 31 | 0.24 | 500 | F. Benedict, priv. comm. 2008 |

A catalogue by Bigourdan (1909) lists trigonometric parallaxes for about 300 stars, a few with up to 40 observations. The consistency of multiple observations indicates a precision (i.e. internal error, therefore in brackets) about 50 mas per observation, and a median precision of 30 mas may be inferred for about 200 stars having more than one observation. Many observations are shown (by bold face) to be the average of several measurements by the same observer, including most of the 100 with only one observation. Russell (1910) presents 52 new photographic parallaxes and claims a standard error about 40 mas.

For the 1952 parallax catalogue by Jenkins a standard error of 15 mas is derived by Hertzsprung

(1952). A round value of 10 mas is given for the median standard error of the last ground-based catalogue (van Altena 1995), but the errors vary greatly, viz. between 1 and 20 mas.

Hipparcos obtained a median standard error of 1.0 mas for parallaxes. A similar or better accuracy has been achieved from the ground and with the Hubble Space Telescope (HST) for several hundred much fainter stars. The number of parallaxes with an error less than a given fraction of the parallax value is given in three cases.

**Acknowledgements:** I am indebted to Adriaan Blaauw for the kind invitation to contribute to the symposium on this vast subject. Without the invitation I would never have engaged myself in this quite large undertaking. Comments to previous versions of the paper from F. Arenou, P. Brosche, A. Chapman, T. Corbin, D.W. Evans, C. Fabricius, F. Mignard, H. Pedersen, P.K. Seidelmann, C. Turon, S.E. Urban, W.F. van Altena, and N. Zacharias are gratefully acknowledged.

Intentionally empty page

Contribution to the history of astrometry  No. 7          25 November 2008

# Astrometric accuracy during the past 2000 years

*Erik Høg, Niels Bohr Institute, Copenhagen*

ABSTRACT: A documentation of the great development of astrometric accuracy since the observations by Hipparchus about 150 B.C. is provided. The development has often been displayed in diagrams, showing the accuracy versus time. These diagrams are discussed, and very significant differences are found, most recently in a diagram from 2007. The diagrams used for Hipparcos up to 1989 are based on a serious misunderstanding of a diagram from 1983. A more correct diagram was constructed in 1995 which was used in the Hipparcos book of 1997. A further improved version is presented here, showing the accuracy of positions and parallaxes in catalogues as based on the included documented data.

## Introduction

The present report, including diagrams in an appendix, shall document and discuss the accuracy of observed positions of stars. The evolution of astrometric observations during the past centuries is shown in three tables of a report (Høg 2008b) to which I will refer: Table 1 for position catalogues, Table 2 for proper motion catalogues, and Table 3 for catalogues of trigonometric parallaxes.

The evolution has often been displayed in diagrams, showing the accuracy versus time. These diagrams have at least one thing in common, the improvement by many powers of ten from the one degree of Hipparchus, the Greek father of astronomy, to one milliarcsec for the diagrams including the Hipparcos Catalogue. But Tycho Brahe and Flamsteed are the only other sources always included, though with quite different numbers. Other differences are pointed out below.

I will present a recommended diagram of astrometric accuracy, including explanations and a list of the sources, in literature or otherwise, for the points as they are plotted.

A detailed history of the various other diagrams will be given. Some of the diagrams give the impression of a smooth, gradual improvement over all the centuries, including the last 500 years. This obscures the historically interesting fact that four jumps can be clearly seen in Fig. 1. A 'jump' means a big improvement within a very short time as the result of great investment of material resources and intellectual efforts. First, the Landgrave of Hesse measured positions ten times more accurately than Hipparchus/Ptolemy and Ulugh Beg, but I know too little about the Landgrave to say more. Tycho Brahe was six times more accurate than the Landgrave thanks to an investment never seen before in the history of science. Third, the Hipparcos satellite gave a factor 100 over the contemporary accuracy of positions obtained from the ground. Fourth, the Gaia satellite mission is expected to yield a factor 100 over Hipparcos.

# Recommended diagram

Figure 1 is a diagram of the development of astrometric accuracy with time, prepared in 2008 for the present report and for Høg (2008d). The diagram is called Høg-2008 since, for convenience, a diagram is designated by "name-year".

## Astrometric Accuracy versus Time

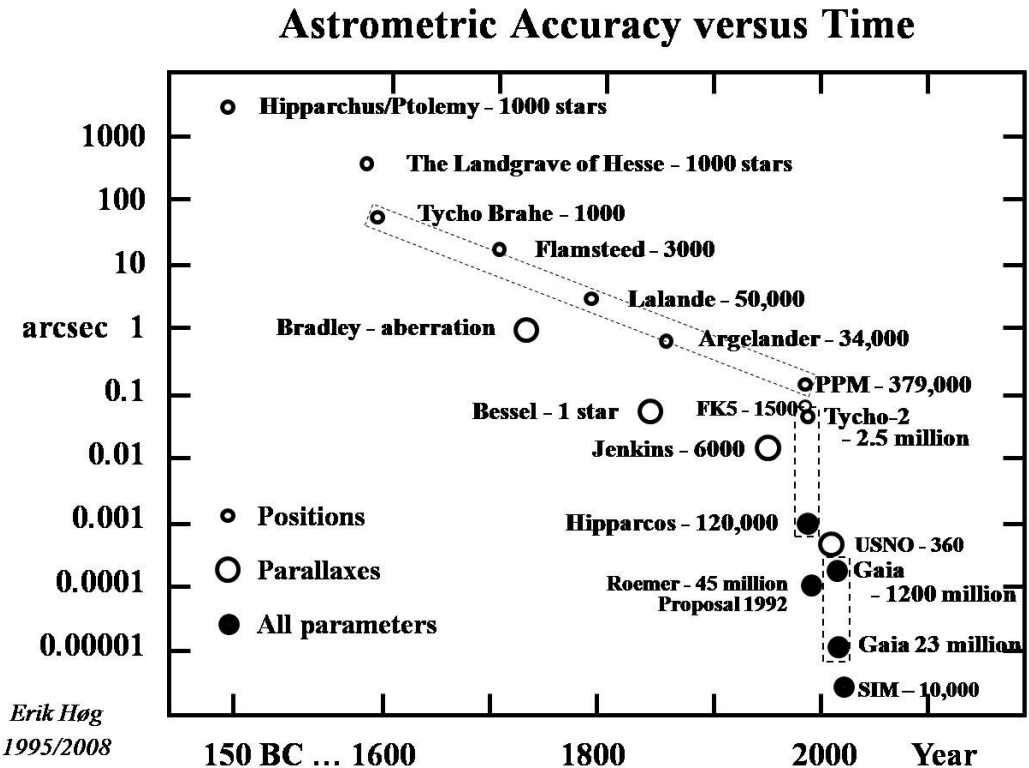

*Fig. 1a.* Astrometric accuracy during 2000 years: Høg-2008. The accuracy was greatly improved shortly before 1600 by Tycho Brahe. The following 400 years brought even larger but much more gradual improvement before space techniques with the Hipparcos satellite started a new era of astrometry

The first version of this diagram is shown in the appendix as Høg-1995. It was drawn in 1995 in correspondence with several colleagues from the Hipparcos Science Team and appears as Fig.1 in Vol.1 of The Hipparcos and Tycho Catalogues. Two principles were followed in this diagram, but apparently not always in the other diagrams: it shows catalogue errors of single stars rather than errors of single observations and it only shows some of the most accurate catalogues of the given time. To be precise: I am plotting the *median external standard error per star in the catalogues, if available.* In most catalogues bright stars are more accurate than faint ones, but since only one number can be accommodated in the diagram, I find a median value most representative which then corresponds to *the error near the faint end of a catalogue.*

Changes in the diagram compared with Høg-1995/2005 are: Hipparchus/Ptolemy 60' instead of Hipparchus 20', The Landgrave of Hesse is the correct English name instead of Hessen, Flamsteed 20" instead of 12", and 3000 stars instead of 4000, Lalande is now included, for Argelander a larger catalogue of 34000 stars at 0.9", PPM, FK5 and Tycho-2 slightly corrected, Roemer proposal 1992 is included because this proposal led to Gaia and the other astrometry satellite projects DIVA, FAME, and JASMINE. Gaia is here plotted with 1200 instead of "many" million stars, and Gaia is shown with two dots in order to give more information. Bradley-aberration is included, USNO updated to 360 stars instead of 100; the dot for SIM has been placed 3 muas with 10,000 stars, although 1300 stars would be more correct at this accuracy, but space in the diagram is limited; see further explanations in the following section on sources.

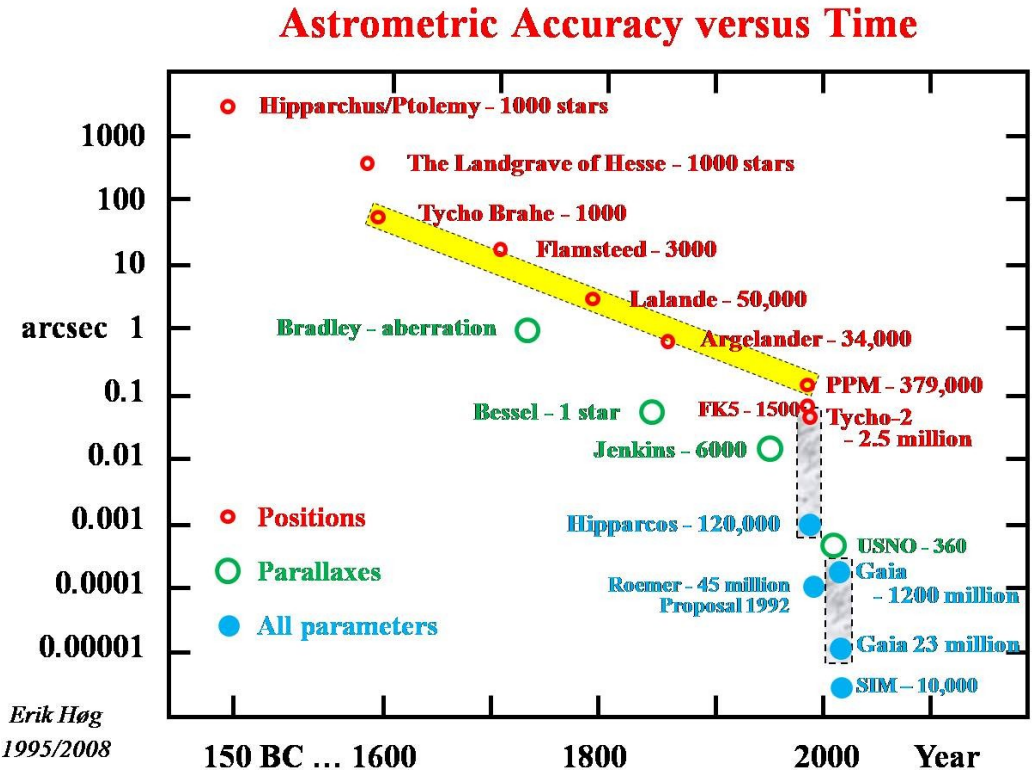

**Figure 1b. Astrometric accuracy during 2000 years, Høg-2008.** Colour version of Fig. 1a

The points are placed at the mean observation epoch, except the compilation catalogues FK5, PPM, and Jenkins which are placed at the year of publication and with the accuracy of the positions in FK5 and PPM in that year. The circles refer to "positions" and "parallaxes", the word "best" from the previous diagram has been omitted as being misleading because we want to show median values of the standard errors in each catalogue, representative for the bulk of stars in a catalogue. It has been suggested to include more information on the most accurate stars in each catalogue, but the diagram would be more complicated and it would be very difficult to collect the information and to present it well in a graph.

## Explanation to the diagram Høg-2008

### Brief explanation

Errors of star position coordinates and parallaxes in accurate catalogues are shown in Fig. 1. Tycho Brahe achieved a jump in accuracy of positions through the first "big science" in history. After four centuries with gradual improvements another much larger jump in accuracy was obtained by the ESA satellite giving the Hipparcos and Tycho-2 catalogues containing a total of 2.5 million stars.

**Detailed explanation**

Errors of star position coordinates and parallaxes in accurate catalogues are shown Fig. 1. This means the *median external standard error per star in a catalogue, if available*. In most catalogues bright stars are more accurate than faint ones The representative median error, dominated by faint stars, is given for most catalogues.

It appears that the Landgrave of Hesse was able to measure positions with errors about six minutes of arc, ten times better than Hipparchus/Ptolemy in the Antique. A few years after the Landgrave and thanks to generous support from the king of Denmark, Frederik II, Tycho Brahe reduced the errors by a further factor of six. The Landgrave and Tycho, both wanted to equal Hipparchus by reaching the same number of 1000 stars. A period of 400 years followed with gradual improvement of the accuracy as astronomers always made use of the best technical possibilities of their time, especially with better time-keeping equipment and accurate manufacturing of mechanics, optics, and with electronics. The accuracy was improved by a factor about 250 in 400 years, i.e. a factor four per century, and the number of stars was greatly increased.

The introduction of space techniques, however, with the Hipparcos mission gave a veritable jump in accuracy by a factor of 100 with respect to FK5, the most accurate ground-based catalogue ever. Hipparcos obtained a median accuracy of 0.001 arcsec for positions, proper motions and parallaxes of 120 thousand stars. The positions even in the Tycho-2 Catalogue with 2.5 million stars are as accurate as the positions in FK5 containing only 1500 bright stars. Tycho-2 includes annual proper motions, derived from Tycho-2 positions and more than 140 ground-based position catalogues, but no parallaxes. The median standard error for positions of all stars in Tycho-2 is 60 mas, and it is 7 mas for stars brighter than 9 mag. The median error of all proper motions is 2.5 mas/yr.

The points marked "parallaxes" might be labelled "small-angle astrometry" or "relative astrometry", and all ground-based measurements of parallaxes are of that kind. This is about ten times more accurate than large-angle astrometry which was required to measure the positions shown in the diagram. The first such point is "Bradley – aberration" shown at 1.0 arcsec, the accuracy which Bradley obtained for the constant of aberration with his zenith sector. The accuracy of ground-based parallaxes begins with Bessel's single star in 1838, followed by a factor 100 improvement in accuracy at the U.S. Naval Observatory in Flagstaff since about 1990 for faint stars.

"All parameters" means that about the same accuracy is obtained for annual proper motions, positions and parallaxes, as was in fact achieved with Hipparcos, for the first time in the history of astronomy. The Roemer proposal of 1992 (Høg 1993) introduced CCDs in integrating scanning mode in a space mission, instead of photoelectric detectors as in Hipparcos. Roemer promised a factor 10 better accuracy than Hipparcos for many more stars, and a development began which led to the Gaia mission due for launch in 2011. For Gaia an improvement by a factor of 100 over Hipparcos is predicted for the 23 million stars brighter than 14 mag, i.e. 10 microarcsec median error. The median accuracy is expected to be 180 microarcsec for the 1200 million stars in the Gaia catalogue brighter than 20 mag, much better than the accuracy of Hipparcos. The two dots for Gaia thus show the expected accuracy for bright and faint stars. Finally, in view of the expected Gaia results, studies are due about the scientific goals for ground-based optical astrometry after Gaia.

# Sources for astrometric accuracy

Here follow the sources and reasoning for the accuracies used in the diagram Høg-2008 of astrometric accuracy, and for Tables 1, 2 and 3 in Høg (2008b) "Selected Astrometric Catalogues", where the references are found, if they are not included in the present report.

## The standard errors

Internal errors of observations are obtained by analysis of repeated observations of the same stars at different times, as is usually done in meridian observation catalogues, e.g. in case of USNO (1920) from the n=10 observations. I have then derived the error for Table 1 by division with sqrt(10) because nothing else is available, but this "internal catalogue error" is not given in the catalogue, and it is certainly too small because of the unknown systematic errors.

The three tables should ideally contain the "external errors" of a catalogue entry as would be obtained from a comparison with a more accurate catalogue. Such comparison could be carried out with any of the older catalogues using the now available Hipparcos and Tycho-2 catalogues, if anyone should wish to do so. This has been done for FK5 by Mignard & Froeschlé (2000), and I have used this comparison to derive below that the errors given in the FK5 of positions at the mean epoch and of the proper motions should be multiplied by a factor about 1.6 to obtain external errors. For any other historical catalogue it would be sufficient to take a representative sample of less than a hundred stars for a comparison, but even that would be no small task. Most interesting would be the following catalogues where the errors in Table 1 may be wrong by a factor two: First priority has Wilhelm of Hesse, Flamsteed, Lalande, and Argelander; please inform me if any such study already exists. A thorough comparison of Bradley/Auwers with Hipparcos by Brosche & Schwan (2007) is mentioned below.

In some cases reliable external errors have been derived, e.g. for the Hipparcos and Tycho-2 catalogues in the publications, and for the parallaxes in Jenkins' catalogue by Hertzsprung (1952). In case of Perth70 it is also believed that a reliable external error is known, as explained below at Perth70. The distinction between external and internal errors of catalogues is important for detailed comparisons, but it is difficult in many cases, if at all possible, to find sufficient information about this matter, and it cannot easily be presented in one line of a table. Internal errors are sometimes placed in brackets.

The standard errors in the tables are sufficient for the original purpose, to show the pace of development of astrometric accuracy over very long periods of time. But much care is needed in comparing within short intervals. I have below in some detail compared four meridian circle catalogues from within one century, i.e. two USNO catalogues from observations around 1907 and 1945 are compared with each other, with the Perth70 catalogue observed about 1970, and with the CMC1-11 catalogues observed about 1991.

It appears that the progress in accuracy and efficiency of meridian circles is rather modest in the first half of the 20[th] century where visual techniques were used, but the progress is very large in the second half thanks to photoelectric techniques and automatic control of micrometer and telescope. This large progress is independent of the Hipparcos mission, but the further progress thanks to the Tycho-2 Catalogue and recording with CCDs is truly tremendous.

The "accuracy of catalogued star positions" is the title of section 3.2.4 in Eichhorn (1974). He discusses theory and practice of this matter in the past where the available means of computation called for simple methods, and in his own time where electronic computers had made rigorous numerical methods feasible. In section 2.2.8 Eichhorn discusses "the accidental accuracy of relative visual positions". He includes three tables adapted from Cohn (1907b), not (1970) as a typo has produced. Some trivial mistakes both in Eichhorn's extract and in Cohn's original paper make the use of the tables cumbersome, as I discuss below at Bradley/Auwers. This is meant as a call for caution.

## Sources for the diagram

The accuracy of an observation catalogue of positions is plotted at the mean epoch, while the catalogues FK5 and PPM, compiled from observations with many instruments, are plotted at the year of publication. The Jenkins compilation of parallaxes is also plotted at the year of publication.

**Hipparchus/Ptolemy 1 degree  at 150 B.C.**
Ulugh Beg also obtained 1 degree accuracy in 1437, but he is not represented in the diagram. The catalogue in the Almagest by Ptolemy is the oldest extant star catalogue. It has been proposed that this catalogue is identical with that of Hipparchus, but this is not supported by Shevchenko (1990). The catalogues of Ptolemy and Ulugh Beg are nearly equivalent in merit, according to Shevchenko. They both have overall systematic longitude errors about one degree, and the systematic error has a scatter about one degree. The root-mean-square errors of the positions of the zodiacal stars in the two catalogues are about 20 arcminutes=1200"=0.33 deg, i.e. within constellations. Shevchenko explains the analogies as due to the fact that the Samarkand astronomers used the equipment and  methods described in Almagest.

Eichhorn (1974) p. 101, says that the rms. errors of ecliptic latitudes and longitudes in Ptolemy's catalogue are 0.58 and 0.37 degrees, respectively, but I will stick to Shevchenko.

For the diagrams we have hitherto always shown Hipparchus with 1200". This is really a local internal error within constellations and I aim at plotting the median external standard error per star which would be 1 degree, and the name should be Ptolemy, not Hipparchus. I have changed the value to 3600" and the name to Hipparchus/Ptolemy; it would be too sad to omit Hipparchus' name entirely.
Very recently, 27 August 2008, F. Mignard informed me of an unpublished study made in 2001 where he compares Ptolemy's catalogue with Hipparcos data. He finds a standard deviation of 0.5 degree using a robust estimator. The following discussion by mail between Mignard and Arenou shows that some issues   deserve a closer study. For the time being I will stick to the one degree error, according to Shevchenko (1990), published recently in a refereed journal.

**Landgrave of Hesse 360" about 1570**
It appears that Tycho Brahe's a little older colleague, Wilhelm IV, called The Wise, Landgrave of Hesse-Kassel (1532-92) was able to measure positions much better than Hipparchus/Ptolemy in the Antique. Eichhorn p. 101 gives an rms error of 6' for the catalogue of 1004 stars, published in 1594 by Wilhelm and Christof Rothmann.

**Tycho Brahe 60" at 1586**
The accuracy of Tycho Brahe's instruments has been studied by Wesley (1978). For the best of Tycho's nine fundamental stars, he finds an accuracy of 25" for individual measurements with some of the six instruments he considered. He says: "For the majority of the stars that appear in Tycho's final catalogue the overall accuracy may be much less; for there were fewer measurements taken with them...". I adopt 60" as still plausible for the median standard error.

**Flamsteed 20" at 1700**
Eichhorn gives an rms. error for Flamsteed of 2", which must be a misprint for 12" since that is what some others assume. Chapman (1983) cites Schuckburgh and Pearson (respectively 1793 and 1819) for an error of 10"-12", here is probably where many others took the values.

Other values are quoted by Nielsen (1968). He quotes Argelander (1822) for finding an internal mean error about 7" and an external about 60". He quotes Piazzi (1813) for a long statement which I condense to: an external mean error of 30" and individual errors exceeding 60". This together, I settle on 20" for the catalogue which differs a lot from the conventional 12", but I cannot avoid it.

**Lalande 3"**
3" from Mineur-1939, 3" from Turon-2007. Arenou (2008) confirms the 3" and calculates the mean epoch to 1795. Lindhagen in 1849AN.....28..129L derives that the number of different stars in Lalande's catalogue is perhaps 40,000, much smaller than the number of entries in the catalogue of about 50,000. The accuracy of 3" can only be valid for the best part of the positions in the catalogue, which is known to contain many errors. F. Mignard notes in a recent mail: "... the Histoire Celeste is a very valuable and extensive description of the sky around 1800 (celebrated as such for example by Olbers), but of low interest in term of astrometric quality. ... In short it is the equivalent in the early 1980 of the SAOC compared to FK4 or GC. ... Histoire Celeste is an astronomical landmark for sideral astronomy, but not for astrometry."

**Argelander 0.9" at 1856**
Eichhorn p. 147, gives a mean error of 0.9" for Argelander's large catalogue of 33811 stars from 1867. On p. 143 Eichhorn explains that he assumes that two observations were always combined to give the published position. In the first versions of my diagram I took Argelander's catalogue of 26425 stars from 1844 for which the error is given as 1.1". I think it is more appropriate to take the larger catalogue, but it makes no significant difference for the diagram.

**FK5 62 mas plotted at 1988**
The catalogue FK5 states on p. 8 an average "mean error" of individual positions at the mean epoch about 23 mas and of proper motions 0.75 mas/yr. This implies an individual standard error in 1991 of 38 mas, but the error is in fact 62 mas, or 1.6 times larger, as may be concluded from a study by Mignard & Froeschlé (2000) who have compared FK5 with Hipparcos. Their tables 3 and 4 show the local systematic differences, averaged over 230 square degrees, between Hipparcos and FK5 positions at the Hipparcos epoch of 1991.25. From the tables we find an rms value of 58 mas. Adding the 23 mas gives 62 mas which we consider to be a reasonable estimate of the individual standard error in 1991 and which is therefore adopted for the last column in Table 2.

We tentatively assume that the above factor 1.6 should be applied to the errors on p. 8 giving 40 mas instead of 23 for the error of positions at the mean epoch which is then adopted for FK5 in Table 1. The individual proper motion error becomes 1.2 mas/yr instead of 0.75 and this is adopted in Table 2.

**PPM 144 mas plotted at 1992**
For Table 2 the standard errors of positions and proper motions are adopted for north and south as given in the catalogue, volumes 1 and 3. This combines to 144 mas for positions for the whole catalogue. It is essential to include PPM in the diagram because it is the last large purely ground-based catalogue before the Hipparcos results appeared. It is therefore more fair to take PPM for comparison with the large catalogues based on space observations, rather than to take the FK5 containing only the very few, very best observed bright stars.

**Tycho-2  60 mas at 1991**

Tycho-2 includes positions and annual proper motions, derived from Tycho-2 positions and more than 140 ground-based position catalogues, but no parallaxes. The median standard error for positions of all stars in Tycho-2 is 60 mas, and for stars brighter than 9 mag it is 7 mas. The median error of proper motions is 2.5 mas/yr.

**Hipparcos  1 mas at 1991**

Hipparcos obtained the median accuracy of 1 mas for positions, annual proper motions and parallaxes of 120 thousand stars.

**Roemer  0.1 mas at 1992**

The Roemer space mission of 1992 (Høg 1993) proposed to use CCDs in TDI mode and promised a factor 10 better accuracy than Hipparcos for many more stars, viz. 0.1 mas as median accuracy for the 45 million stars brighter than 15 mag, and an error better than Hipparcos for the 400 million stars brighter than 18 mag. It is included in the diagram because the Roemer idea led to the Gaia mission, and to the studies of DIVA and FAME. The use of CCDs as modulation detectors was proposed by Høg & Lindegren (1993) but this idea was not further pursued after the superiority of CCDs in scanning mode had been realized.

**Gaia  10 and 180 microarcsec at 2015, two dots plotted**

```
Table A. Median astrometric accuracy for Gaia as
function of magnitude. Courtesy of Jos de Bruijne.
    ====================================================
      (1)         (2)      (3) (4) (5)
    ----------------------------------------------------
    G=06.0-13.0  10.200    8   6   4
    G=13.0-14.0  12.700   11   8   6
    G=14.0-15.0  24.567   17  13   9
    G=15.0-16.0  50.340   27  20  13
    G=16.0-17.0  94.486   42  32  21
    G=17.0-18.0 170.625   67  51  34
    G=18.0-19.0 308.589  112  84  56
    G=19.0-20.0 562.010  196 147  98
    ====================================================
    Column (1) G magnitude range.
    Column (2) Number of stars in the G magnitude range (unit is million
stars); the sum of column (2) is 1233.517 which is the total number of
stars used in the Gaia galaxy model (1.2 billion).
    Column (3) Median parallax error for all stars up to the faint
magnitude of the magnitude range (unit is muas).
    Column (4) Median proper-motion error for all stars up to the faint
magnitude of the magnitude range (unit is muas per year).
    Column (5) Median positional error for all stars up to the faint
magnitude of the magnitude range (unit is muas).

    Example: "G=17.0-18.0", "column (4) = 51 muas per year" means that
the median proper-motion standard error for all stars brighter than
G=18 mag (all stars in the range G = 6-18 mag) is 51 muas per year.
```

The Gaia mission will be launched in 2011 and a factor of 100 over Hipparcos is predicted for the 23 million stars brighter than 14 mag, i.e. 10 microarcsec median error. The median accuracy for parallaxes and annual proper motions of the 1200 million stars in the final Gaia catalogue is expected to be about 180 microarcsec, much better than the accuracy of Hipparcos. This appears from the following Table A, including explanations by J. de Bruijne.

**SIM  3 microarcsec**
The dot for SIM has been placed at 3 muas with 10,000 stars, although 1300 would be more correct at this accuracy, but space in the diagram is limited. In fact, a dot at 10 muas with 10,000 stars and another dot at 3 muas with 1300 stars would be more correct.

The NASA interferometric mission (Unwin et al. 2008, Shao 2008) is expected to give global astrometry with few microarcsec accuracy after a five year mission down to 20 mag for more than 10,000 stars. Table 7 in Unwin et al. (2008) gives expected performances, especially 4-20 muas for 10,000 stars of -1.4-20 mag in key projects and 3 muas for 1300 stars of 9-10.5 mag in the astrometric grid.

Narrow angle accuracy of 1 microarcsec per 20 minutes integration is predicted for stars of 6-9 mag. The SIM project has passed all milestones in over ten years of design and development, but is not yet an approved mission and the launch will be after 2014-15.

**Bradley-aberration  1" at 1728**
The points marked "parallaxes" might be labelled "small-angle astrometry" or "relative astrometry", and all ground-based measurements of parallaxes are of that kind. This is about ten times more accurate than large-angle astrometry required for the stellar positions in the diagram. The first such point is "Bradley – aberration" shown at 1.0 arcsec, the accuracy which Bradley obtained for the constant of aberration with his zenith sector. According to Arenou (2008) using Flamsteed observations (1689-1697) the precision of aberration can be found within 1.1". This information is from F.G.W. Struve, *Ueber Doppelsterne nach den auf der Dorpater Sternwarte mit Fraunhoffers grossem Fernrohre von 1824 bis 1837*, 1837, page 95:
http://books.google.com/books?id=MEMJAAAAIAAJ&pg=RA2-PA95&lpg=RA2-PA95&dq=flamsteed+aberration+1689+1697&source=web&ots=0YS4rHY2eg&sig=1Kh53ZgrXvLb1BGUeLjbLXbYhhc&hl=fr&sa=X&oi=book_result&resnum=1&ct=result

**Bessel  60 mas at 1838**
The accuracy of ground-based parallaxes begins with Bessel's single star in 1838. The 60 mas is based on the analysis below for Table 3. Previous diagrams had, e.g., 60 mas in Høg-1995 and 300 mas in Mineur-1939.

**Jenkins  15 mas plotted at 1952**
This accuracy for the parallaxes in Jenkins' catalogue was derived by Hertzsprung (1952).

**USNO  0.6 mas at 2008**
At the U.S. Naval Observatory in Flagstaff, relative parallaxes for 357 faint stars has been obtained with a standard error of 0.6 mas, according to W. van Altena/ C. Dahn (2008 priv. comm.).

*The above sources are usually NOT repeated below at the three tables!*

## Sources for Table 1

### Hevelius 20"
Eichhorn (1974) does not give a value for the accuracy of Hevelius. Chapman (1983) p. 136, gives the values 15" to 20" with a reference to Schuckburgh and Pearson from respectively 1793 and 1819 which I have not read. But I adopt the value 20" for my Table 1. Chapman in fact plots a value at 25".

### Rømer 4"
Ole Rømer's only surviving observations with meridian circle in 1706, written in the so called Triduum (three nights), were published by Horrebow (1735). They are discussed by Nielsen (1968) where further references are given. On three nights, 250 transits were observed of 88 stars, the Sun, the Moon, and all the planets known at that time, from Mercury to Saturn. Nielsen has determined the errors of a subset of the star positions by comparison with newer observations and finds external errors in RA of 3.4" and in Dec. 4.5", which I combine to the one number 4". This seems to agree with a statement by Piazzi (1813), according to Nielsen.

### Lacaille 6"
6" from Mineur-1939; unfortunately I know no primary source. See more below under Piazzi.

### Bradley/Auwers 1.1"
Turon-2007 shows 2" for Bradley/Bessel. This is in accordance with the following analysis.

Rather than Bessel's version the one by Auwers should be used, thus Bradley/Bessel/Auwers, which has probably been used for the German fundamental catalogues from Auwers' FC to FK5. Bradley's precision was in general 1", if one should believe
http://www.flamsteed.info/fasbradley_files/page0002.htm.

Eichhorn's table II-1 on p.66 gives internal errors of a *single observation*, which is not stated by Eichhorn, but it is by Cohn (1907b) on p.269. The errors are 0.16 s and 1.92" for Greenwich in 1755, i.e. Bradley. But table II-3 gives 0.16 s and 1.3" for one observation by Bradley. Using the formulae in the footnote to table II-1 give however 0.18 s = 2.7" for Dec=0 and 1.92" for zenith distance =0. Rounded to 2" for Bradley/Bessel in accordance with Turon2007. The value is for a single Bradley observation, which may apply to the bulk of the 3222 stars in the catalogue. He did probably make many more observations per star for those few hundred used in the German fundamental catalogues.

It is not clear from Cohn (1907b) or Eichhorn whether this accuracy refers to Auwers' reduction of Bradley/Bessel, and this makes a difference. The version Bradley/Bessel/Auwers obtains an increase of weights compared with Bradley/Bessel of the factors 1.75 in RA and 1.4 in Dec, according to Auwers as quoted by Cohn (1907b), p.269. This would lead to 2"/sqrt(1.6)=1.6". This is an example how difficult it can be to get a half-way reliable standard error for a catalogue position in Table 1.

Very recently, however, I received Brosche & Schwan (2007) from the first author. It contains a direct comparison of Bradley/Auwers and Hipparcos. For 2450 catalogue values out of the 3268 entries the rms values are 1.2" and 1.0" for respectively RA and Dec. This gives 1.1" for my Table 1, in reasonable agreement with the above 1.6". The weight has then been calculated using for simplicity the N=3222 in the preceding column, although a smaller number would be more correct since only N=2450 were good enough for the comparison.

**Piazzi  1.5"**
1.5" from Mineur-1939; unfortunately I know no primary source. F. Mignard wrote in a recent mail: "The most interesting report I found [on Lacaille and Piazzi] is by R. Grant (History of physical astronomy (London 1852) in chap. XIX on the Catalogues of fixed stars from Hipparchus to his time. He praised very much Lacaille care in obtaining absolute measurements on few reference stars. Same opinion about Piazzi work in Palermo using again the 36 fundamental stars of Maskelyne before and building himself a fundamental catalogue of 120 stars before forming his catalogue of 7600 stars. Every stars have been observed several times and "this work is justly considered to be one of the most important that has ever been executed by a single individual"."

**Küstner 1908, AC, Stoy 1968, SAOC 1965**
Standard errors are taken from Eichhorn p.157, p.279, p.162, p.209.

**USNO 1920 and USNO 1952,  about 0.15 internal errors**
Standard errors are taken from the references in Høg (2008b). Only internal errors are given in the publications as derived from the repeated observations of the same star on different night. These internal errors are divided by sqrt(n) for inclusion in Table 1, because no external error is available. The details for these catalogues are as follows.

USNO (1920) gives the typical internal errors of one observation for RA and Dec on p. A79 and A139 as 0.50" and 0.48", respectively, which combine to 0.49". The probable errors used by USNO in those year are converted to standard errors by multiplication with 1.50. With n=10 the 0.15" in Table 1 is obtained.

USNO (1952) gives the typical internal errors of one observation for RA and Dec on p. 375 and 377 as 0.32" and 0.45", respectively, which combine to 0.37" (as average of the weight from each coordinate). With n=6 the 0.15" in Table 1 is obtained.

These two catalogues are based on respectively 45,000 and 31,000 meridian observations, both obtained in eight years in Washington DC around 1907 and 1945. The development in this period improved the internal error of an RA observation from 0.50" to 0.32" while an observation of Dec stayed about 0.46".

**GC    0.15" and 10 mas/yr**
According to Eichhorn (1974) p. 204: "... in the General Catalogue the accidental rms. errors of the positions vary strongly from one star to the next. However, at the epoch they are on the average about 0.15" in both coordinates, and rise to an average of at least 0.70" in 1965 because of the uncertainties of the proper motions (Schlesinger and Barney 1939a)."

Since the (mean) epoch for GC is 1900 this implies a standard error of the proper motions in GC of sqrt(0.7^2-0.15^2)/65 = 0.0105"/yr. The value of 10 mas/yr is adopted for Table 2, but is not stated by Eichhorn; it is however in accordance with the error given by Scott (1963). For Table 1 the value 0.15" is adopted.

**Perth70  0.15" external error**
Standard errors are taken from the reference in Høg (2008b). Internal standard errors of one observation reduced to zenith is 0.17" and 0.27" for RA and Dec, respectively, cf. Eq. 15, and 0.10 mag for the photoelectric photometry in the visual band. External errors have been derived from observations of circumpolar stars, taking asymptotic errors into account. The typical standard errors of a catalogue position for a program star with four observations are accordingly 0.12" and 0.20" in respectively RA and Dec. This combines to an error per coordinate of 0.15", adopted for the Table 1.

These internal errors of one Perth70 observation obtained about 1970 are about half the size of those in USNO (1952) and 100,000 such observations were obtained in 5 years in Perth, Western Australia, compared with the 31,000 in 8 years in Washington DC. Thus, a considerable progress in meridian observations were achieved in those years using the photoelectric semi-automatic instrument of the Hamburg-Perth Expedition.

The error of a catalogue coordinate is given as 0.15" in both cases, but they cannot be compared directly because the USNO error is an internal error, the Perth70 error is external.

### CMC1-11 1999 and CMC14 2005
Information from the web supplemented by correspondence with D. Evans is shown in Table 1 and explained in Høg (2008b). The CMC1-11 catalogues were obtained with a photoelectric slit micrometer, similar to the one used for Perth70, but with automatic control of micrometer and telescope giving a much higher efficiency. Observed in the better seeing on La Palma and during 14 years instead of 5 years for Perth70 the weight of the catalogue is larger by a factor 30. This is the last meridian circle catalogue in the table where large-angle astrometry is performed. The CMC14 is observed with CCDs in drift-scan mode and the reference stars of the Tycho-2 Catalogue are used for the resulting small-angle astrometry.

### USNO-B1.0 2002, UCAC2 2003, GSCII 2005
Information from the web supplemented by correspondence with S. Urban.

### 2MASS
The 2MASS all-sky catalogue was obtained by two highly automatic telescopes with 1.3 m aperture equipped with HgCdTe detectors sensitive in the J,H,K bands (1-2 microns) with a limit of 17 mag in J. An accuracy of 0.5" for positions was expected, in fact 0.08" was achieved according to N. Zacharias.

## Sources for Table 2

### Auwers' FC and NFK
For lack of better knowledge, the values are estimated, based on FK3 and N30, therefore the question mark after each of the values.

### FK3, GC and N30
Scott (1963) gives an overview, including the proper motion errors for FK3, GC, and N30.

### FK4 and FK5
The individual proper motion error becomes 1.2 mas/yr for FK5 instead of 0.75, as derived above under FK5. The error given for FK4 is simply set a bit larger, 2 mas/yr, for lack of better knowledge.

### SPM3, UCAC2, USNO-B
All data were received from N. Zacharias in October 2008.

**More on proper motions from Arenou**

Arenou (2008) mentions two important catalogues: "One led to the discovery of the astrometric binaries: I think that Bessel had 38 stars among which 36 zodiacal stars from Bradley as first epoch (1755) or Maskelyne?. Then, I understand that Argelander had proper motions for 560 stars in 1835 (see 1837MNRAS...4...82A) of which he used 390 to confirm the solar motion."

**More on proper motions from Zacharias**

"Traditionally proper motions of stars have been determined by comparing absolute positions (on a fundamental system) at different epochs. With the improvement of the photographic technique in the middle of the 20th century it became possible to image distant galaxies in a sufficient number to determine absolute proper motions field by field with differential, small angle measures of pairs of plates taken many years apart, covering large areas of the sky for galactic dynamics studies (Wright 1950). This lead to the Northern Proper Motion (NPM) program using the Lick 50 cm double-astrograph (Klemola et al. 1987) and its southern counterpart, the SPM, using the Yale / San Juan instrument of similar design (Girard et al. 1998). These plates, spanning an epoch difference of about 25 years were initially measured with slow but accurate PDS machines for selected stars. By the turn of the century all applicable plates were measured with the PMM at the Naval Observatory Flagstaff station to obtain positions of all stars to 18th magnitude. Reductions are still in progress as part of the UCAC3 effort. Even after no photographic emulsions are any longer in use in astrometry, the development of plate measure machines progressed in the late 20th and early 21st century to allow extraction of all astrometric (and photometric) information available in those data materials."

**On reference catalogues**

The fundamental catalogues, Auwers FC to FK5, contained too bright and too few stars, FK5 only 1535, to serve directly as a reference catalogue for the reduction of photographic plates. Special observing campaigns were therefore organized to provide denser nets of reference stars for the various photographic surveys, e.g., the AGK3R of 21,499 stars was observed with meridian circles in the 1950s while the AGK3 survey of the northern sky was made. Subsequently, a list of 20,495 Southern Reference Stars was defined and these stars were observed in an international collaboration agreed at the IAU Assembly in Moskau 1958. The resulting SRS catalogue combined with the AGK3R was called International Reference Stars (IRS) which was completed in the 1990s.

The more detailed history of the IRS and the larger ACRS, Astrographic Catalog Reference Stars, is told in the recent message from T. Corbin which I have slightly edited.

"The IRS project originated in the 1960's when T. Corbin was asked to derive proper motions for the observed positions being compiled from the AGK3R observing program. This was to allow the AGK3R positions to be brought to the epochs of the individual AGK3 plates. Only meridian circle catalogs were to be used in order to avoid the color and magnitude terms that older astrograph catalogs would introduce. Catalogs that had been observed using screens were employed to extend the FK4 system to fainter magnitudes, and that extension provided the reductions for the other catalogs. The same thing was done for the SRS.

The IRS then resulted from combining the AGK3R and SRS, each reduced to FK5, and, using the same approach for reducing the older catalogs, computing new mean positions and proper motions on the FK5 system. The FK5 Part II was compiled by combining the FK5 based positions and motions for both FK4 Sup stars selected at Heidelberg and IRS selected for the list at USNO.

ACRS grew from a USNO collaboration with P. Herget in the early 1970's to get improved plate constants for the Astrographic Catalog. The Bordeaux zone was selected, and Corbin compiled a more dense catalog for this part of the sky by combining IRS data with astrograph programs. Herget obtained a significant improvement in the plate solutions, and this showed that compiling such a catalog on a global scale for the reductions of all AC zones would be worth the effort.

The ACRS (Astrographic Catalog Reference Stars) is basically an extension of the IRS. Particular attention was given to minimizing the systematics in order that the 320,211 stars would represent the FK5 system at the CdC epochs. The PPM was being compiled at Heidelberg at about the same time. PPM includes the AC data, and this is the main difference between it and the ACRS. Both catalogs are based on IRS.

S. Urban used the ACRS database, in combination with Tycho-1 to create a new version of ACRS that then gave an improved set of results for the AC zones. This was all combined to produce the ACT catalog which was quickly superseded by a new version of the proper motions using Tycho-2 results. These were combined with the Tycho-2 observed positions to give the final Tycho-2 Catalogue.

IRS contains 36,027 stars, 124 catalogs were used
errors of proper motions - 4.3 mas/yr in RA and 4.4 mas/yr in DEC
errors of positions - 0.22 arcsec in both coordinates

ACRS contains 320,211 stars, 170 catalogs were used
errors of proper motions - 4.7 mas/yr in RA and 4.6 mas/yr in DEC
errors of positions - 0.23 arcsec in both coordinates at 2000
"

## Sources for Table 3

**The three first parallaxes**
This is here at first retold after Stephen Webb (1999) p.71, and then after F.W. Bessel (1838 and 1840), in both cases abbreviated, followed by my conclusions about the standard errors of the three values as adopted for Table 3.

**Quoting Webb (1999):** The parallaxes were:
Bessel 0.31" for 61 Cygni (modern value from Hipparcos: 0.287")
Henderson 1.26" for alfa Cen (Hipparcos : 0.742")
Struve 0.2619" for Vega (Hipparcos: 0.129").
(Webb gives the same modern values for the first two stars, but 0.125 for Vega!)

Struve studied Vega with a wire micrometer on the big refractor in Dorpat. Struve made 17 observations during 1836 which gave a parallax of 0.125" with an uncertainty of 0.05". This was published in 1837. He promised to make more observations and published in 1840 the results of 96 observations made up to 1838. The parallax he obtained this time was 0.2619, more than twice the original result, which cast doubt on both values.

Bessel, meanwhile, studied 61 Cygni with a Fraunhofer Heliometer in Königsberg, using two nearby companions. He began observations in September 1834, but this was interrupted by other work. He returned to the task in 1837 and made 16 or more observations every clear night. As result of his analysis at the end of 1838 he announced a parallax of 0.31" with an error of 0.02".

Henderson studied alfa Cen with a mural circle from Cape. He completed his observations in 1833, and analysed them upon his return to Scotland later that year. He arrived at a parallax of 1.16" with an error 0.11". Before publishing his results, however, he asked a colleague to check his work. In the end he published several weeks after Bessel.

**More now from Bessel :** Bessel (1838) explains his observations and reductions and gives first the annual parallax derived from the star *a* at 8' distance and from star *b* at 12'. They are respectively 0.3690" +-0.0283" and 0.2605" +-0.0278". The combined solution from *a* and *b* gives 0.3136" +-0.0202".

Knowing today the very accurate modern values for all three stars, considering them to be the true values, we can derive the true residuals. For Bessel (1838) it is O-C= +0.026", in good accordance with Bessel's mean error of 0.0202". That would have led to 20 mas for Table 3, but recently I learnt (Arenou 2008) that two years later, Bessel (1840) gives the value 0.3483" with the mean error 0.0141", at 0.061" or more that 4 sigma from the true value. I therefore finally adopt 60 mas for the Table 3, also because Bessel's final value will have been the most trusted at his time. In previous diagrams are found 60 mas in Høg-1995 and 300 mas in Mineur-1939.

**Struve and Henderson:** For Struve's final value O-C= 0.2619"-0.129"=0.133". This is our best estimate of his standard error, and this estimate has a relative standard error of 1/sqrt(2f)=0.71 since there is f=1 degree of freedom. I adopt 100 mas for Table 3.

Henderson's value gives O-C=1.26"-0.742"=0.518", much larger than his own claimed error of 0.11". I adopt the error of 500 mas for Table 3.

**Review and catalogue by Oudemans in 1889**
I quote Mignard from a mail in Aug. 2008: "I came across the attached reference of interest for your current investigations. This compilation of parallaxes was mentioned in the 'Traite d'Astronomie Stellaire' of Ch. André published in 1899. This is given by him as the Catalogue of the known stellar parallaxes. An interesting point is that in 1899, the analysis of a large number (55) of determinations for 61 Cyg led to pi= 0"44." - end of quote from Mignard. This catalogue by Oudemans, see Mignard (2008b), is dated 1889, and the "best" value for 61 Cyg was 0.40", if I read from Tabelle II, i.e. 0.11" too large.

**Bigourdan  50/30 mas and Russell  40 mas**
The values for both are placed in brackets because they are internal, formal errors. A catalogue by Bigourdan (1909) lists trigonometric parallaxes for about 300 stars, a few with up to 40 observations. The consistency of multiple observations indicates a precision about 50 mas per observation, and a median precision of 30 mas may be inferred for the about 200 stars having more than one observation. Many observations are shown (by bold face) to be the average of several measurements by the same observer, including most of the 100 with only one observation.

The catalogue by Bigourdan is very complete for its time, and may be of  interest for further analysis. It is made available in a file, collected by Mignard (2008a).

Russell (1910) presents 52 new photographic parallaxes and claims a standard error about 40 mas.

**Schlesinger 15 mas**
This is my estimate, based on the value for Jenkins.

**Jenkins  15 mas**
The 15 mas are from Hertzsprung (1952). It is perhaps interesting to note that this catalogue from 1952 containing photographic parallaxes of 5800 stars has nearly the same accuracy as claimed in 1840 (see above) by Bessel for 61 Cygni, with heliometer. But of course, Bessel observed only one star, with utmost care and with an excellent instrument, and later observations with heliometers gave a much larger parallax. To reach 15 mas and much smaller systematic errors for thousands of stars required an enormous effort in development and implementation. Strand (1963) gives an overview of parallaxes at that time.

**Van Altena  10 mas**
Bill van Altena has seen the whole Table 3, made no remarks to the rest of it either, and has thus agreed to the information about modern photographic parallaxes.

**Hipparcos 1 mas**
Hipparcos obtained a median standard error of 1.0 mas for parallaxes.

**USNO 0.6 mas and Hubble 0.24 mas**
A better accuracy than 1 mas has been achieved from the ground and with the Hubble Space Telescope for several hundred much fainter stars. This informations was received in correspondence with W. van Altena and the informers are named in the table.

**Parallaxes according to Westfall (2001)**
The numbers of well-measured stars by (year) are about: (1839) 3, (1850) 6, (1862) 10, (1888) 25, (1901) 38. The same source mentions a 1912 catalogue with the parallaxes of 244 stars, determined as follows: 8 with filar micrometers, 83 with meridian transits, 39 by photography, 3 by spectroscopy, and 111 with heliometers.

**More on parallaxes from Arenou (2008)**
"About the number of parallaxes and the reference by Westfall (2001), one can find that in 1846, Peters has 8 parallaxes (Polaris, Capella, i Ursae maj, Groombridge 1830, Arcturus, Véga, alfa Cygni, 61 Cygni), observations between 1842 and 1843, cf FGW Struve, "Études d'astronomie stellaire", 1847, p 94 (vs Westfall: 1850: 6). In 1889, Oudemans, 1889AN....122..193O, there are 46 stars (vs Westfall: 1888: 25). And then, "The Parallaxes of 3650 Stars of different galactic latitudes, derived from photographic plates", 1908PGro...20....1D, Donner et al."

**Present-day catalogues  for astrometric data**
A list of presently widely used or well known catalogues for astronomical and especially astrometric data is provided by Zacharias et al. (2004). The list is intended to give users some basic information with regards to the content and usefulness of each. Within each section the catalogues are listed with progressively more and fainter stars but generally with decreasing accuracy.

# Other diagrams of accuracy

Here follows a series of diagrams, placed in the sequence I have first seen them. This is the sequence in which the reader can most easily follow the development of the diagram ending with the above Fig. 1. But it is not the sequence in which the diagrams have been published. The first two of the kind were published in 1939 and 1983, but they only came to my knowledge in respectively March and May 2008. Only then did I understand what had made the confusion;

these two diagrams by respectively H. Mineur and A. Chapman are shown as Figures 9 and 10.

The diagram Hipparcos-1985 puzzled me in 1985 because a nearly linear development is indicated over 450 years from Copernicus to Hipparcos, even the last piece of 150 years from Simms to Hipparcos fits this line! This cannot be correct, but we had other more urgent tasks in 1985 than to dig deeper here. Four years later the same diagram was used, Hipparcos-1989. In general in these diagrams, one should never draw lines from one point to the next since this indicates that one could interpolate. But it is appropriate to draw a longer line in order to indicate a trend, as has been done in later diagrams, e.g. Høg-1995 and ESA-1998.

Then I saw the diagram Kovalevsky-1990 presented a year later, very different, but again I was puzzled. I wanted to dig deeper, but five years passed before I found the time to make Høg-1995 which was immediately accepted in the Hipparcos Science Team. The jumps in accuracy at Tycho Brahe and at our Hipparcos satellite are clearly seen. Two more versions are shown here as ESA-1998 and Høg-1995/2005.

At the symposium in Shanghai in 2007 Catherine Turon showed the diagram Turon-2007. The smooth curve could give the, I think erroneous, impression that the development had no jumps, but was completely gradual over 550 years from Ulugh Beg to Gaia, though starting to become steeper about 1950.

The diagrams are shown in the sequence they came to my eyes, in the appendix found at:

www.astro.ku.dk/~erik/AccuracyAppendix.pdf

Mignard (mail of August 2008) gives references to further diagrams: "In the book of Walter and Sover (Astrometry of Fundamental Catalogues, Springer, 2000) there is one more diagram of accuracy vs. time on p. 5. The reference is given to: Schmeidler F., 1980, Die Geschichte des FundamentalKataloge, in Astrometrie und Dynamische Astronomie, W. Fricke, Th. Schmidt-Kaler, W. Seggewiss (eds), Mitteilungen der Astron. Gesell. 48, 11-23."

**Acknowledgements:** I am grateful to W.F. van Altena, F. Arenou, P. Brosche, J. de Bruijne, T. Corbin, D.W. Evans, D.W. Hughes, J. Kovalevsky, F. Mignard, M.A.C. Perryman, H. Pedersen, C. Turon, and N. Zacharias for sending information without which the present report could not have been written and/or for comments to previous versions of the report. The documentation has been collected with much support from colleagues, but I must take the responsibility for imperfections which may still be present in spite of considerable efforts on my part.

**Appendix to**
**Astrometric accuracy during the past 2000 years**

**by Erik Høg**

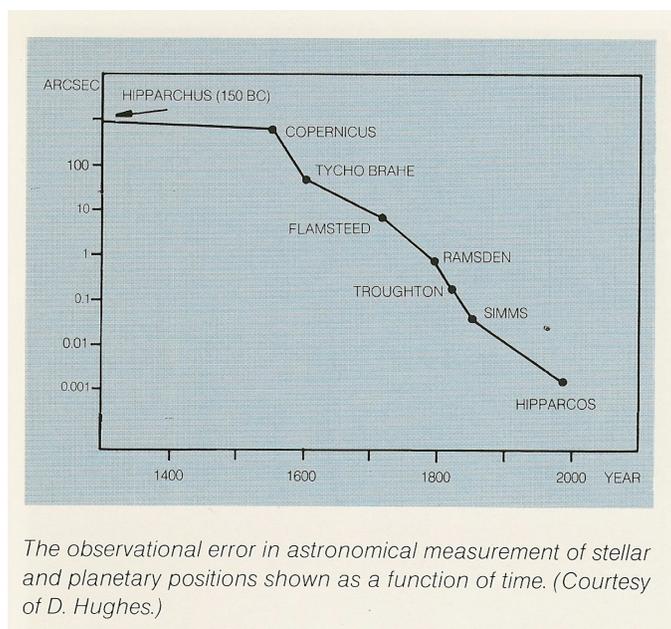

*The observational error in astronomical measurement of stellar and planetary positions shown as a function of time. (Courtesy of D. Hughes.)*

*Fig. 2. Accuracy diagram: Hipparcos-1985*

**Hipparcos-1985**

This diagram appeared in June 1985 in the brochure ESA BR-24, called Ad Astra Hipparcos. I recently wrote to David Hughes as quoted below under Mineur-1939. I then asked Michael Perryman who answered: "I recall seeing such a plot by David Hughes (as he confirms in his mail) although I do not remember where (New Scientist, perhaps?). The first time we used it in Hipparcos that I am aware of was in "Ad Astra" (BR-24, June 1985, p8), but perhaps before. The credit there is given as D. Hughes, but I have no recollection whether the editor (Norman Longdon) had any correspondence with Hughes in preparing that version." See the discussion of this diagram at Chapman-1983.

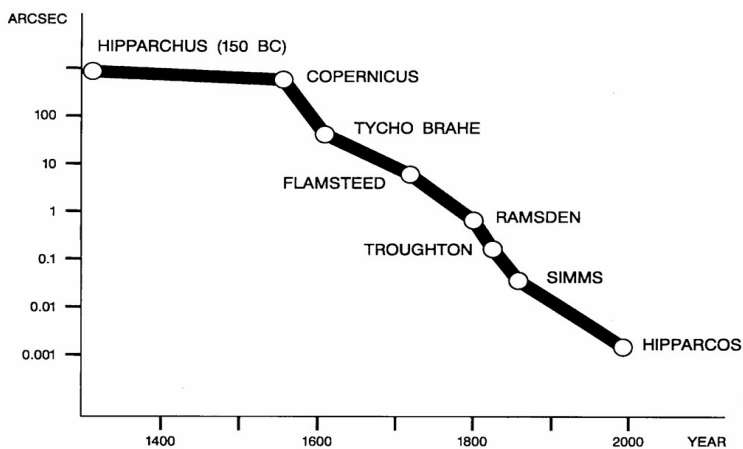

*Fig. 3. Accuracy diagram in ESA SP111.*
*Hipparcos-1989.*

**Hipparcos-1989**

This diagram appeared in ESA SP-1111 as FIG. 1.1 on p. 3, and it is nearly identical to Hipparcos-1985.

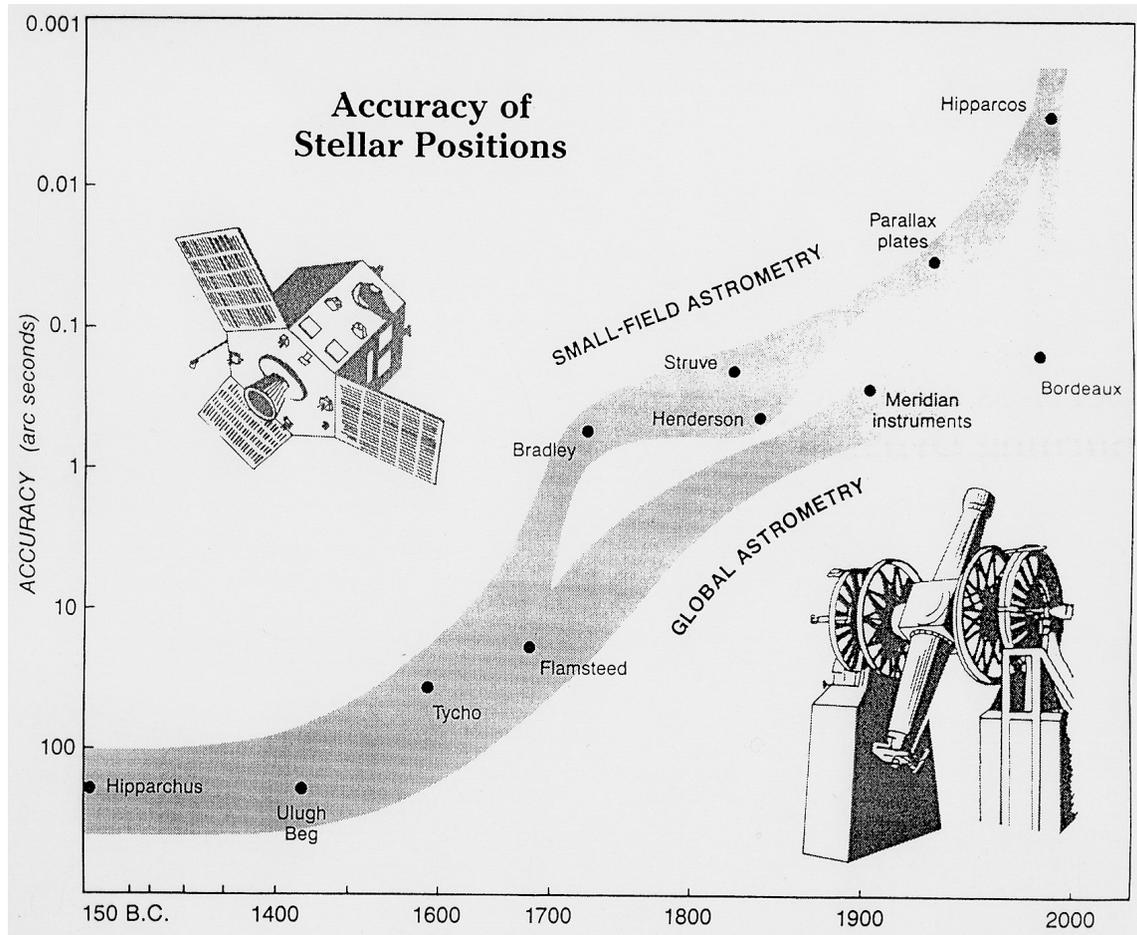

*Fig. 4. Accuracy diagram: Kovalevsky-1990.*

**Kovalevsky-1990**

I believe that Jean Kovalevsky presented the diagram at the Cospar meeting in Den Haag about 1990. I thought at that time that it needed some check and improvement, but only in 1995 did I study this matter carefully and elaborated the new diagram, Høg-1995.

Jean sent me his diagram in October 2007 as I had asked him. He wrote: "I found the one that I append. but I do not know when I projected it, and even whether I draw it or borrowed it from somebody else." I have scanned the viewgraph and I must apologize for any deterioration hereby introduced.

The values plotted in the diagram Kovalevsky-1990 are: Hipparchus and Ulugh Beg both with 200", Tycho 60", Flamsteed 20", Bradley and Henderson 0.5", Meridian instruments 0.5", Struve and Bordeaux 0.2", Parallax plates 0.03", and Hipparcos 0.004". The values for Hipparchus, Ulugh Beg, and Hipparcos deviate a lot from those in other diagrams.

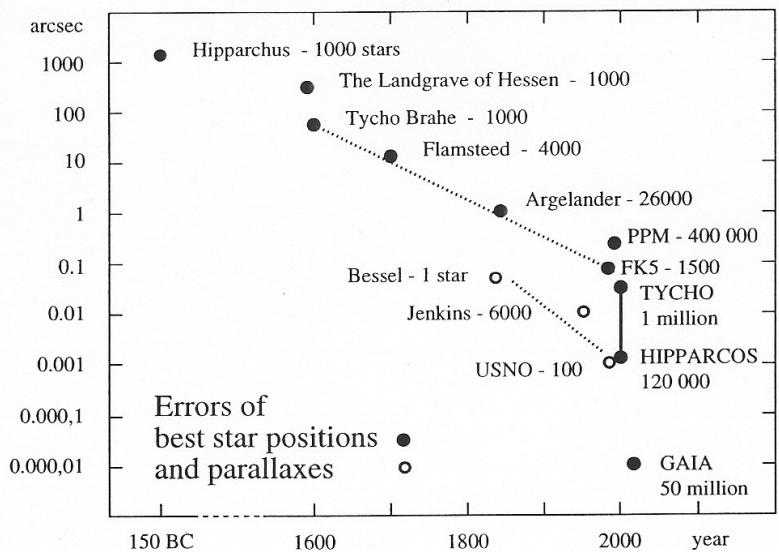

*Fig. 5. Accuracy diagram: Høg-1995.*

**Høg-1995**

This diagram was made in 1995, I profited thereby from correspondence with Michael, Lennart and Uli. It appeared in 1997 (with courtesy E. Høg) as Fig. 1 in Volume 2 of ESA SP-1200, the Hipparcos and Tycho Catalogues from where Fig. 5 has been scanned.

The diagram is included in the Gaia information sheet by Jos de Bruijne, dated 2006-02-13, but there the UCAC2 has been added with 58 million stars at 0.04", and the 50 million stars for Gaia is changed to 1000 million. A change of Tycho into Tycho-2 with 2.5 million stars could also have been made. Furthermore, I found on the internet that the median accuracy would be about 0.07" for UCAC2 while the 0.04" shown in Jos' diagram may apply to brighter stars.

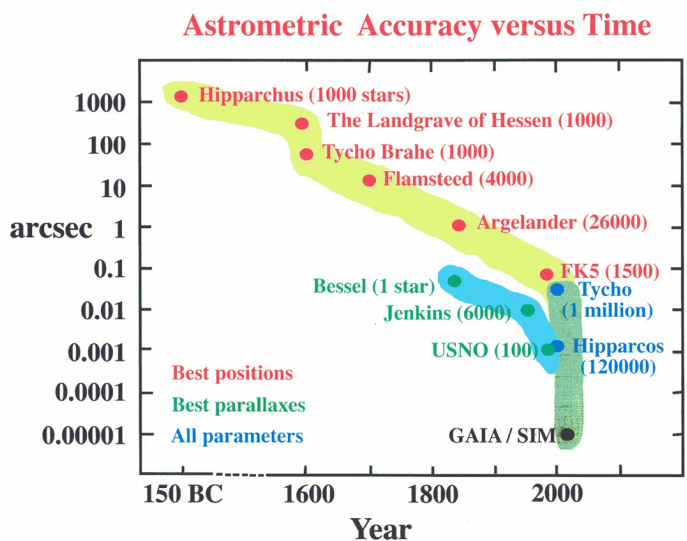

*Fig. 6. Accuracy diagram: ESA-1998*

**ESA-1998**

This diagram with colours was produced by Michael Perryman for a technical presentation around 1998.

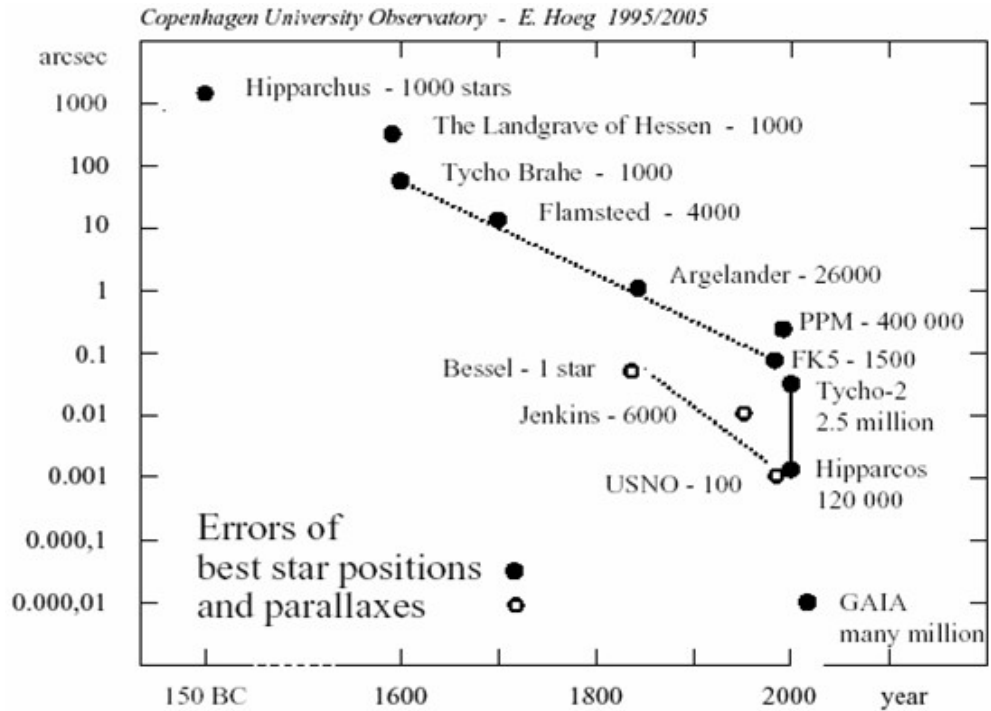

Errors of star positions in the most accurate catalogues. Tycho Brahe achieved a jump in accuracy through the first "big science" in history. After four centuries with more gradual improvement another much larger jump in accuracy is obtained by the ESA satellite giving the Hipparcos and Tycho-2 Catalogues containing a total of 2.5 million stars. - Parallaxes are also measured by Hipparcos and GAIA with the same accuracy as positions.

*Fig. 7. Accuracy diagram: Høg-1995/2005*

**Høg-1995/2005**
This is a modification from 2005 of the original from 1995, Tycho-2 is now included.

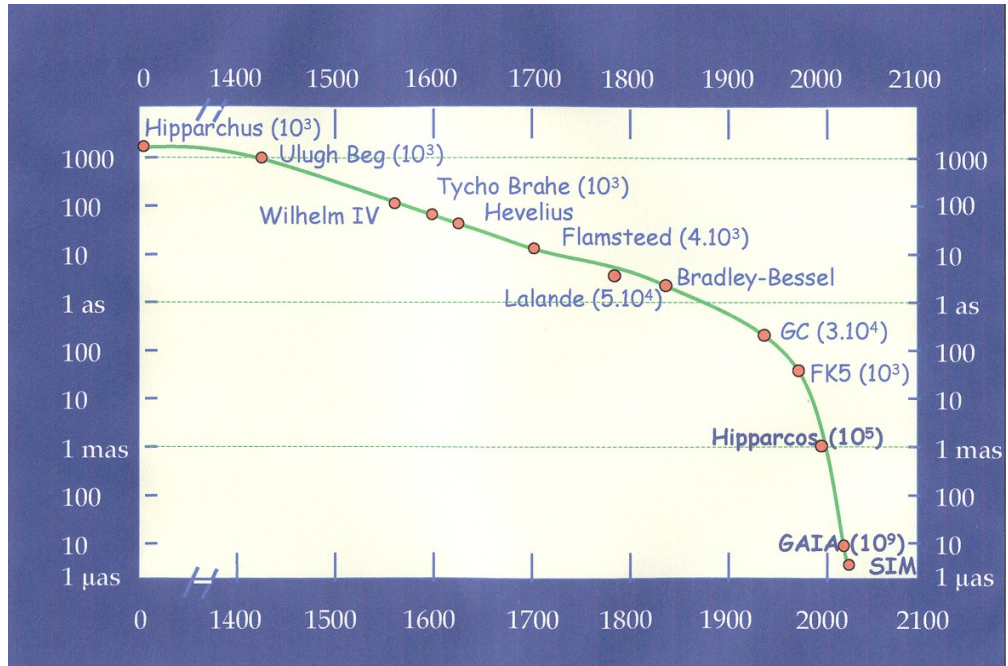

*Fig. 8. Accuracy diagram: Turon-2007.*

**Turon-2007**
This diagram was used by C. Turon in Shanghai 2007 during the presentation Turon & Arenou (2008). She kindly gave me the diagram and she wrote on 8 Nov 2007: "The diagram of p22 is one which has been used and modified by so many people .... I do not know whose original idea it is. If my memory is correct (I cannot check as I do not have this document at home), one version of this graph is in the "Hipparcos phase A report". An updated version is in the Gaia information sheet "Astrometric Accuracy Assessment" (with no reference either). It is why I did not put any "courtesy by". And I do not want that it is quoted as "Courtesy by C. Turon" as this is clearly a collective work. We will re-check the position of each of the points." Later on Turon has add that this graph, originally by Mignard, should be put into its context in Shanghai where it was briefly shown *only* for illustration, just to show the drastic improvement provided by space astrometry, not as a careful historical work.

According to Arenou (2008), the same diagram has been used in presentations by F. Mignard (18 May 2004) and by S.A. Klioner (31 March 2006), and it has originally been made by Mignard. The two presentations are here:
http://www.ari.uni-heidelberg.de/gaia/arc-of-current-t/mignard_gaia_ari.ppt
http://www.jb.man.ac.uk/ska/gravmeeting06/talks/klioner.ppt

The accuracies are read from the plot as follows: Hipparchus 1500", Ulugh Beg 1000", Wilhelm IV 120" Tycho Brahe 60", Hevelius 40", Flamsteed 12", Bradley-Bessel 2", Lalande 3", GC 0.2", FK5 0.04", Hipparcos 1 mas, GAIA 0.01 mas, and SIM 0.005 mas.

Some inconsistencies are noted: The point for Hevelius is misplaced at 1620 but belongs at 1670, Bradley-Bessel is misplaced at 1830 but belongs at 1755 when the observations were made. GC is placed at about the date of publication 1937 with 0.2", but at that time the error was 0.37". FK5 is placed at 1970, but the 0.04" corresponds to the mean epoch which was 1950. If these things were corrected the dots would move away from the smooth curve on which they are presently lying.

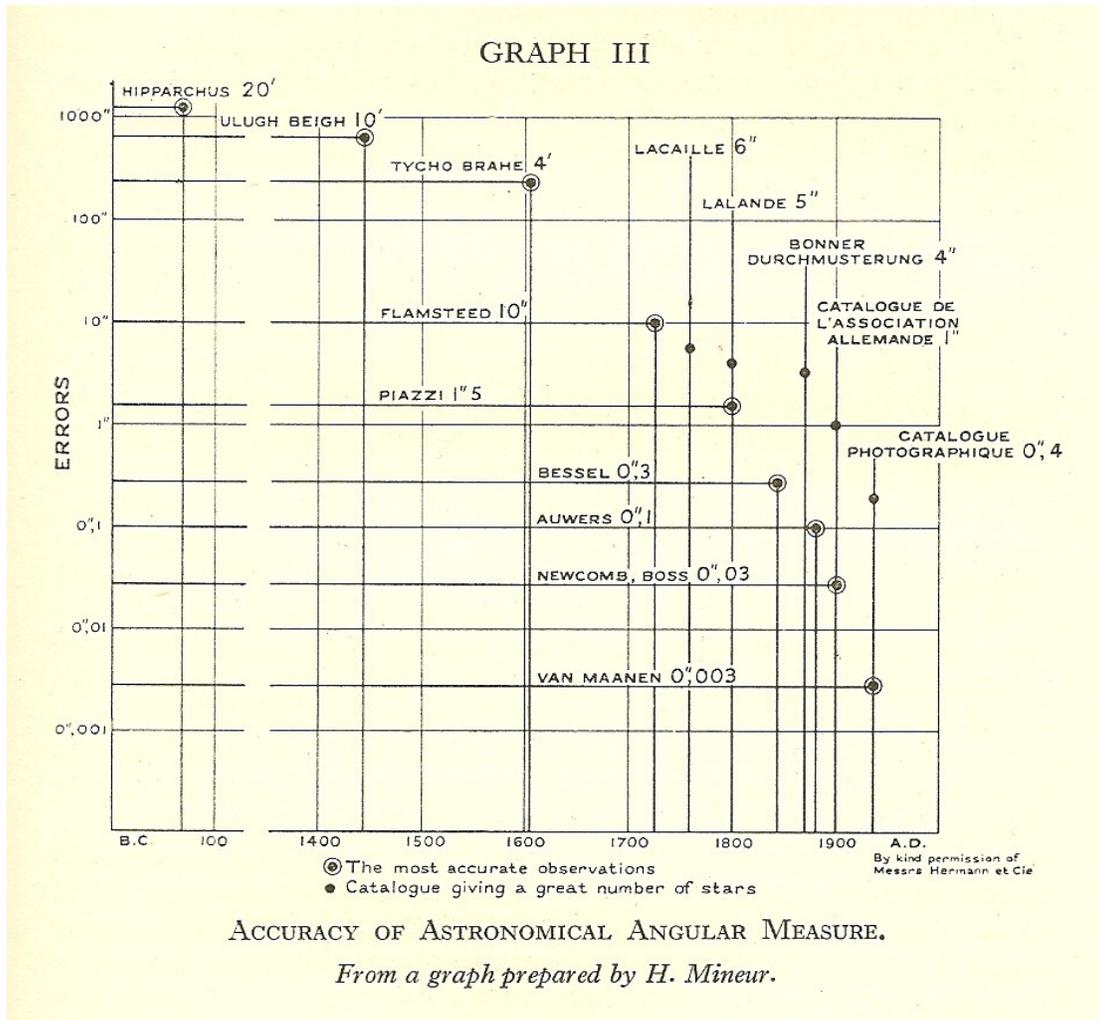

*Figure 9. Accuracy diagram: Mineur-1939.*

**Mineur-1939**

I am grateful to Professor David W. Hughes for drawing my attention to this diagram in Pledge (1939) in a mail of March 2008. I had asked him whether he knew the diagram shown here as Hipparcos-1989. David Hughes has no recollection of this specific graph from the Hipparcos publication, but he did produce a very similar graph for his history of astronomy students at the University of Sheffield, and this might be where the author (of Hipparcos-1989) got the idea from, he concludes.

The diagram is shown on p.291 in Pledge (1939), facing a page where the first measurement of stellar parallaxes about 1838 is mentioned on just six lines. No reference to the diagram is made in the text and no explanation is given other that on the graph itself. The reader can see the general trend towards better accuracy by a factor of 100,000 since Tycho Brahe, but some numbers are rather strange. The value 0.3" for Bessel is plotted as "most accurate observation" and probably means his parallax, but his measurement of the parallax had 5 times smaller error. Hipparchus is included with 20' which is his error within constellations while the error over the sky is 1 degree (see above). For van Maanen is given 0.003" which must be for relative astrometry, perhaps for his infamous measurements of proper motions in the Andromeda galaxy. I do not consider any specific number in this diagram as very trustworthy, but it should be credited as the first known attempt to make such a diagram.

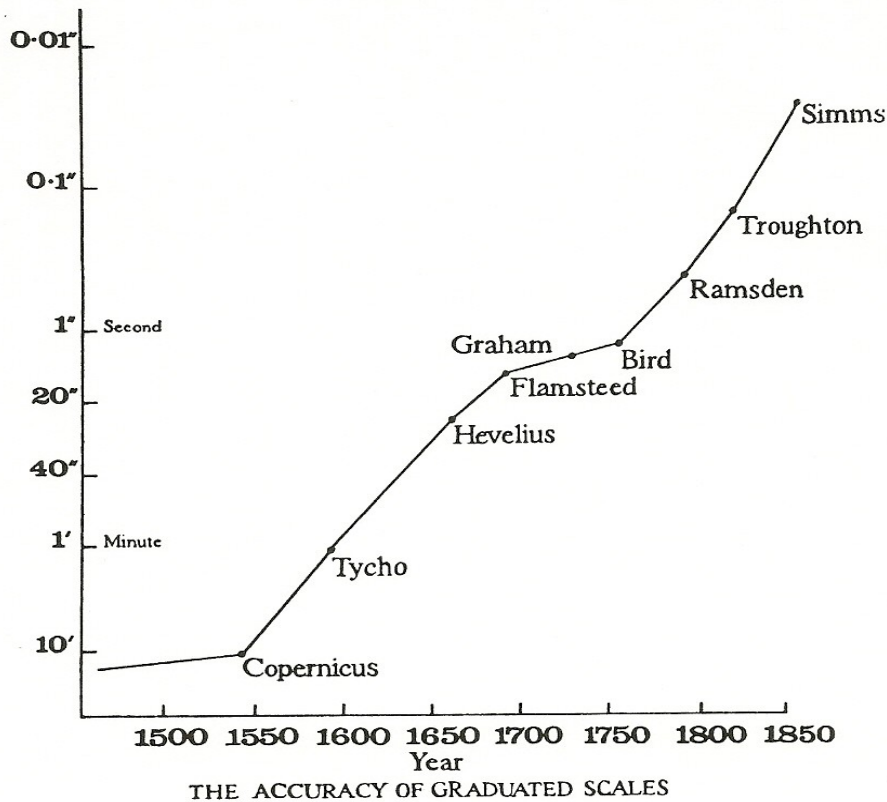

THE ACCURACY OF GRADUATED SCALES

*Fig. 10. Accuracy of graduated scales: Chapman-1983.*

**Chapman-1983**
I found of this diagram in May 2008, and that suddenly made me understand the whole confusing story about the diagrams. I have now understood where the misunderstanding came in, but I do not know who made the mistake, and I do not try to find out.

The diagram by Chapman (1983) was prepared independently of the one by Mineur, Chapman writes. It shows the accuracy of *graduated scales,* as stated on the graph and explained in the accompanying text.

Nevertheless, this very diagram has been taken to mean the accuracy of star positions in the first diagrams used for the Hipparcos mission in 1985 and onwards, but many other errors than that of the graduated scale enter in an astrometric observation. For the use with Hipparcos the diagram has been turned upside down. More essential changes are that Hipparchus is included in Hipparcos-1985, the points for Hevelius, Graham and Bird have been omitted, and of course Hipparcos is included with 0.002" which was the expected accuracy in 1983.

The value of 0.025" for Simms at 1850 represent, according to Chapman, the precision of reading a divided circle with six microscopes and taking the average. This does probably not take division line errors into account which can be much larger. By 1850 the error of a position in a catalogue was about 1", e.g. Argelander in Høg-2008, thus 40 times larger than the error from reading the circle.

In Hipparcos-1985 the point for Ramsden is placed at 0.9" while it is here at 0.4". But there can be no doubt where the points in Hipparcos-1985 came from.

Contribution to the history of astrometry No. 8            25 November 2008

# 400 Years of Astrometry: From Tycho Brahe to Hipparcos

*Erik Høg, Niels Bohr Institute, Copenhagen*

ABSTRACT: Galileo Galilei's use of the newly invented telescope for astronomical observation resulted immediately in epochal discoveries about the physical nature of celestial bodies, but the advantage for astrometry came much later. The quadrant and sextant were pre-telescopic instruments for measurement of large angles between stars, improved by Tycho Brahe in the years 1570-1590. Fitted with telescopic sights after 1660, such instruments were quite successful, especially in the hands of John Flamsteed. The meridian circle was a new type of astrometric instrument, already invented and used by Ole Rømer in about 1705, but it took a hundred years before it could fully take over. The centuries-long evolution of techniques is reviewed, including the use of photoelectric astrometry and space technology in the first astrometry satellite, Hipparcos, launched by ESA in 1989. Hipparcos made accurate measurement of large angles a million times more efficiently than could be done in about 1950 from the ground, and it will soon be followed by Gaia which is expected to be another one million times more efficient for optical astrometry.

## Introduction

The prospects for astrometry looked bleak at the middle of the 20[th] century. If an astrometrist retired the vacancy was usually filled with an astrophysicist, and astrophysics was moving towards the exciting new extragalactic astronomy. But I did not feel any pressure from this trend when I studied in Copenhagen (1950-56) where both my teachers at the observatory, Bengt Strömgren and Peter Naur, were very familiar with astrometry, and it was natural to follow their advice. As a boy, I had read about Tycho Brahe and Ole Rømer, the two Danish heroes in astronomy, who both in fact worked on what is now called astrometry. Astrometric catalogues on the library shelves like the Albany General Catalogue (GC), the AGK2, and the Jenkins catalogue of parallaxes attracted me, though I did not of course know that 40 years later I should lead the construction of the Tycho-2 Catalogue (Høg et al. 2000). This catalogue has replaced all previous reference catalogues with its positions and proper motions derived from observations with the Hipparcos satellite and 100 years observations with ground-based telescopes. This presentation is focused on optical astrometry over the past 400 years. More details about the recent history of astrometry are given in (Høg 2008a, b, and c).

The term astrometry does not apply to astronomical measurement in general as the word suggests, but only to the measurement of positions on the sky of stars and other celestial objects. The position of a star changes with time due to its proper motion, to the parallactic motion created by the motion of the Earth around the Sun, and to the orbital motion in the case of a binary star. The radial velocity of a star along the line of sight as measured by the shift of lines in the spectrum is the third component of the space velocity, which is needed for many applications, but it does not belong to astrometry. It should be noted that the radial velocity can affect the proper motion of nearby stars to such extent that accurate measurement of the radial velocity can be obtained from observations of positions over long intervals of time, according to e.g. Lindegren et al. (2000). The term astrometry came into use to distinguish it from astrophysics, especially after the introduction of stellar spectroscopy 150 years ago and of atomic theory later on, which were used to analyse the spectra. For the two millennia prior to that, astrometry had in fact been the main task of astronomy. Astrometric observational data have been the basis for a deep astronomical understanding of stars, star systems, planetary motions, and the underlying physical laws.

I will follow the history of astrometric instruments from the introduction of telescopic sighting and wire micrometers in the 17[th] century, via the transit instrument and the meridian circle in the 18[th] to photographic astrometry in the late 19[th] and photoelectric astrometry in the 20[th] century, including satellite astrometry, and finally CCD detectors. Some of the main astronomical results obtained by optical astrometry during the centuries will be outlined.

Today, positions and proper motions are given as celestial coordinates in the main astrometric reference system which is the International Celestial Reference System (ICRS). This system, adopted by the International Astronomical Union, is defined by the radio positions of 212 quasars, supposed to have negligible proper motions. It is represented in the optical by positions and proper motions in the Hipparcos Catalogue, officially called the Hipparcos Celestial Reference Frame (HCRF). ICRS is a coordinate system in right ascension and declination (RA and Dec) very close to the previous adopted system which was called J2000.0. Tycho Brahe used a system of ecliptic longitude and latitude. Later on RA-Dec systems tied to the Earth's rotation axis were used although the change with precession and nutation had to be taken into account. With the adoption of the ICRS for all catalogues the astronomical application of catalogues is simplified since precession and nutation only need be taken into account in connection with the pointing of ground-based telescopes as a computed coordinate transformation. The transformations are computed by means of vectors, no longer by spherical trigonometry as was usual some thirty years ago.

The history of other aspects of astrometry deserves equal appreciation. Such aspects are mathematical methods of data reduction, computing techniques, electronic control of instruments, electronic data acquisition, accurate clocks, and machines for measuring photographic plates, but these could not be included here. Briefly, however, on computing: Tycho Brahe had used a method called *prosthaphaeresis,* which had been invented by the Arab mathematician Ibn Jounis in the 11[th] century. It replaced multiplication with the addition of trigonometric functions. Logarithmic tables came into use after they had been introduced by John Napier in 1614 and had been enthusiastically supported by Kepler. In the subsequent period of over 300 years astronomical formulae were developed in logarithmic form to facilitate calculations, and books appeared with logarithms of seven and more decimals of trigonometric and many other functions. The time of logarithms had run out about 1950. We had used logarithms in school, but we were using mechanical and electrical calculators in astronomy in the 1950s. The first electronic computer, an IBM 650 with punched card in- and output, came to Copenhagen in 1954 and I took a programming course. Already two years earlier Peter Naur had told of his experience with the electronic computer EDSAC in Cambridge at a lecture in the university. The room was overfull, I was sitting with others on the floor, and even Niels Bohr had come to listen and to wonder at the fantastic punched tapes with rows of holes for numbers, which Naur rolled out on the floor and then gave us freely. I was soon engaged to interpolate in Naur's ephemerides of the minor planet 51 Nemausa computed with EDSAC, using Leslie Comrie's interpolation tables and a mechanical calculator.

## Tycho Brahe's legacy: Instruments and Newton's theory

Tycho Brahe had set a new standard for astronomical measurements by his twenty years of work on Hven. He improved the then classical instruments, the sextant and the quadrant, to give positions of stars with errors some five times smaller than those of his contemporary colleague Wilhelm of Hesse, i.e. about one minute of arc. Tycho made many improvements and had the means to carry them out thanks to a lavish support by the Danish king, Frederik II. The king wanted to promote science and saw that Tycho Brahe complied with this ambition. Tycho received the island of Hven close to Copenhagen in 1572 and then enjoyed a support equivalent to one or two per cent of the king's annual income, altogether the value of "a barrel of gold". With this basis he could make instrumental improvements and obtain observations in amounts and qualities never seen before.

Tycho had learned from his experience with the large quadrant, which he built in Augsburg 1570, that size alone was not a safe way to better accuracy. The quadrant was not stable, exposed to rain and wind as it stood, and it tipped over in a storm after a few years. Moderate size would lead to more accurate manufacture which was another of Tycho's achievements. For instance, on Hven he developed the sighting device slits-and-plate (Fig. 1) which became the preferred tool for his followers until the telescopic sight could take over in about 1660.

In Danzig, Johannes Hevelius (1611-1687) continued to use the slits-and-plate sighting device until his observatory was destroyed by fire in 1679. His observations, accurate to about 15"-20", could compete easily with any of the early telescopic observations, and were not seriously rivalled until Flamsteed and Rømer had developed their own observing techniques after Hevelius' death (Chapman 1990).

Tycho left a catalogue with positions of 1000 stars, the same number of stars as in the famous catalogue of Ptolemy. It remained unsurpassed for a century, until Hevelius' catalogue with 1564 stars appeared posthumously in 1690. Tycho's observations of the planets, especially of Mars, came to make the strongest impact on science and civilization. Johannes Kepler derived the three famous laws of planetary motion: about the elliptical form of the orbits, the speed in the orbit, and the size and period of all orbits in the system of planets. The three laws, completed in 1619, found an explanation in Isaac Newton's *Principia* of 1687 in terms of the universal gravitational attraction between masses and the laws of mechanics concerning force, velocity, and acceleration. These laws became the basis for the subsequent theories of celestial mechanics for planets and satellites, and for the technical revolution with engineering of machines and buildings.

## 17[th] century: Telescopic sight

The Galilean telescope magnified the object: it increased the angular resolution beyond the several minutes of arc of the unaided eye. This enabled Galilei in 1609-10 to see mountains on the Moon, satellites around Jupiter, the changing form of Saturn, spots on the Sun, phases of Venus, and to resolve the Milky Way into individual stars. The angular resolution opened up a new view of the physical nature of the heavens, simply by watching what you could see inside the quite small field of view. But this type of telescope was unsuited as a sighting device; it could therefore not be used in classical astrometry concerned with the measurement of large angles between stars.

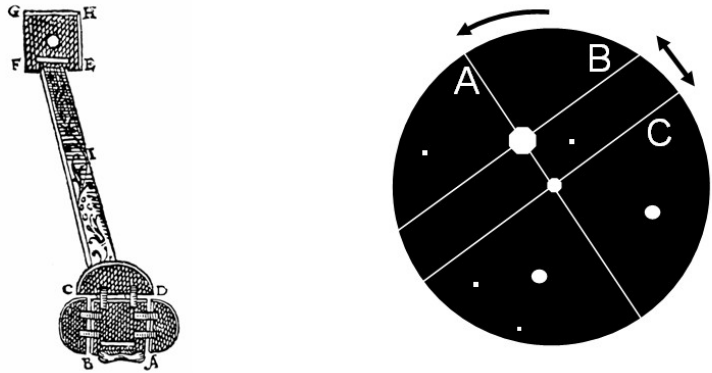

**Fig. 1 Left:** Tycho Brahe's sighting device, after 1570, slits-and-plate. **Right:** Wire micrometer, after 1660. The observer centers one of the stars on the A-C cross of fixed wires by moving the telescope. He turns the micrometer to place wire A on both stars, a scale shows the position angle. He moves wire B to the star and obtains the separation

This problem could be solved with the telescope invented by Kepler in 1611 in which a convex lens is used at the eye, instead of the concave lens in Galilei's telescope. The convex eye-lens is placed with its focus where the star images are formed by the front lens. This allows a cross hair to be placed at the common focus of the objective and the eyepiece lens, thus defining a line and then one has a sighting device. This invention was described in a letter by the English amateur astronomer William Gascoigne in 1641, in which he also tells how the cross hair can be made visible at night by illumination (Chapman 1990). Gascoigne developed the filar micrometer on the principle shown in Fig. 1, and he made a few observations in 1640 of the angular diameters of Jupiter, Mars, and Venus and of the angular separation of stars in the Pleiades cluster. But Gascoigne's work on the telescopic sight (and on the filar micrometer) fell into neglect after the inventor's death in the Civil War in 1644.

Christiaan Huygens independently invented the filar micrometer in about 1660 and made accurate measurements of the diameters of the Moon and all the planets. The invention was made independently in Italy and by 1675 the telescopic sight with a cross wire was in common use. The micrometer became a standard tool for measuring angles within the field of view during more than 300 years, manufactured in ever more accurate versions. This was the birth of s*mall-angle astrometry* which soon achieved accuracies of one second of arc and even better, while the accuracy of *large-angle astrometry* with quadrants and later on with meridian circles improved more slowly.

The regular swing of a pendulum was known to Galilei. To use it to control a mechanical clock was not simple, but Huygens succeeded in 1658. The pendulum clock in ever better versions was used by astronomers until the quartz clock could take over by 1950. An accurate clock was required in connection with observation of the transit time over the north-south meridian. In practice, the time was recorded when the star crossed one or several wires parallel to the meridian in the telescope field of view. Thanks to the regular rotation of the Earth this measurement corresponds to the right ascension coordinate of the star.

The other coordinate, the declination, could be measured when the telescope was precisely set to let the star follow a wire in the field. Reading the altitude angle of the telescope (cf. Fig. 2) on a finely divided circle was then required. The manufacturing of ever more accurate divided circles was a high art (Chapman 1990) and crucial for astrometry up to the 1990s when the reference systems of stars provided in the Hipparcos and Tycho-2 catalogues made the accurate circles virtually obsolete.

Flamsteed in Greenwich used mainly a sextant with two telescopic sights from 1676 to 1690. He then built a mural "quadrant", but of 140 degrees arc so that he could directly observe the Pole Star. This instrument served him for the Great Catalogue and the last recorded observation was made in 1719, shortly before his death.

## 18th century: Quadrants prevail

Astronomers of the new century would learn that the Earth's motion around the Sun makes all stars write an ellipse on the sky in one year and that the Earth's axis wobbles. They realized that the fixed stars are moving, that the Sun moves among the stars, and that some stars orbit each other. They learned that comets return, and a new planet, Uranus, was found. These discoveries were results of measuring and mapping the stars ever more accurately and in ever larger number, and of recording the motion of the objects in the solar system. The latter observations were studied in celestial mechanics, i.e. developments of the theory of gravitation, and the scientists were challenged by the planets, and especially by the intricate motion of the Moon.

In 1718, Edmund Halley had discovered that fixed stars are moving since he found that three bright stars had changed their position since the ancient Greeks. On the same occasion Halley stated that the stars were at least 20,000 or 30,000 times as distant as the Sun. That could be concluded from the lack of positive evidence of a parallax in the most accurate position observations of the time. Already in 1748 James Bradley could extend this lower limit to at least 400,000 times since the failure to measure annual parallax with his precision instrumentation showed that it must be less than half a second of arc.

Astronomers had been attempting to measure the annual parallax of a star since Copernicus in 1543 had proposed that the Earth orbits the Sun in one year. Bradley was one of those and had a telescope built, a zenith sector, which could measure the position of stars near the zenith. The small angles to be measured and the stable mounting gave a high accuracy about one second of arc, though only in the north-south direction. His measurements in 1725 showed a shift, but not in a direction that could be explained by the shift of the Earth relative to the Sun. Bradley gave the true explanation which was that the velocity of the Earth in its orbit causes the direction to a star to be shifted forward in the direction of the Earth velocity. This was a new effect due to the very large, but finite velocity of light which had been discovered in 1675. The effect is called aberration and amounts to 20 seconds of arc for a given star at some times of the year. This is a large amount, and aberration was taken into account in all subsequent astrometry, resulting in much better accuracy of star positions.

With the higher accuracy Bradley was able to discover a smaller effect; the stars wobble with a period of 18 years and an amplitude of nine seconds of arc. In 1748 he explained the effect as a wobble of the Earth rotation axis due to a variation of the Moon's orbit.

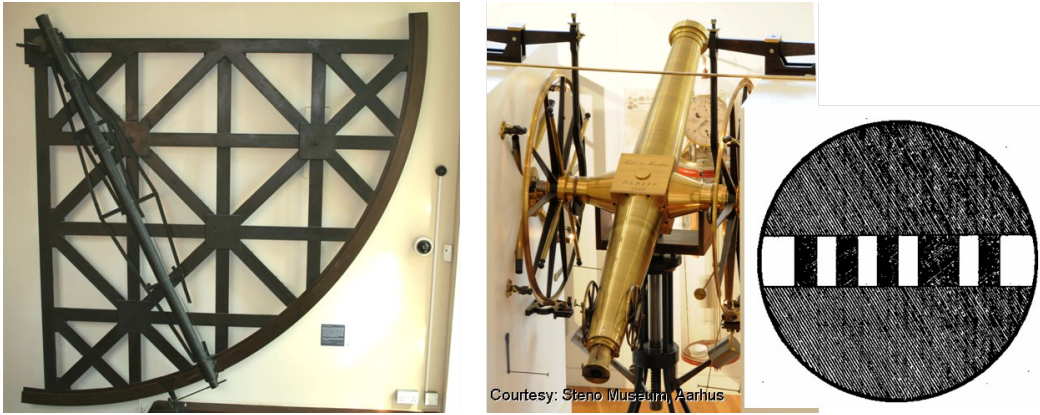

Courtesy: Steno Museum, Aarhus

**Fig. 2   Left:** The John Bird quadrant of 1773, Greenwich. **Right:** The Copenhagen meridian circle from 1859 and a grid used for experiments in 1925 with photoelectric recording of star transits

The quadrants were still dominating astrometric observations in the 18th century, but a different type of astrometric instrument was being developed. Already in 1675 Ole Rømer had introduced the transit instrument, which he set up in his Copenhagen residence in 1691. It consists of a telescope mounted perpendicularly on an axis, and the ends of the axis are placed on pivots in east and west. The telescope is pointed at a star before it crosses the meridian, and the transit time is measured, just as at a quadrant with telescopic sight. In 1704, 30 years after he had the first idea (Herbst 1996), Rømer set up a similar instrument, but now with a divided circle also mounted on the axis. Rømer introduced fixed microscopes to read the circle, thus obtaining the declination coordinate. This was *Rota Meridiana,* the meridian circle (cf. Fig. 2), which is called a transit circle in English-speaking countries, and it became the most accurate instrument for measuring large angles on the sky up until 1990. But it took a century before the meridian circles took over, for reasons explained in the following section. Rømer was motivated by the search for parallaxes and believed for a while that he had succeeded in finding one.

All Rømer's observations were destroyed in the Copenhagen city fire in 1728, except those from three nights, 20-23 October 1706, called *Triduum,* of which several copies had been made and placed at different locations. The external standard errors of the mean positions were 3.4" in RA and 4.5" in declination (Nielsen 1968). Tobias Mayer in Göttingen chose 80 of the stars from *Triduum* for re-observation and derived proper motions by comparing with Rømer's observations in 1706, with his own and with Lacaille's observations in 1750 and 1756. His aim was to see if there was a systematic motion indicating that the Sun was moving towards any specific part of the sky. In 1760 he concluded that this was not the case. In 1783, however, William Herschel concluded from the motions of only 13 stars that the Sun was moving towards Hercules. Innumerable studies of this phenomenon, the Solar Apex, and of other systematic motions of stars were to follow whenever a new set of proper motions became available. - Incidentally, William Herschel used the motions from Mayer, without mentioning Mayer or Rømer, but only that he had the motions from a book by Jerome Lalande (F. Mignard 2008, private comm.).

Halley computed parabolic orbits of 24 comets from the preceding three centuries and in 1705 drew attention to the fact that three of them had nearly identical orbits in space and that they were therefore successive appearances of the same comet. He predicted its next return in 1758. In the popular mind comets had always been portents of disaster, while even to astronomers the nature of comets and their role in the cosmic order were still shrouded in mystery. The year came and Halley's Comet reappeared, spectacularly for everybody to see, and a triumph for astronomy, showing that the events of the world are predictable, as many in fact believed.

Bradley left the raw data of a large number of observations. They had been made from 1750 until his death in 1762 with a transit instrument for the right ascensions and a mural quadrant for the declinations. Since Bradley had carefully recorded temperatures and various calibrations the observations were considered to be very valuable and they were printed in full between 1798 and 1805. Then Friedrich Bessel took the reduction in hand and produced a good catalogue. Later in the 19$^{th}$ century A. Auwers improved the reduction further, resulting in a catalogue of over 3000 stars. Brosche & Schwan (2007) have shown by direct comparison with the Hipparcos catalogue that the uncertainty in both coordinates for a subset of 2450 entries is only 1.1 arcsec. This catalogue was used to derive proper motions even for the General Catalogue published in 1937 and for FK5 published 1988. Much larger star catalogues were published from observations later in the 18$^{th}$ century, especially the one of 50,000 stars by Jerome Lalande in *Histoire Céleste Francaise*.

William Herschel wanted to measure parallaxes and thought he could do it by measuring pairs of a bright and a faint star with his wire micrometer. Their separation would change with the time of year since the brightest star was closer to Earth than the faint one, so he believed to begin with. He knew that to measure the small separation in the field of view would be much more accurate than to measure large angles with a quadrant as others were doing. Hence he started to examine all the bright stars attentively with his 7-foot (focal length) telescope, to see whether they had faint companions nearby. On a night in 1781 he noticed a bright star that appeared larger than the others, about 4" diameter, which he could see because of the high magnification and the excellent quality of his self-made mirror telescope. He suspected it to be a comet, but it was a new planet, later called Uranus. A new planet beyond Saturn was a world sensation. It was followed by astrometric position measurements in the following years, and it was found that its position had already been measured in 1756 and 1690, but it had been taken for a star.

Herschel's survey of companions to the bright stars begun in 1779 turned out to be important in itself. The large number of cases in which bright stars had a close companion, faint or bright, far surpassed what could be expected through chance distribution of stars. This had already been pointed out in 1767 by the Rev. John Michell. Herschel soon thought of these stars as real binary systems. Twenty years later he again measured the relative position of some of his stars and found fifty pairs where the position angle had changed by between 5 and 51 degrees. Thus began double star astrometry by which the masses of stars were to be determined in great number as time allowed the stars to revolve so much that an orbit could be calculated. The first orbit was obtained for the binary *Xi* Ursae Majoris by Felix Savary and published in 1827.

## 19$^{th}$ century: Finally parallaxes and meridian circles

Three men in the previous century, Rømer, Bradley, and Herschel, have been mentioned as motivated by the parallax question, but without the final success. Their efforts had however far-reaching consequences: The transit instrument and the meridian circle were invented; first-epoch positions were obtained in 1706 from which proper motions were derived in such number that the solar apex motion could be discovered in 1783; and double-star astrometry was begun in 1781 leading to the discovery of physical binaries and then orbits.

By 1840 parallaxes had been measured by three men, simultaneously and independently: Friedrich Bessel, Wilhelm Struve and Thomas Henderson, who had all been observing stars with large proper motions since that was now considered to be an indication of nearness. They published credible parallaxes for 61 Cygni, Vega, and α Centauri, respectively. John Herschel said: "It is the greatest and most glorious triumph which practical astronomy has ever witnessed", speaking as President of the Royal Astronomical Society when he awarded the gold medal to Bessel. Bessel received the medal because he had shown the reality of the parallax most convincingly by his analysis.

The instruments used were respectively a heliometer, a wire micrometer, and a mural circle which had reached the necessary perfection through decades of technological developments, not least through a co-operation with the astronomer, who by his demands drove the technician to persevere with improving the instruments. The heliometer was originally developed in 1753 by London's John Dollond, the same man who had marketed the world's first achromatic refractors. Its purpose was to measure the diameter of the Sun, Helios in Greek, hence the name of the device. It was also called a "divided-lens micrometer" since the objective of the telescope was cut along a diameter into two semi-circles which could be positioned very accurately when they were shifted along the diameter. The observer would see two superposed images of the star field, one from each half, and he could shift the images, thus measuring separations of stars along the diameter. Bessel had used an outstanding example of a heliometer built by Joseph Fraunhofer in Munich. Struve used the largest refractor in the world, also from Fraunhofer. Henderson had only an ordinary mural circle, far less accurate; but his observations afforded a large parallax for α Cen of 0.91" which has later been reduced, and it is 0.742" from Hipparcos observations in our time.

This initial success spurred further activity, and results for the first dozens of parallaxes allowed some general conclusions to be drawn, e.g. that there is such a large diversity of luminosity among the stars that one can be millions of time brighter than another. But further progress was hampered by systematic errors, as could be seen when results from observations with different instruments by different observers were compared. Introduction of photography towards the end of the century did not bring a significant improvement to begin with, also because of systematic errors.

Measuring the distance to the Sun, needed to convert the stellar parallax measures into kilometres, renders an equally fascinating story as for the stars. Kepler wrote in 1620 that the distance must be at least three times larger than the antique (van Helden 1985). Ptolemy's value was 1160 times the Earth's radius. Kepler's value corresponds to a solar parallax of one minute of arc, i.e. the "horizontal parallax", the angle subtended by the Earth radius as seen from the Sun. Publication by Flamsteed and Cassini in 1673 gave a value between 9.5 and 10", based on observations from a favourable opposition of Mars in the year before. This is approximately correct as we now know, and implies a distance to the Sun 20 times larger than the antique value; what a widening of the cosmos within a century! The following centuries brought a "struggle for the next decimal" by expensive expeditions to observe transits of Venus across the solar disc and by equally expensive observations of the minor planet Eros at opposition, as reported by Pannekoek (1961). The value was given in 1942 as 8.790" with a claimed accuracy of 0.001, and 30 years later 8.794,18" with an uncertainty of 0.000,05.

The new century began with a remarkable discovery by Guiseppe Piazzi of Palermo, who was using his alt-azimuth circle to assemble a star catalogue of greater accuracy than any of his predecessors. On New Year's Day 1801 he had measured a star that appeared to have changed its position when he measured again on a subsequent night. It turned out to be an object moving in an orbit between Mars and Jupiter, where a vast empty interval of distances from the Sun often had made astronomers wonder. By 1807 three more objects, Pallas, Juno, and Vesta, had been found in the belt of asteroids, as they were later called. By 1891, more than 300 asteroids had been found, and the pace of discovery then greatly increased with the application of photography.

Uranus and Ceres had been found quite unexpectedly during the study of stars. But in 1846 the large planet Neptune was discovered in an orbit beyond Uranus as a result of calculations based on many observed positions of Uranus. These positions deviated from the ones predicted by the mathematicians from Newton's law of gravity. J.C. Adams and U.J.J. Le Verrier were able to predict that an unknown planet could be responsible, and Neptune was found by J.G. Galle and H.L. d'Arrest in the predicted area of the sky on the first night. The discovery of Neptune was a spectacular and widely publicized victory of mathematical astronomy (e.g. Hoskin 1999).

Rømer's meridian circle of 1704 remained unique for a century during which mural circles were preferred for the measurement of star positions. But systematic differences between positions from different instruments of more than 10" were an increasing problem. The history of the meridian circle and how it became the preferred instrument by about 1820 has been studied by Herbst (1996) and Chapman (1990). The potential advantage of the full divided circle over the quarter circle had been realised by the astronomers, and the technical ability to manufacture and accurately divide the circle had come with the technical evolution. Achromatic objective lenses could now be made, but many technical issues played a role, and a rethinking of the astronomers took time. A transition period from 1780 to 1820 can be recognised; Piazzi's alt-azimuth of 1789, built in England, was the first instrument with a full circle, but it was in Germany that superlative meridian circles were first manufactured. It began with Repsold in 1802, then Reichenbach and Ertel, and now with full mechanical symmetry about the meridian plane which Rømer's instrument lacked. By 1820 continuous production of meridian circles was going on, and from 1850 the meridian circle had become the main instrument of an astronomical observatory.

Between 1859 and 1863, F.W. Argelander of Bonn, published a three-volume catalogue of 325,000 stars, known as the *Bonner Durchmusterung (BD)*, and a forty-plate atlas. It was the work of a tiny group of Bonn observers using a small refractor, and this survey was to prove immensely useful for observers for more than a century, but the positions were necessarily inaccurate. In 1867, therefore, Argelander proposed to the Astronomische Gesellschaft that a project be organized to measure, this time with great accuracy, the positions of the BD stars down to the ninth magnitude, and the work should be shared among observatories, each observing a zone of about five degrees in declination. The same idea had been expressed fifty years earlier by Bessel (1822), whose assistant Argelander had been at the time. The work started almost immediately, but proceeded slowly in some places, the last results appearing in 1910 for the northern declinations and in 1954 for the southern.

The slow progress with the meridian circles and the new photographic technique (the dry plate was invented in 1871) in 1885 led the director of the Paris Observatory, Admiral E.B. Mouchez, to suggest the possibility of a great photographic star chart, which became the *Astrographic Catalogue (AC)*. The long history of this great project includes the happy ending with a catalogue containing over 4.5 million star positions, published as AC 2000.2 by Urban et al. (2001) which was used as the major source of old positions to derive the proper motions of the 2.5 million stars in the Tycho-2 Catalogue.

## 20th century: Photography, radio astrometry, and a satellite

Photography was a very powerful astrometric technique during most of the 20th century, ending observationally about 1995 when photographic plates of the required quality were no longer manufactured. But the accumulated plates remain a valuable resource which is being exploited in catalogues with up to 1000 million stars. Combined also with CCD astrometry at new epochs, accurate proper motions are obtained. The idea of deriving absolute proper motions by differential, small-angle measures with respect to galaxies was proposed by Wright (1950) which lead to the NPM and SPM programs (Klemola et al. 1987). An account of the history up to 2008 of the observations and the resulting catalogues of positions and proper motions is given in Høg (2008b), including tables of selected catalogues. Present-day catalogues of astronomical and especially astrometric data are listed by Zacharias et al. (2004).

Meridian circles were able to measure large arcs on the sky and were therefore used to provide ever more accurate fundamental catalogues with positions and proper motions for about one thousand stars. Hertzsprung (1905) used proper motions from Auwers' Fundamental-Catalog (1879) as a measure of stellar distances, so-called secular parallaxes, when he discovered the dichotomy of red stars into giants and dwarfs; distances from trigonometric parallaxes were not accurate enough at the time.

Further reference stars were tied to the given fundamental catalogue by meridian circle observations, and by means of these stars the photographic plates could be reduced to a proper celestial reference system. Visual observation of the stars dominated meridian circle work during most of the century. It appears from Table 1 in Høg (2008b) that the error of a position in an observation catalogue improved from 0.9" in 1856 to 0.25" about 1910, and to 0.15" with the photoelectric observations in the Perth70 Catalogue (Høg & von der Heide 1976). The weight of this catalogue was ten times higher than that of Küstner's large catalogue of 1908, partly because Perth70 was observed in the many clear nights of Western Australia.

The CMC1-11 (1999) catalogues with the Carlsberg Meridian Circle on La Palma were observed with a photoelectric slit micrometer similar to the one used for Perth70, but with automatic control of micrometer and telescope, giving a much higher efficiency. Observed in the better seeing on La Palma and during 14 years instead of 5 years for Perth70 the weight of the catalogue is larger by a factor 30.

The accuracy of the CMC1-11 catalogues comes close to the limit set by atmospheric image motion for the measurement of large angles between stars, according to a formula by Høg (1968). Small angles can be measured much more accurately from the ground than large angles, according both to experience and to a formula by Lindegren (1980). The CCD astrometry, explained in Høg (2008b), gives better accuracy because the stars are referred to the dense set of reference stars in Tycho-2 within the small field of the CCD. The CMC14 catalogue was observed with CCDs, resulting in 0.034" accuracy from only two images of stars brighter than 13th mag. Altogether, the catalogue weight is 500 times higher with CCDs on the same meridian circle on La Palma than with the slit micrometer, in spite of the shorter observing period.

Photoelectric techniques came to revolutionize astrometry with meridian circles, and even more with the Hipparcos satellite by ESA, observing in the years 1989-93. The evolution of photoelectric astrometry from the experiments by Bengt Strömgren in 1925 (cf. Fig. 2) to the Hipparcos satellite has recently been presented by Høg (2008a). The satellite is described in ESA (1997), the optical system is shown in Fig. 3, and results are summarized in Høg (2008b) and in the following section.

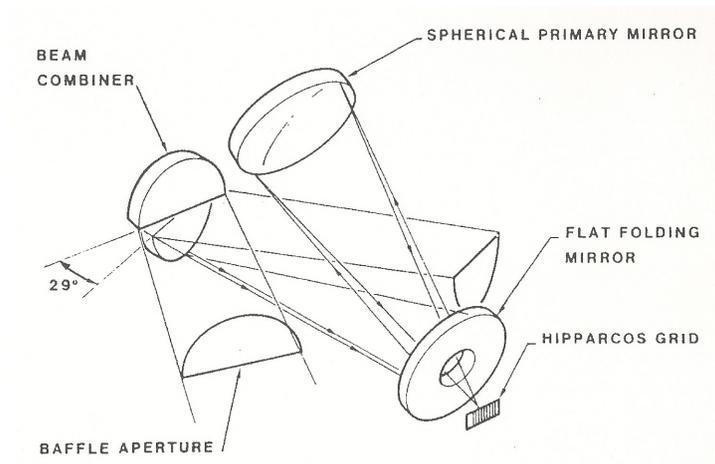

**Fig. 3** Hipparcos: Schmidt system with 29 cm diameter aperture, 1.4 m focal length and two viewing directions, all mirrors silver coated for maximum reflectivity. The satellite rotation makes the stars cross the modulating grid and the Tycho star mapper slits

Radio astrometry with high precision was introduced in about 1970 by Ryle (1972) and others. It has since played an important role through observation of quasars because they are extragalactic and therefore have very small proper motions, if any, i.e. the quasars represent a non-rotating celestial coordinate frame. A non-rotating frame has traditionally, including the fundamental catalogue FK5, been established dynamically by the requirement that position observations of the Sun and planets must obey Newton's laws. Radio astrometry of the static frame of quasars is ideal for following the complicated rotational movements of the planet Earth. A selected set of 212 quasars defines the International Celestial Reference System (ICRS) to which the stars in the Hipparcos catalogue have been tied (Kovalevsky et al. 1997).

## Astrometry during 400 years

Errors of star positions and parallaxes in accurate catalogues are shown in Fig. 4. This means the *median external standard error* per star and coordinate in a catalogue*, if available*. In most catalogues the positions of bright stars are more accurate than those of faint ones. The representative median error, dominated by faint stars, is given for most catalogues.

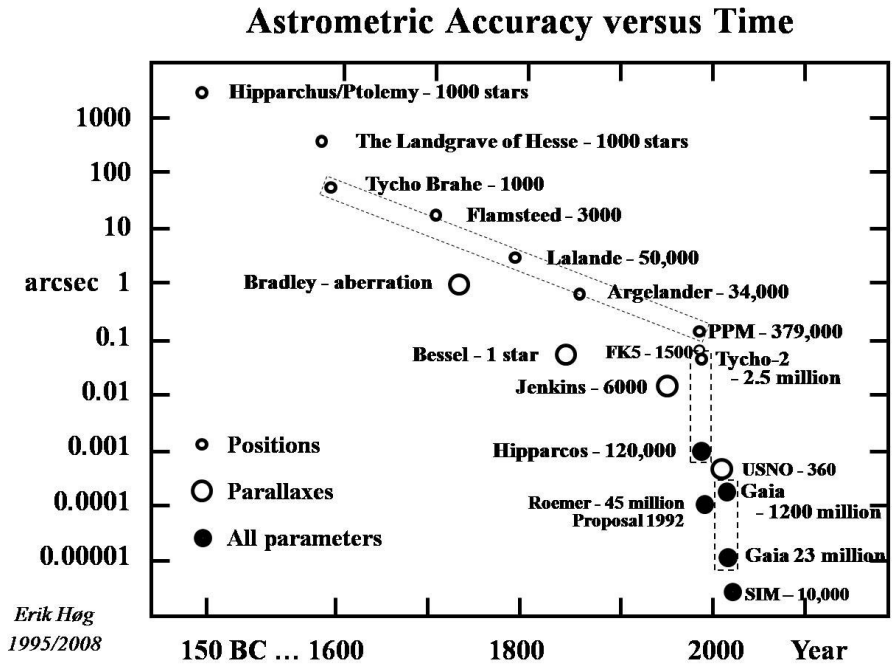

**Fig. 4**  Astrometric accuracy during the past 2000 years. The accuracy was greatly improved shortly before 1600 by Tycho Brahe. The following 400 years brought even larger but much more gradual improvement before space techniques with the Hipparcos satellite started a new era of astrometry

It appears that the Landgrave of Hesse was able to measure positions with errors about six minutes of arc, ten times better than Hipparchus/Ptolemy in Antiquity. A few years after the Landgrave, and thanks to generous support from the king of Denmark, Frederik II, Tycho Brahe reduced the errors by a further factor of six. The Landgrave and Tycho, both wanted to equal Hipparchus by reaching the same number of 1000 stars. A period of 400 years followed with gradual improvement of the accuracy as astronomers always made use of the best technical possibilities of their time, especially with better time-keeping equipment and accurate manufacturing of mechanics, optics, and with electronics. The accuracy was improved by a factor of about 250 in 400 years, i.e. a factor four per century, and the number of stars was greatly increased.

The introduction of space techniques, however, with the Hipparcos mission (Perryman et al. 1997 and ESA 1997) gave a veritable jump in accuracy by a factor of 100 with respect to FK5, the most accurate ground-based catalogue ever. Hipparcos obtained a median accuracy of 0.001 arcsec for positions, annual proper motions and parallaxes of 120 thousand stars. The positions even in the Tycho-2 Catalogue with 2.5 million stars are as accurate as the positions in FK5 containing only 1500 bright stars. Tycho-2 includes proper motions, derived from Tycho-2 positions and more than 140 ground-based position catalogues, but no parallaxes. The median standard error for positions of all stars in Tycho-2 is 60 mas, and it is 7 mas for stars brighter than 9 mag. The median error of all proper motions is 2.5 mas/yr.

The points marked "parallaxes" might be labelled "small-angle astrometry" or "relative astrometry", and all ground-based measurements of parallaxes are of that kind. This is about ten times more accurate than large-angle astrometry which was required to measure the positions shown in the diagram. The first such point is "Bradley – aberration" shown at 1.0 arcsec, the accuracy which Bradley obtained for the constant of aberration with his zenith sector. The accuracy of ground-based parallaxes begins with Bessel's single star in 1838, followed by a factor 100 improvement in accuracy at the U.S. Naval Observatory in Flagstaff since about 1990 for faint stars.

"All parameters" means that about the same accuracy is obtained for annual proper motions, positions and parallaxes, as was in fact achieved with Hipparcos, for the first time in the history of astronomy. The Roemer proposal of 1992 (Høg 1993) introduced CCDs in integrating scanning mode in a space mission, instead of photoelectric detectors as in Hipparcos. Roemer promised a factor 10 better accuracy than Hipparcos for many more stars, and a development began which led to the Gaia mission due for launch in 2011. For Gaia an improvement by a factor of 100 over Hipparcos is predicted for the 23 million stars brighter than 14 mag, i.e. 10 microarcsec median error. The median accuracy is expected to be 180 microarcsec for the 1200 million stars in the Gaia catalogue brighter than 20 mag, much better than the accuracy of Hipparcos. The two dots for Gaia thus show the expected accuracy for bright and faint stars. Finally, in view of the expected Gaia results, studies are due about the scientific goals for ground-based optical astrometry after Gaia.

**Acknowledgements:** I am indebted to Adriaan Blaauw for the kind invitation to contribute to the symposium on this vast subject. Without the invitation I would never have engaged myself in this quite large undertaking. Comments to previous versions of the paper from F. Arenou, P. Brosche, A. Chapman, T. Corbin, D.W. Evans, C. Fabricius, F. Mignard, H. Pedersen, P.K. Seidelmann, C. Turon, S.E. Urban, W.F. van Altena, and N. Zacharias are gratefully acknowledged.

# 650 Years of Optics: From Alhazen to Fermat and Rømer

*Erik Høg, Niels Bohr Institute*

ABSTRACT: Under house arrest in Cairo from 1010 to 1021, Alhazen wrote his Book of Optics in seven volumes. (The kaliph al-Hakim had condemned him for madness.) Some parts of the book came to Europe about 1200, were translated into Latin, and had great impact on the development of European science in the following centuries. Alhazen's book was considered the most important book on optics until Johannes Kepler's "Astronomiae Pars Optica" from 1604. Alhazen's idea about a finite speed of light led to "Fermat's principle" in 1657, the foundation of geometrical optics.

## Ibn al-Haytham

The opinions in the Antique about light and how we see the objects around us followed Platon who understood light as rays emitted from the eye towards the surrounding. Since we see remote objects immediately upon opening our eyes the rays must propagate with infinite speed. Euklid said a hundred years later, around 280 BC, that light moves along a straight line, and he formulated the laws about reflection in a mirror. Heron from Alexandria proposed about 60 AD the general hypothesis that light takes the shortest path between two points, and on this basis he was able to reach the same results as Euklid.

In the Middle Ages the centre of natural sciences moved to the Arabic world where Alhazen formulated views about light which made great impact in Europe. Therefore his work and the effect on science in Europe should be mentioned at the present celebration of 400 years of telescopes. Alhazen's Book of Optics in a printed version from 1572 is found in the Leiden University Library in the Latin translation from the Kitab al-Manazir (Book of Optics). The ideas of light by Alhazen and the European "perspectivists" are explained and some pages from the book in Leiden are shown in the Figures 1 and 2.

Alhazen realized that light has its origin outside the observer, that the rays on their way hits the objects and we see an object when the rays from the object enter our eyes, an idea already proposed by Aristotle. Alhazen described the eye and its functioning, and he made mathematical descriptions of the properties of light. He proposed that light moves with finite speed, and that it moves more slowly in dense media. His astronomy was a theoretical attempt to fit the spheres of the celestial bodies into each other, a task he criticized his predecessors, especially Ptolemy, for not having solved.

Alhazen with the full Arabic name Abu 'Ali al-Hasan ibn al-Hasan ibn al-Haytham, or just Ibn al-Haytham, was born in Basra about 965 and travelled to Egypt and Spain. He worked in Cairo and died there in the year 1040. According to Steffens (2007) and Wikipedia (2008), he conducted research in optics, mathematics, physics, medicine and development of scientific methods. His main work, The Book of Optics, was written while under house arrest in Cairo during eleven years 1010-1021. According to Wikipedia, in his over-confidence about the practical application of his mathematical knowledge, he had assumed that he could regulate the floods caused by the overflow of the Nile. Ordered by the sixth Fatimid caliph, al-Hakim, to carry out this operation, he quickly perceived the insanity of what he was attempting to do, and retired in disgrace. Fearing for his life, he feigned madness and was placed under house arrest until al-Hakim died. During and after the arrest he devoted himself to his scientific work until his death.

According to medieval biographers, Ibn al-Haytham wrote more than 200 works on a wide range of subjects of which at least 96 of his scientific works are known. Most of his works are now lost, but more than 50 of them have survived to some extent. Nearly half of his surviving works are on mathematics, 23 of them are on astronomy, and 14 of them are on optics, with a few on other subjects. Not all of his surviving works have yet been studied.

## *Alhazen in Europe*

Some parts of the Book of Optics came to Europe about 1200, were translated into Latin, and had great impact on the development of European science in the following centuries. Alhazen's book was considered the most important book on optics until Johannes Kepler's "Astronomiae Pars Optica" from 1604. Surprisingly, Alhazens's book was almost unknown in the Islamic world until the 1320s, according to Denery (2005) from whom I am quoting in the following.

The Latin translation of the Book of Optics exerted a great influence, for example, on the work of Roger Bacon, who cites him by name, and on Kepler and Fermat. It brought about a great progress in experimental methods. His research in catoptrics centred on spherical and parabolic mirrors and spherical aberration. He made the important observation that the ratio between the angle of incidence and refraction does not remain constant, and investigated the magnifying power of a lens. His work on catoptrics also contains the important problem known as Alhazen's problem. Alhazen has sometimes been called the "father of optics" and "the first scientist".

The scientists discussing optics in Europe at those times are called "perspectivists" after Roger Bacon's book "Perspectiva" from about 1270, but the word perspective has here a very different meaning from that in the Renaissance art of painting. Perspective meant the science itself about seeing, and the perspectivists thought that optics gave a deep insight into how we get to know anything about the world. It begins with the emission of light and colours from the objects through air to the eyes and then to the brain. This view is based on Alhazen's book.

Roger Bacon, a Franciscan theologian, describes emission of light and colours as a "multiplication of species", which is in fact the title of one of his books, and an idea going back to Aristotle. The word species is explained as force or likeness, and species are considered as the source of all natural action and causation. The species is the real source to our sensing and intellectual understanding of the world. Through the species the surrounding medium is assimilated to the object. For example, a flame creates species in the surrounding air. These species heat the air and assimilate therefore the air to the nature of fire, but the air does not become fire, etc. etc., according to Denery (2005) pp.86-96. Bacon says that the visible object in its true nature and essence enters the eye and reproduces itself inside the eye.

A great hurdle for the understanding of how we see the objects was, with our words: the image formation in the eye. Light comes to our eyes from all directions, according to Alhazen, but how can the eye distinguish the directions to the various objects? Alhazen gave an answer which we know is wrong, but which Bacon adopted: the species multiply themselves in all directions, but only the species arriving perpendicularly to the surface of the eye are really sensed. The other species are refracted in the lens of the eye and neutralize each other. Clearly, the species must be understood in their remote historical context, and not as an "anticipation" of the modern photons.

Alhazen's ideas were not bettered until the time of Kepler and Snell; Willebrod Snell van Royen and René Descartes formulated the law of refraction mathematically in respectively 1621 and 1637. Pierre de Fermat, however, could not accept Descartes' justification or demonstration of Snell's law which was based on an analogy with mechanical phenomena. Based on Heron's ideas about the shortest path and Alhazen's assumptions about a finite speed of light, smaller in dense media, he formulated in 1657 "Fermat's principle" which expresses that light follows the path which takes the shortest time. Fermat then derived the law about light moving on straight lines, as well as the laws of reflection and refraction, thus perfecting the geometrical optics.

The legacy of Ibn al-Haytham in the Book of Optics about the different finite speeds of light spanned six centuries culminating with "Fermat's principle" in 1657 and, finally, the measurement of the finite speed in cosmic space by Ole Rømer in 1676.

**Acknowledgement:** I am grateful to John Heilbron for comments to previous versions of this paper.

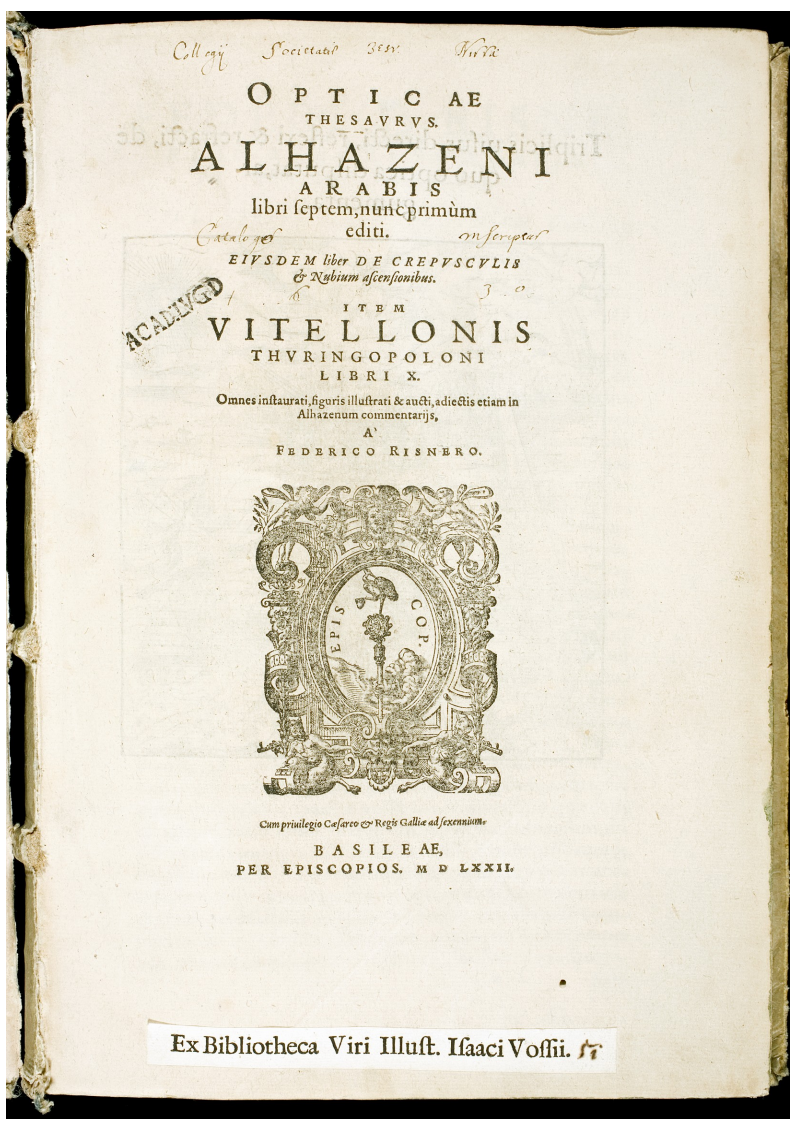

**Fig. 1** The title page of Kitab al-Manazir (Book of Optics) in Latin translation, a version printed in 1572 and found in the Leiden University Library

nes semicirculi protractas, id est ad lineas tales semidiametro propinquiores. Pòst secetur tabula circa semicirculum maiorem, ut solum remaneat semicirculus: & secetur tabula sub centro, ut centri locus acuatur quasi punctum: hoc tamen modo, ut in eadem superficie remaneat cum semiciculo & alijs lineis. Pòst sumatur tabula lignea plana excedens aeneam in longitudine duobus digitis : & sit quadrata : & eius altitudo siue spissitudo septem digitorum. Signetur ergo in hac tabula punctum medium: & super ipsum fiat circulus excedens maiorem circulum tabulæ æneæ, quantitate digiti magni: & fiat super idem centrum circulus, æqualis circulo minori tabulæ æneç: & diuidatur circulus maior in partes, in æqualitate respondentes partibus semicirculi tabulæ æneç: ut scilicet prima respondeat primæ, secunda secundæ, & sic de alijs : & circumquaque secetur tabula lignea, ut solum remaneat maior circulus: & fiet hæc sectio usitato secandi modo. Secetur etiam pars minore circulo contenta: & modus sectionis erit: ut huic tabulæ associetur alia tabula, ita ut linea à centro huius ad centrum illius transiens, sit perpendicularis super illam: & adhibito tornatili instrumento centris earum, fiat sectio partis circularis iam dictæ: ( est autem alterius tabulæ associatio, ut fixa stet in sectione) igitur restabit tabula quasi annulus circularis, cuius latitudo erit duorum digitorum : longitudo quatuordecim : altitudo septem.    Et sit hæc altitudo

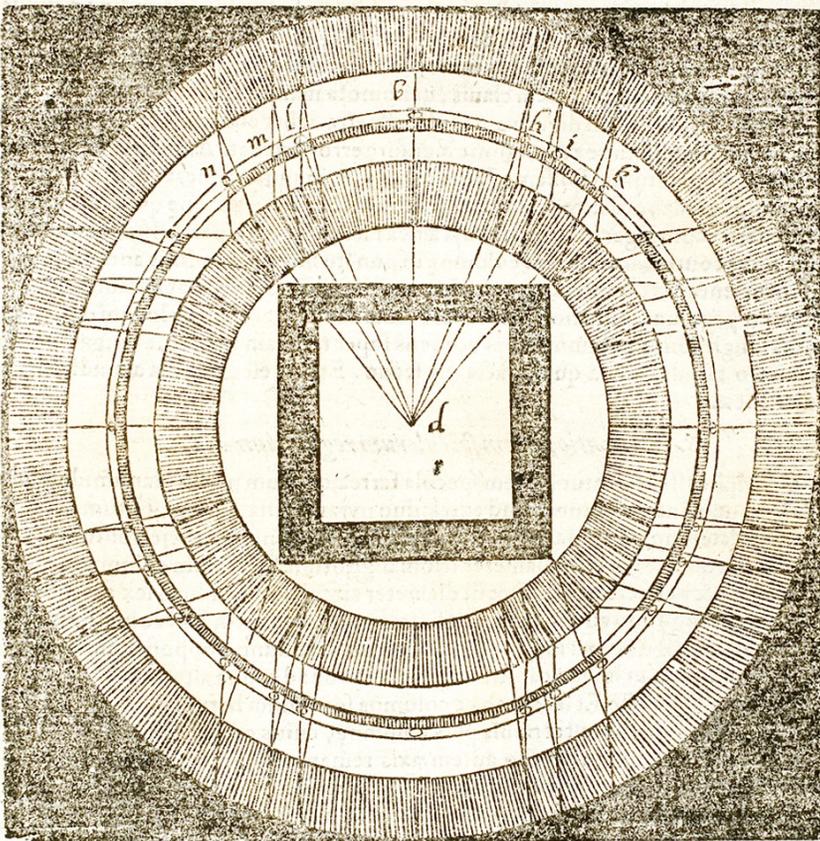

optimè circulata ad modum columnç: remanét autè in latitudine huius annuli lineç dividentes circulu eius secundum divisionè semicirculi tabulæ æneç. A' capitibus autem linearum harú producantur lineæ in superficie altitudinis exterioris, perpédiculares super superficiem latitudinis: & poterit hoc modo fieri. Quæratur regula bene acuta, cuius capiti lineæ adhibeantur, & regula moueatur, donec transeat superficiè altitudinis, in qua libet parte acuminis: Signa eius capita, & fac lineam, quoniam illa erit perpendicularis, quam quæris. Aliter poterit hoc idem fieri. Ponatur pes circini super terminú lineæ dividentis circulú, & fiat semicirculus secúdú altitudinè annuli, qui dividatur ne per æqualia, & protrahatur à punctò in punctú linea, & ita de singulis. Pari modo à termis

**Fig. 2** A page from the Book of Optics, selected mainly for its nice figure, the meaning of which, however, we do not understand. According to an expert in Latin, Dr. Christian Marinus Taisbak, the text speaks in great detail about a bronze plate and a wooden plate in which certain lines shall be drawn with great care, but for what purpose we could not deduce